\documentclass[12pt,a4paper]{article}

\usepackage{amsmath,amssymb,amsthm}

\usepackage[margin=1in]{geometry}
\usepackage{setspace}
\usepackage{pdflscape}
\usepackage{mathrsfs}
\usepackage{needspace}
\usepackage{multirow}
\usepackage{subcaption}

\usepackage{booktabs}
\usepackage{enumitem}
\usepackage{threeparttable}
\usepackage{graphicx}

\usepackage[round]{natbib}
\usepackage{xcolor}
\usepackage[colorlinks=true,linkcolor=blue,filecolor=blue,urlcolor=blue,citecolor=blue]{hyperref}

\hypersetup{
  pdftitle={Identification and Inference of Bi-directional Causal Effects in Mendelian Randomization without Instrument Selection},
  pdfauthor={Liyuan Mo and Qingyuan Zhao}
}

\newtheorem{theorem}{Theorem}
\newtheorem{proposition}{Proposition}
\newtheorem{lemma}{Lemma}
\newtheorem{assumption}{Assumption}

\theoremstyle{definition}

\numberwithin{equation}{section}

\title{Identification and Inference of Bi-directional Causal Effects in Mendelian Randomization without Instrument Selection}
\author{%
  Liyuan Mo\textsuperscript{*},
  Qingyuan Zhao\textsuperscript{\(\dagger,\ddagger\)}
}
\date{}

\begin{document}

\maketitle

\begingroup
\renewcommand{\thefootnote}{\fnsymbol{footnote}}

\footnotetext[1]{Department of Mathematical Sciences,
Tsinghua University, Beijing 100084, China.}

\footnotetext[2]{Department of Pure Mathematics and Mathematical Statistics,
Centre for Mathematical Sciences, University of Cambridge,
Wilberforce Road, Cambridge CB3 0WB, United Kingdom.}

\footnotetext[3]{Corresponding author.
E-mail: \texttt{qyzhao@statslab.cam.ac.uk}.}

\endgroup

\begin{abstract}
\small
\onehalfspacing

Mendelian randomization (MR) uses genetic variants as instruments to estimate causal effects. Bidirectional MR usually uses two reversed unidirectional analyses, but feedback can invalidate instruments in at least one direction and induce correlated pleiotropy. We model both traits as a stable cyclic system and estimate both directional effects from GWAS summary data without instrument selection or a prespecified valid-instrument subset. After LD score normalisation, marginal associations form a heteroscedastic noisy independent component analysis model, assuming the variants' direct effects on the two traits are independent. If those effects are non-Gaussian, this identifies the total-effect matrix up to scaling and permutation; no-self-feedback and stability of the system further resolve these ambiguities. We develop an estimator based on noise-adjusted second- and fourth-order moments, avoiding inversion of the high-dimensional linkage disequilibrium matrix and having computational cost linear in the number of variants. Under weak local dependence, we establish consistency, asymptotic normality, and validity of a chromosome-respecting moving block bootstrap. Simulations show negligible bias, generally reliable interval coverage, and higher rejection rates for nonzero directional effects than focusing-based tests.
Three real-data applications illustrate how joint analysis of feedback systems can yield causal decompositions differing from two separate unidirectional analyses.

\vspace{0.8em}

{\sloppy\noindent\textbf{Keywords:} Correlated pleiotropy; Feedback systems; GWAS; Independent component analysis; LD score.\par}

\end{abstract}

\section{Introduction}
\label{sec:introduction}

Mendelian randomization (MR) uses genetic variants as instrumental variables, or
``natural experiments,'' to infer causal effects from observational data
\citep{DaveySmithEbrahim2003,DaveySmithHemani2014,BurgessThompson2015,Sanderson2022}.
The increasing availability of genome-wide association study (GWAS) summary
statistics has made MR an important tool for studying complex traits and has
led to its widespread use in health and social science
\citep{Hemani2018,Richmond2022}. Over the past decade, substantial methodological
work has developed estimators robust to horizontal pleiotropy under assumptions
such as InSIDE (INstrument Strength Independent of Direct Effect) or
majority/plurality conditions
\citep{bowden2015egger,bowden2016weightedmedian,Kang2016,Guo2018,zhao2020statistical}.
MR has also been extended to multivariable settings
\citep{sanderson2019mvmr,wang2021causal}, mediation and time-varying settings
\citep{wu2024causal}, nonlinear exposure-outcome relationships
\citep{StaleyBurgess2017}, and within-family designs that reduce bias from population structure and familial effects
\citep{Brumpton2020,Tudball2022}.

Despite these advances, most MR methods prespecify an exposure-outcome ordering and select variants associated with the designated exposure as instruments. However, the causal direction for many pairs of biological or behavioural traits is uncertain a priori, and reciprocal effects are often scientifically plausible
\citep{DaveySmithHemani2014,RichmondDaveySmith2019,Sanderson2022}. A common
practice is to reverse the role of the two traits and fit two separate
unidirectional analyses, but this does not estimate one coherent
feedback system in general. The two analyses may use different selected variants,
significance thresholds and effective linkage disequilibrium (LD) structures. More importantly, a variant selected for one direction may primarily affect the outcome and be associated with the exposure via reverse causal pathway, resulting in an invalid instrumental variable and non-identifiability due to correlated pleiotropy \citep{morrison2020mendelian,wang2021causal,darrous2021simultaneous,LiYe2025}.
Direction-specific variant selection also poses methodological and practical 
challenges. Estimates may be affected by winner's curse or post-selection bias,
and conclusions can depend on the GWAS dataset used for SNP selection, significance threshold and
LD-clumping parameters
\citep{WangHan2021,MaWangWu2023,Jiang2023WinnerCurse,Swerdlow2016,BoehmZhou2022}.
For highly polygenic traits, selection can discard many weak but
collectively informative signals
\citep{boyle2017expanded,Visscher2017,zhao2019powerful}. Moreover, selection on observed marginal associations conditions changes the distribution of true genetic effects and can invalidate identifying assumptions like InSIDE in bidirectional analyses.

Several methods have considered causal orientation or explicitly bi-directional
relationships. Steiger-type procedures orient a putative one-directional
relationship by comparing explained variation
\citep{hemani2017steiger,XuePan2020}. \citet{wang2021causal} developed a test for causal direction by investigating a bimodal structure in a pseudo-likelihood function. \citet{darrous2021simultaneous} proposed a parametric model with a latent heritable confounder and estimated the structural parameters by maximizing a complex likelihood function. \cite{LiYe2025} developed a test for bi-directional causal effects by selecting a ``focused" set of instruments for each direction. Other work has studied identification and inference under  plurality-type, covariance heterogeneity, or valid-instrument-set conditions
\citep{XuePan2022,li2024discovery,Xie2024}. These approaches make
important progress, but identification typically remains linked to a
parametric effect distribution or to the existence of variants that behave as valid instruments in at least one direction.

\subsection{Our approach and contributions}

We briefly review the model used in this article and refer the reader to Section \ref{sec:setup} for further details. We model the two traits as a stable simultaneous system with structural coefficients matrix
\[
\mathbf B=
\begin{pmatrix}0&\beta_{12}\\ \beta_{21}&0\end{pmatrix},
\]
where \(\beta_{jk}\) is the direct causal effect of trait $k$ on trait $j$. We will refer to \(\mathbf A=(\mathbf I_2-\mathbf B)^{-1}\), which admits a Neumann-series expansion \(\mathbf A = \mathbf I_2 + \mathbf B + \mathbf B^2 + \cdots\) if $\mathbf B$ is (Schur) stable, as the ``total-effect" matrix. Let $N$ be the number of genetic variants. We assume the marginal GWAS summary data $\widehat{\boldsymbol \Gamma} \in \mathbb{R}^{N \times 2}$ ($\widehat{\Gamma}_{jk}$ is the GWAS effect of variant $j$ on trait $k$) satisfy 
\begin{equation} \label{eq:gamma-hat}
\widehat{\boldsymbol\Gamma}
 = \mathbf R\boldsymbol\alpha\mathbf A^\top+\mathbf E,
\end{equation}
where \(\mathbf R \in \mathbb{R}^{N \times N}\) is the linkage disequilibrium (LD)
correlation matrix, and \(\boldsymbol\alpha \in \mathbb{R}^{N \times 2} \) contains variant-specific direct
genetic effects, \(\mathbf E\) is the GWAS sampling error with a known multivariate normal distribution. A similar formulation was used by \citet{darrous2021simultaneous}.
The formulation \eqref{eq:gamma-hat} allows every variant to have direct effects on both traits and
does not require any particular variant, or any pre-specified subset, to satisfy
the conventional exclusion restriction. Instead, for identification we rely on the key assumption that the genetic variants have independent direct effects on the two traits (two columns of $\boldsymbol \alpha$ are independent); we term this the INdependent DIrect Effect (InDiE) assumption. Such independence does not exclude widespread pleiotropy but rules out systematic alignment of the direct genetic effects. It is related in spirit to the InSIDE assumption \citep{bowden2015egger} (and indeed reduces to InSIDE in uni-directional MR), but is imposed on the joint distribution of all variant-level direct genetic effects rather than on a selected set of instruments.

Our first main contribution is linking bi-directional MR with independent
component analysis (ICA) \citep{Comon1994,HyvarinenOja2000}. Under the InDiE assumption, the model in
\eqref{eq:gamma-hat} admits a noisy ICA representation, with $\mathbf A$ as
the mixing matrix and the LD-aggregated direct genetic effects as the latent
source signals. It is well known that the mixing matrix in ICA is identifiable up to
column rescaling and permutation if 
the source components are independent and at most
one is Gaussian. Building on the ICA-based
approach to identifying cyclic structural equation models developed by
\citet{Lacerda2008}, we show that these ambiguities can be resolved in our
bi-directional MR setting: the unit-diagonal constraint on
$\mathbf A^{-1}=\mathbf I_2-\mathbf B$ fixes the scaling, while stability of
the simultaneous system selects the correct permutation.

Note that \eqref{eq:gamma-hat} is not a standard ICA model: LD creates local dependence across variants and GWAS sampling produces measurement error. This creates new challenges: direct inversion of the
high-dimensional LD matrix $\mathbf{R}$ is numerically unstable, and existing ICA methods do not allow heteroscedastic noise. We develop novel solutions to these challenges by using LD-score normalization to stabilize the LD-aggregated genetic signals and extending the popular JADE algorithm \citep{cardoso1993blind} to this noisy ICA setting. Because we do not attempt to remove the local dependence of GWAS marginal associations, our algorithm does not require the full LD matrix and has a computational complexity that is linear in the number of variants. 

To quantify the uncertainty of our estimator of the structural coefficients in $\mathbf B$, we use a  moving block bootstrap \citep{lahiri2003resampling} that respects the genomic structure of the problem. 
We further develop an asymptotic theory that establishes the consistency and central limit theorem of the estimator and justifies moving block bootstrap under weak local dependence. In simulations, the proposed method 
has negligible bias and generally reliable coverage across a range of feedback, LD and signal-strength
settings. It also shows higher statistical power for directional effects than the directional focusing
procedures of \citet{LiYe2025} and is substantially faster than LHC-MR \citep{darrous2021simultaneous}.
Three real-data applications further demonstrate how joint estimation of
the simultaneous system can yield different conclusions from two separately
reversed analyses.

The rest of the paper is organized as follows.
Section~\ref{sec:setup} introduces the structural and summary-data models.
Section~\ref{sec:ica_formulation} establishes identification of the
bi-directional causal effects and develops the proposed estimator, including
recovery of the causal matrix from the ICA representation.
Section~\ref{sec:theory} gives the asymptotic theory and bootstrap
inference.
Sections~\ref{sec:simulation} and~\ref{sec:real_data} present the simulation
studies and real-data applications, respectively, and
Section~\ref{sec:discussion} concludes.
Additional numerical results, methodological details and proofs are provided
in the Supplementary Material.

\section{Bi-directional summary-data model}
\label{sec:setup}

Let $\mathbf T=(T_1,T_2)^\top$ denote the two traits and
$\mathbf Z=(Z_1,\ldots,Z_N)^\top$ the vector of centred and
standardised genotypes at $N$ harmonised variants, with the
variants ordered by chromosome and physical position.
We assume $\mathbb E(\mathbf Z)=\mathbf 0$ and
$\operatorname{Cov}(\mathbf Z)=\mathbf R$, where $\mathbf R$
is the LD correlation matrix.
The observed data consist of the marginal GWAS association
estimates for the two traits at these variants and their reported standard errors.

\subsection{Cyclic structural model}
\label{subsec:structural_target}

We model the traits by the simultaneous system
\begin{equation}
\label{eq:structural_model}
    \mathbf T
    =
    \mathbf B\mathbf T
    +\boldsymbol\alpha^\top\mathbf Z
    +\mathbf U+\boldsymbol\varepsilon,
    \qquad
    \mathbf B=
    \begin{pmatrix}
        0 & \beta_{12}\\
        \beta_{21} & 0
    \end{pmatrix}.
\end{equation}
The zero diagonal of \(\mathbf B\) encodes the no-self-feedback constraint, excluding direct causal effects of either trait on itself. The coefficient \(\beta_{12}\) is the causal effect from trait 2 to trait 1 and
\(\beta_{21}\) is the effect from trait 1 to trait 2. The matrix
\(\boldsymbol\alpha=(\alpha_{jk})\in\mathbb R^{N\times2}\) contains the direct
genetic effects before causal propagation, while
\(\mathbf U+\boldsymbol\varepsilon\) collects non-genetic and trait-specific
structural disturbances, respectively. 

The formulation in \eqref{eq:structural_model} is a simultaneous-equation model for a feedback system. We impose the following stability condition.

\begin{assumption}[Stability]
\label{assum:stability}
The bi-directional causal effects satisfy \(|\beta_{12}\beta_{21}|<1\).
\end{assumption}

Since $\mathbf B^2=\beta_{12}\beta_{21}\mathbf I_2$,
Assumption~\ref{assum:stability} ensures that
$\mathbf I_2-\mathbf B$ is invertible and that
the Neumann series converges:
\[
    (\mathbf I_2-\mathbf B)^{-1}
    =\sum_{m=0}^{\infty}\mathbf B^m.
\]
Consequently, the cumulative effects of successive
feedback remain finite, and the system has a unique
stable equilibrium
\citep{lauritzen2002chain,
hyttinen2012learning}.

The stability condition also distinguishes between
observationally equivalent structural representations.
When both $\beta_{12}$ and $\beta_{21}$ are nonzero,
the equations in \eqref{eq:structural_model} can be
rearranged to give an alternative representation in which
$\beta_{12}$ is replaced by $\beta_{21}^{-1}$ and
$\beta_{21}$ by $\beta_{12}^{-1}$.
This rearrangement involves exchanging and rescaling
the two structural equations. However, the product of the
bi-directional causal effects in the alternative
representation is $(\beta_{12}\beta_{21})^{-1}$, whose
absolute value exceeds one. The alternative
representation is therefore excluded by
Assumption~\ref{assum:stability}.

Solving \eqref{eq:structural_model} for $\mathbf T$
yields the reduced-form representation
\begin{equation}
\label{eq:reduced_form}
\mathbf T =
\mathbf A\boldsymbol{\alpha}^{\top}\mathbf Z
+
\mathbf A(\mathbf U+\boldsymbol{\varepsilon}),
\qquad
\mathbf A:=(\mathbf I_2-\mathbf B)^{-1}=
\frac{1}{1-\beta_{12}\beta_{21}}
\begin{pmatrix}
1 & \beta_{12}\\
\beta_{21} & 1
\end{pmatrix}.
\end{equation}
The matrix \(\mathbf A\) is the total-effect matrix, aggregating the direct and indirect effects propagated through the feedback system.

\subsection{Marginal association model under LD}
\label{subsec:summary_model_ld}

Assume genetic exogeneity with respect to the non-genetic disturbance,
\(\mathbb E\{\mathbf Z(\mathbf U+\boldsymbol\varepsilon)^\top\}=\mathbf 0\).
The matrix of marginal variant--trait associations then satisfies
\begin{equation}
\label{eq:population_gamma}
    \boldsymbol\Gamma=\operatorname{Cov}(\mathbf Z,\mathbf T)
    =\mathbf R\boldsymbol\alpha\mathbf A^\top.
\end{equation}
Thus marginal associations combine LD aggregation of the direct genetic effects
with propagation through the causal system.

Let \(\widehat{\boldsymbol\Gamma}\) and \(\mathbf E\) denote the corresponding
matrix of GWAS summary estimates and summary-statistic estimation errors. The
observed model is
\begin{equation}
\label{eq:summary_model}
    \widehat{\boldsymbol\Gamma}
     =\boldsymbol\Gamma+\mathbf E
    =\mathbf R\boldsymbol\alpha\mathbf A^\top+\mathbf E,
    \qquad \mathbf A=(\mathbf I_2-\mathbf B)^{-1}.
\end{equation}
The following working condition specifies the summary-statistic error model.

\begin{assumption}[Summary-statistic error]
\label{assum:summary_error}
The error \(\mathbf E\) is independent of \(\boldsymbol\alpha\). If
\(\mathbf e_j^\top\) is its \(j\)th row, then
\(\mathbf e_j\sim\mathbb N(\mathbf0,\boldsymbol\Sigma_j)\). For
non-overlapping two-sample studies,
\(\boldsymbol\Sigma_j =\operatorname{diag}(\sigma_{j1}^2,\sigma_{j2}^2)\),
where \(\sigma_{jk}\) is the reported standard error. With sample overlap,
a nonzero cross-trait error covariance may be incorporated when available.
\end{assumption}

The Gaussian approximation is standard for GWAS summary statistics
\citep{zhao2020statistical}. The covariance \(\boldsymbol\Sigma_j\) may vary
across variants, and nearby summary-statistic errors may remain weakly dependent
through LD. 

The target in \eqref{eq:summary_model} is causal matrix \(\mathbf B\), whose estimation is complicated by latent nuisance quantities, LD-induced dependence, and heteroscedastic summary-statistic errors.

\section{Identification and estimation}
\label{sec:ica_formulation}

We identify the causal effects by exploiting the distributional
structure of the direct genetic effects.
LD-score normalisation first converts
\eqref{eq:summary_model} into a noisy ICA model with mixing
matrix \(\mathbf A\). Standard ICA arguments identify this matrix up to scaling
and permutation, while the
no-self-feedback and stability constraints of the feedback system 
resolve these remaining ambiguities. We estimate the mixing matrix using noise-adjusted second- and fourth-order moments, and then recover the causal effects using the structural constraints of the feedback system.

\begin{assumption}[Direct-effect distribution]
\label{assum:direct_sources}
For
\(\boldsymbol\alpha_{j\cdot}=(\alpha_{j1},\alpha_{j2})^\top\):
\begin{enumerate}[label=(\roman*),leftmargin=2.2em,itemsep=0pt,topsep=2pt]
\item the vectors \(\{\boldsymbol\alpha_{j\cdot}\}_{j=1}^N\) are independent
and identically distributed across variants, and their two
components are independent, with
\(\mathbb E(\boldsymbol\alpha_{j\cdot})=\boldsymbol 0, \quad \operatorname{Cov}(\boldsymbol\alpha_{j\cdot}) =\mathbf D_\alpha =\operatorname{diag}(\sigma_{\alpha,1}^2,\sigma_{\alpha,2}^2)\),
where \(0<\sigma_{\alpha,k}^2<\infty\) for \(k=1,2\);
\item at most one component distribution is Gaussian.
\end{enumerate}
\end{assumption}

The cross-trait independence in part~(i) is the InDiE condition introduced in
Section~\ref{sec:introduction}. It is imposed on the primitive direct genetic effects,
not on the observed marginal associations, and allows both direct effects of a
variant to be nonzero. Part~(ii) is the standard non-Gaussianity requirement for
ICA identification and accommodates many sparse and mixture-like genetic architectures. Further discussion is given in Supplementary
Section~\ref{supp:source_selection_details}.

\subsection{LD-score normalisation and the noisy ICA representation}
\label{subsec:ldscore_noisy_ica}

In the observed model \eqref{eq:summary_model}, LD aggregates direct genetic
effects across variants through $\mathbf R$, while the total-effect matrix
$\mathbf A$ propagates these effects through the feedback system. Our interest
is in $\mathbf A$, rather than the direct effects at individual variants.
LD aggregation makes the covariance of the aggregated effects depend on the
local LD structure and induces dependence between neighbouring variants.
Undoing LD aggregation by premultiplying the summary associations by
$\mathbf R^{-1}$ would require inversion of a high-dimensional, potentially
ill-conditioned matrix and would also transform the summary-statistic errors.
We instead retain the marginal associations and use the covariance structure
of the LD-aggregated effects to construct a normalised model.

For variant $j$, let
$\mathbf g_j=\sum_{r=1}^N R_{jr}\boldsymbol\alpha_{r\cdot}$ denote its
LD-aggregated direct-effect vector. Under
Assumption~\ref{assum:direct_sources},
$\operatorname{Cov}(\mathbf g_j)=\ell_j\mathbf D_\alpha$,
where $\ell_j=\sum_{r=1}^N R_{jr}^2$ is the LD score of variant $j$.
Thus the variation in this covariance across variants is entirely captured
by the scalar $\ell_j$. Scaling each marginal association by
$\ell_j^{-1/2}$ removes this factor from the covariance of the
LD-aggregated effects without changing the mixing matrix $\mathbf A$.

Let $\mathbf D_\ell=\operatorname{diag}(\ell_1,\ldots,\ell_N)$.
The LD-score-normalised summary-data model is
\begin{equation}
\label{eq:ldscore_transformed_X}
    \mathbf X
    =\mathbf D_\ell^{-1/2}\widehat{\boldsymbol\Gamma}
    =\mathbf S\mathbf A^\top+\widetilde{\mathbf E},
    \qquad
    \mathbf S=\mathbf D_\ell^{-1/2}\mathbf R\boldsymbol\alpha,
    \quad
    \widetilde{\mathbf E}=\mathbf D_\ell^{-1/2}\mathbf E.
\end{equation}
For each variant $j=1,\ldots,N$, the corresponding row model is
\begin{equation}
\label{eq:row_noisy_ica}
    \mathbf x_j=\mathbf A\mathbf s_j+\widetilde{\mathbf e}_j,
    \qquad
    \mathbf s_j
    =\ell_j^{-1/2}\sum_{r=1}^N R_{jr}\boldsymbol\alpha_{r\cdot}.
\end{equation}
This is a noisy independent component analysis (ICA) representation with
heteroscedastic Gaussian errors: $\mathbf A$ is the common mixing matrix,
$\mathbf s_j$ is the latent source vector, and $\widetilde{\mathbf e}_j$
is the transformed summary-statistic error. The two coordinates of
$\mathbf s_j$ are independent because they are formed from the two
independent columns of $\boldsymbol\alpha$; the error is independent of
the source vector. Moreover,
\begin{equation}
\label{eq:effective_source_covariance}
    \operatorname{Cov}(\mathbf s_j)=\mathbf D_\alpha,
    \qquad
    \operatorname{Cov}(\widetilde{\mathbf e}_j)
    =\boldsymbol\Sigma_j^{\mathrm{ld}}
    :=\ell_j^{-1}\boldsymbol\Sigma_j.
\end{equation}
Consequently, the signal component $\mathbf A\mathbf s_j$ has the common
covariance $\mathbf R_s:=\mathbf A\mathbf D_\alpha\mathbf A^\top$, whereas
$\operatorname{Cov}(\mathbf x_j)=\mathbf R_s+\boldsymbol\Sigma_j^{\mathrm{ld}}$
may vary across variants. This common signal covariance provides the basis
for noise-adjusted whitening in Section~\ref{subsec:estimating_A}.

LD-score normalisation yields a common signal covariance across variants, but does not eliminate dependence across variants or ensure that the transformed observations are identically distributed. Neighbouring source vectors may
share direct genetic effects through overlapping LD patterns, and their
higher-order moments may differ despite their common covariance. The
independence of the source components at each variant supports the ICA
identification argument in Section~\ref{sec:causal_recovery}, while
estimation requires noise-adjusted moments and inference must accommodate
local dependence. Further comparison with ordinary ICA is given in
Supplementary Section~\ref{app:relation_to_ordinary_ica}.

\subsection{Identification of the causal matrix}
\label{sec:causal_recovery}
\label{subsec:mixing_identification}
\label{subsec:ica_indeterminacy_normalisation}
\label{subsec:stability_permutation_selection}

At any fixed variant index, the Gaussian error in
\eqref{eq:row_noisy_ica} has known covariance and an everywhere nonzero
characteristic function. Its contribution can therefore be deconvolved from the
distribution of \(\mathbf x_j\), identifying the distribution of
\(\mathbf A\mathbf s_j\). The coordinates of \(\mathbf s_j\) are independent;
moreover, if \(s_{jk}\) were Gaussian, Cramér's decomposition theorem and \(R_{jj}=1\) would imply that the corresponding primitive direct-effect distribution is Gaussian. Assumption~\ref{assum:direct_sources}(ii) therefore ensures that at most one effective source is Gaussian. Thus, the usual ICA identifiability condition holds, and \(\mathbf A\) is identified up to column scaling and permutation \citep{Comon1994,Hyvarinen2001}. A formal statement is given in
Supplementary Theorem~\ref{thm:ica_identification_A}.

It remains to use the structural form of the feedback system to remove the usual ICA ambiguities. Let
\[
    \mathbf W=\mathbf A^{-1}=\mathbf I_2-\mathbf B
    =
    \begin{pmatrix}
        1 & -\beta_{12}\\
        -\beta_{21} & 1
    \end{pmatrix}.
\]
If ICA returns \(\mathbf A\mathbf P\mathbf D\), with permutation matrix
\(\mathbf P\) and nonsingular diagonal matrix \(\mathbf D\), its inverse
differs from \(\mathbf W\) by row permutation and scaling. Since \(\mathbf W\) has unit diagonal, we consider the two possible
row orderings and normalise each to have unit diagonal, provided
that both diagonal entries are nonzero.
When both causal effects are nonzero, exchanging the sources and rescaling them
leaves the observed ICA model unchanged but gives the alternative causal matrix
\[
    \mathbf B^{\mathrm{swap}}
    =
    \begin{pmatrix}
        0 & 1/\beta_{21}\\
        1/\beta_{12} & 0
    \end{pmatrix},
    \qquad
    \beta_{12}^{\mathrm{swap}}\beta_{21}^{\mathrm{swap}}
    =(\beta_{12}\beta_{21})^{-1}.
\]
Under Assumption~\ref{assum:stability}, the alternative product has absolute
value greater than one and is therefore excluded by stability. If either causal
effect is zero, 
the swapped row ordering has a zero diagonal entry before
row-normalisation and is inadmissible. Thus the structural constraints select a
unique causal matrix from the ICA equivalence class.

The resolution of the ICA scaling and permutation ambiguities
builds on the identification approach for cyclic structural models
of \citet{Lacerda2008}. Combining this structural recovery with
identification of the mixing matrix in the noisy ICA representation
of LD-score-normalised GWAS summary statistics yields the following
result for bi-directional MR.

\begin{theorem}[Identification of the bi-directional causal effects]
\label{thm:B_identification}
Under Assumptions~\ref{assum:stability}-\ref{assum:direct_sources}, the causal matrix \(\mathbf B\), and hence
\((\beta_{12},\beta_{21})\), is uniquely identified from the distribution of
the summary statistics together with the summary-error covariance information.
\end{theorem}

The identification uses the joint distribution of all variant-level direct genetic effects rather than a
pleiotropy-free subset. This has a further consequence for variant selection.

\subsection{Variant selection and source independence}
\label{subsec:selection_source_independence}

The InDiE condition ensures independence between the two
source components in \eqref{eq:row_noisy_ica}. However,
instrument selection based on marginal associations conditions
on mixtures of these components rather than on the direct
genetic effects themselves. To examine its implications,
consider the noise-free component
\(\mathbf x_j^{(0)}=\mathbf A\mathbf s_j\).
If a variant is selected when
\(\mathcal E_j=\{\mathbf x_j^{(0)}\in\mathcal C\}\),
the corresponding event in source coordinates is
\(\{\mathbf s_j\in\mathbf A^{-1}\mathcal C\}\).
Such conditioning can induce dependence between the source
components.

\begin{proposition}[Selection and source independence]
\label{prop:selection_breaks_independence}
Let \(\mathbf s_j=(s_{j1},s_{j2})^\top\) have independent components and let
\(\mathbf x_j^{(0)}=\mathbf A\mathbf s_j\), where \(\mathbf A\) is nonsingular.
Consider a selection event
\(\mathcal E_j=\{\mathbf x_j^{(0)}\in\mathcal C\}\)
with \(0<\mathbb P(\mathcal E_j)<1\).
If \(\mathbf A^{-1}\mathcal C\) is not a Cartesian product
of two measurable sets, up to a null set under the distribution
of \(\mathbf s_j\), then
\(s_{j1}\not\!\perp\!\!\!\perp s_{j2}\mid\mathcal E_j\).
\end{proposition}

For example, selecting variants according to
\(\mathcal E_j=\{|x_{j1}^{(0)}|>c\}\) amounts to conditioning on
\(\{|A_{11}s_{j1}+A_{12}s_{j2}|>c\}\).
When both source components enter the marginal association, the
selection region is generally not a Cartesian product, and
conditioning can induce dependence between the components.
For noisy GWAS summary statistics, selection on observed
associations similarly reweights the source distribution by
a generally non-factorisable selection probability; see
Supplementary Section~\ref{supp:source_selection_details}.
Thus, direction-specific instrument selection can violate the
source independence underlying ICA identification, beyond simply discarding weak but collectively informative variants or introducing sensitivity to the significance threshold. 

\subsection{H-JADE estimation and causal recovery}
\label{subsec:estimating_A}
\label{subsec:estimator_B}

We estimate the total-effect matrix $\mathbf A$ using a
heteroscedastic-noise-adjusted version of JADE, referred to as H-JADE.
The common signal covariance $\mathbf R_s$ established in
Section~\ref{subsec:ldscore_noisy_ica} provides the starting point.
The observed covariance includes the variant-specific GWAS estimation
error, which must be removed before whitening. Write
$\mathbf x_j^c=\mathbf x_j-\bar{\mathbf x}$ and define
\begin{equation}
\label{eq:signal_cov_A_estimation}
    \widehat{\mathbf R}_s
    =\frac1N\sum_{j=1}^N
    \left\{
        \mathbf x_j^c(\mathbf x_j^c)^\top
        -\boldsymbol\Sigma_j^{\mathrm{ld}}
    \right\},
    \qquad
    \widehat{\mathbf H}=\widehat{\mathbf R}_s^{-1/2}.
\end{equation}
The whitened observations are
$\mathbf z_j=\widehat{\mathbf H}\mathbf x_j^c$.
To characterise the signal after whitening, let
\[
    \mathbf H=\mathbf R_s^{-1/2},
    \qquad
    \mathbf O=\mathbf H\mathbf A\mathbf D_\alpha^{1/2}
    =(\mathbf o_1,\mathbf o_2),
    \qquad
    \mathbf t_j=\mathbf D_\alpha^{-1/2}\mathbf s_j.
\]
Since $\mathbf R_s=\mathbf A\mathbf D_\alpha\mathbf A^\top$,
$\mathbf O\mathbf O^\top=\mathbf I_2$. The signal component after
whitening is $\mathbf y_j=\mathbf O\mathbf t_j$, where $\mathbf O$ is
orthogonal and the coordinates of $\mathbf t_j$ are independent with unit
variance. The remaining task is to estimate the orthogonal mixing directions.

LD aggregation also has a tractable effect on the fourth-order
information. For a centred variable $Y$ with a finite fourth moment,
write \(\operatorname{cum}_4(Y)=\mathbb E(Y^4) -3\{\mathbb E(Y^2)\}^2\). For $k=1,2$, let
$\kappa_{\alpha,k}=\operatorname{cum}_4(\alpha_{jk})$ denote the
fourth-order cumulant of a primitive direct effect, and define
\begin{equation}
\label{eq:ld-induced_factor}
    a_{j,N}
    =\frac{\sum_{r=1}^N R_{jr}^4}
           {\left(\sum_{r=1}^N R_{jr}^2\right)^2},
    \qquad
    a_N=\frac1N\sum_{j=1}^N a_{j,N}.
\end{equation}
Independence of the direct effects across variants yields
\begin{equation}
\label{eq:ld_attenuated_fourth_contrast}
    \operatorname{cum}_4(t_{jk})
    =\frac{\kappa_{\alpha,k}}{\sigma_{\alpha,k}^4}\,a_{j,N}.
\end{equation}
Thus LD-score normalisation makes the source variances common, while
LD aggregation changes their fourth-order cumulants. The factor
$a_{j,N}$ is shared by the two coordinates, so the variant-specific
fourth-order cumulant matrices differ only by a common scalar factor.
This structure also matters for the block-bootstrap argument in
Section~\ref{subsec:theory_bootstrap_ld}.

Since $\operatorname{Cov}(\mathbf y_j)=\mathbf I_2$, the average
fourth-order cumulant matrix of the whitened signal component, indexed by
a symmetric $2\times2$ matrix $\mathbf M$, is
\[
    \mathbf Q_N(\mathbf M)
    =\frac1N\sum_{j=1}^N
      \mathbb E\!\left\{
        (\mathbf y_j^\top\mathbf M\mathbf y_j)
        \mathbf y_j\mathbf y_j^\top
      \right\}
      -\operatorname{tr}(\mathbf M)\mathbf I_2-2\mathbf M.
\]
By independence of the source coordinates and
\eqref{eq:ld_attenuated_fourth_contrast},
\begin{equation}
\label{eq:population_joint_diagonalisation_main}
    \mathbf O^\top\mathbf Q_N(\mathbf M)\mathbf O
    =a_N\operatorname{diag}\!\left\{
        \frac{\kappa_{\alpha,k}}{\sigma_{\alpha,k}^4}
        (\mathbf o_k^\top\mathbf M\mathbf o_k)
    \right\}_{k=1}^2.
\end{equation}
Thus the same orthogonal matrix $\mathbf O$ diagonalises
$\mathbf Q_N(\mathbf M)$ for every symmetric $\mathbf M$. Since the
Gaussian summary-statistic errors are independent of the sources and have
zero fourth-order cumulants, the noise-corrected fourth-order cumulant matrices have
the same diagonalising directions. The factor $a_N$ changes the magnitude
of the fourth-order contrast, but not these directions. The non-degeneracy
condition required for their estimation is given in
Section~\ref{subsec:theory_asymptotics}; the derivation of
\eqref{eq:population_joint_diagonalisation_main} is in Supplementary
Section~\ref{app:corrected_cumulants}.

To estimate this family of matrices, the covariance terms in the
fourth-order moments must be removed variant by variant. Under the
theoretical whitening transformation, the covariance of $\mathbf H\mathbf x_j$ is
$\mathbf I_2+\mathbf H\boldsymbol\Sigma_j^{\mathrm{ld}}\mathbf H^\top$.
Accordingly, define
\[
    \widehat{\mathbf C}_j
    =\mathbf I_2+
    \widehat{\mathbf H}\boldsymbol\Sigma_j^{\mathrm{ld}}
    \widehat{\mathbf H}^\top.
\]
For each symmetric $\mathbf M$, the noise-corrected sample cumulant matrix
is
\begin{equation}
\label{eq:corrected_cumulant_main}
    \widehat{\mathbf Q}(\mathbf M)
    =\frac1N\sum_{j=1}^N
    \left\{
      (\mathbf z_j^\top\mathbf M\mathbf z_j)
         \mathbf z_j\mathbf z_j^\top
      -\operatorname{tr}(\mathbf M\widehat{\mathbf C}_j)
         \widehat{\mathbf C}_j
      -2\widehat{\mathbf C}_j\mathbf M\widehat{\mathbf C}_j
    \right\}.
\end{equation}
The subtraction consists of the three covariance pairings in a
fourth-order cumulant (the last two coincide for symmetric $\mathbf M$).
A correction based on the average error covariance would generally
be insufficient, since the fourth-order correction involves
products of variant-specific covariance matrices.

Choose a finite collection $\mathcal M$ spanning the symmetric
$2\times2$ matrices. The orthogonal directions are estimated by
approximate joint diagonalisation:
\begin{equation}
\label{eq:joint_diagonalization_A}
    \widehat{\mathbf V}
    =\arg\min_{\mathbf V\in\mathcal O(2)}
    \sum_{\mathbf M\in\mathcal M}
    \left\|
        \operatorname{off}\{\mathbf V^\top
        \widehat{\mathbf Q}(\mathbf M)\mathbf V\}
    \right\|_F^2,
\end{equation}
where \(\mathcal O(2)\) denotes the set of \(2\times2\) orthogonal matrices,
$\operatorname{off}(\cdot)$ sets diagonal entries to zero, and
$\|\cdot\|_F$ is the Frobenius norm. The total-effect matrix is then
estimated, up to ICA column scaling and permutation, by
\begin{equation}
\label{eq:Ahat_section3}
    \widehat{\mathbf A}
    =\widehat{\mathbf H}^{-1}\widehat{\mathbf V}.
\end{equation}

The two causal effects are recovered from $\widehat{\mathbf A}$ using
the identification result in Section~\ref{subsec:mixing_identification}.
We invert the estimated mixing matrix, consider its two row orderings,
and normalise each admissible ordering to have unit diagonal. The
negatives of the resulting off-diagonal entries give the candidate
effects $\widehat\beta_{12}$ and $\widehat\beta_{21}$. We select
the admissible ordering satisfying
$|\widehat\beta_{12}\widehat\beta_{21}|<1$, as established by
the stability argument in Section~\ref{subsec:mixing_identification}.

The estimator uses LD scores and variant-specific noise covariance
matrices but does not invert the full LD matrix. For fixed trait
dimension, the corrected second- and fourth-order moments require one
pass over the variants and the joint diagonalisation is fixed-dimensional.
The resulting computational cost is linear in $N$.

\section{Theoretical properties}
\label{sec:theory}

We study estimation and inference under local dependence of the LD-score normalised
rows. Two departures from standard fourth-order ICA drive the asymptotic analysis. First, heteroscedastic Gaussian noise requires variant-specific correction of the sample second- and fourth-order moments. Second, LD induces dependence between nearby variants, so the limiting covariance is a long-run covariance rather than an i.i.d. variance. The resulting ICA estimator is subsequently mapped to the causal effects through the stability-based recovery rule. Technical details and proofs are given in Supplementary Sections~\ref{supp:theory_details} and~\ref{supp:proofs}.

\subsection{Asymptotic distribution under local LD}
\label{subsec:theory_asymptotics}

Recall that
\(\mathbf x_j=\mathbf A\mathbf s_j+\widetilde{\mathbf e}_j\), with
\(\operatorname{Cov}(\widetilde{\mathbf e}_j)=
\boldsymbol\Sigma_j^{\mathrm{ld}}\). The following condition controls the
moments and short-range dependence of the corrected H-JADE moment
contributions.

\begin{assumption}[Moment and local-dependence conditions]
\label{assum:moment_mixing}
For some \(\eta>0\):
\begin{enumerate}[label=(\roman*),leftmargin=2.2em,itemsep=0pt,topsep=2pt]
\item
\[
    \mathbb E\|\boldsymbol\alpha_{j\cdot}\|^{8+\eta}<\infty,
    \qquad
    \sup_{N\geq1}\max_{1\leq j\leq N}
    \|\boldsymbol\Sigma_j^{\mathrm{ld}}\|_{\mathrm{op}}<\infty;
\]
\item ordered by chromosome and physical position, the triangular array
\(\{(\mathbf x_j,\boldsymbol\Sigma_j^{\mathrm{ld}}):1\leq j\leq N\}\)
is strongly mixing, with coefficients \(\alpha_{\mathrm{mix}}(h)\) satisfying
\begin{equation}
\label{eq:mixing_summability}
    \sum_{h=1}^{\infty}
    \alpha_{\mathrm{mix}}(h)^{\eta/(8+\eta)}<\infty;
\end{equation}
\item the long-run covariance of the finite-dimensional corrected moment
vector used by H-JADE converges to a finite matrix
\(\boldsymbol\Omega\).
\end{enumerate}
\end{assumption}

The \(8+\eta\) moment condition is matched to the fourth-order estimator:
the corresponding moment contributions have a \(2+\eta/4\) moment, yielding
the exponent in \eqref{eq:mixing_summability}. Neighbouring LD-weighted
source vectors can share primitive genetic effects, but the overlap weakens
as their genomic separation increases. For finite-range LD, the signal part
is itself finite-range dependent; the mixing assumption also covers any
remaining short-range dependence in the summary-statistic errors. 
Part~(iii) specifies the limiting long-run covariance
$\boldsymbol\Omega$ of the corrected moment contributions,
accounting for covariance between variants under local dependence.
This matrix enters the asymptotic covariance of the causal-effect
estimator through the H-JADE estimation and structural recovery.
Supplementary Section~\ref{supp:moment_dependence_details}
gives the triangular-array formulation of $\boldsymbol\Omega$
and relates the mixing condition to overlap between neighbouring
LD-weight vectors.

The H-JADE estimator in Section~\ref{subsec:estimating_A}
relies on fourth-order cumulants, which may vanish even for
non-Gaussian sources. Equations~\eqref{eq:ld-induced_factor}
and \eqref{eq:ld_attenuated_fourth_contrast} show that LD
aggregation attenuates the standardised fourth-order cumulants
of the direct effects by the factors $a_{j,N}$. We therefore
impose the following condition to ensure sufficient fourth-order
contrast for joint diagonalisation as $N\to\infty$.

\begin{assumption}[Non-degenerate fourth-order contrast]
\label{assum:primitive_fourth_contrast}
At most one of \(\kappa_{\alpha,1}\) and \(\kappa_{\alpha,2}\) is zero, and
\(a_N\to a>0\) as \(N\to\infty\).
\end{assumption}

Under Assumption~\ref{assum:primitive_fourth_contrast},
the fourth-order cumulant matrices in
\eqref{eq:population_joint_diagonalisation_main}
have a common diagonaliser that is unique up to signed permutation.
Further details are given in Supplementary
Section~\ref{supp:fourth_contrast_details}.

Define
\(\boldsymbol\beta=(\beta_{12},\beta_{21})^\top\)
and
\(\widehat{\boldsymbol\beta} =(\widehat\beta_{12},\widehat\beta_{21})^\top\).
The preceding conditions yield root-\(N\) inference for both directional
causal effects.

\begin{theorem}[Asymptotic distribution of the causal-effect estimator]
\label{thm:B_asymptotic}
Suppose Assumptions~\ref{assum:stability}-~\ref{assum:primitive_fourth_contrast} hold.
Then, as \(N\to\infty\),
\[
    \widehat{\mathbf B}\xrightarrow{p}\mathbf B,
    \qquad
    \sqrt N\left(
        \widehat{\boldsymbol\beta}-\boldsymbol\beta
    \right)
    \xrightarrow{d}
    \mathbb N(\mathbf0,\boldsymbol\Sigma_\beta),
\]
where
\(\boldsymbol\Sigma_\beta =\mathbf J_\beta\boldsymbol\Omega\mathbf J_\beta^\top\),
and \(\mathbf J_\beta\) is the Jacobian of the locally oriented H-JADE and causal
recovery map.
\end{theorem}

The discrete causal-ordering step is the main addition to ordinary ICA
asymptotics. The H-JADE estimate first converges to one representative of the
mixing matrix up to scaling and permutation. Stability separates the correct
row ordering from the alternative: when both causal effects are nonzero, the two
cycle products are reciprocals; when either effect is zero, the candidate under
the swapped ordering is not row-normalisable. The correct branch is therefore
selected with probability tending to one, after which inversion and row
normalisation are smooth. This also shows explicitly that local LD affects the limiting variance through the long-run covariance \(\boldsymbol\Omega\); it does not require
inversion of the full LD matrix.

\subsection{Chromosome-respecting block-bootstrap inference}
\label{subsec:theory_bootstrap_ld}

Direct estimation of \(\boldsymbol\Sigma_\beta\) would require estimating the
long-run covariance of the corrected moments and propagating it through
whitening, joint diagonalisation and causal recovery. We instead use a moving
block bootstrap that respects chromosome boundaries
\citep{goncalveswhite2002,lahiri2003resampling}.

The resampling unit is the pair
\((\mathbf x_j,\boldsymbol\Sigma_j^{\mathrm{ld}})\), so that each transformed association remains attached to the covariance used in its heteroscedastic noise
correction. Overlapping blocks are sampled only within chromosomes, and each
bootstrap sample repeats the complete H-JADE and causal-recovery procedure.
Let \(\widehat{\boldsymbol\beta}^{*}\) denote the resulting estimate and define
\[
    \widehat{\boldsymbol\Sigma}_{\beta,\mathrm{MBB}}
    =
    \operatorname{Var}^{*}
    \left[
        \sqrt N\left(
            \widehat{\boldsymbol\beta}^{*}
            -\widehat{\boldsymbol\beta}
        \right)
    \right],
\]
where \(\operatorname{Var}^{*}\) denotes conditional variance given the
observed transformed data.

\begin{proposition}[Consistency of bootstrap covariance]
\label{prop:mbb_cov_consistency}
Suppose the conditions of Theorem~\ref{thm:B_asymptotic} hold. Let
\(N_c\) denote the number of variants on chromosome \(c\), with
\(N=\sum_{c=1}^{C}N_c\). Suppose \(C\) is fixed and, as \(N\to\infty\),
\(N_c/N\to\pi_c\in(0,1)\) for each chromosome, while the
block length satisfies \(l_N\to\infty\) and \(l_N=o(N^{1/2})\).
Then
\(\widehat{\boldsymbol\Sigma}_{\beta,\mathrm{MBB}} \xrightarrow{p} \boldsymbol\Sigma_\beta\).
\end{proposition}

The bootstrap requires an additional argument because the expected
fourth-order cumulant contributions vary across variants.  In a generic heterogeneous array, such
heterogeneity can introduce an additional bootstrap variance term under
block resampling \citep{goncalveswhite2002}.  Here, however, the
noise-adjusted second-order moment is common across variants,
while the population counterpart of the $j$th corrected
fourth-order cumulant contribution satisfies
$\mathbf Q_{j,N}(\mathbf M)=a_{j,N}\mathbf Q_{0,N}(\mathbf M)$,
where $\mathbf Q_{0,N}(\mathbf M)$ is common across variants
and $a_{j,N}>0$ is the LD-induced attenuation factor defined in \eqref{eq:ld-induced_factor}.
Block resampling may therefore alter the overall fourth-order scale, but
not the common joint diagonaliser. Since joint diagonalisation
and causal recovery are invariant to this rescaling, this scale
fluctuation has no first-order effect on the causal estimator.
The remaining block-edge effects are asymptotically negligible
under the stated block-length condition. Details are given
in Supplementary Section~\ref{supp:bootstrap_details}
and the proof of Proposition~\ref{prop:mbb_cov_consistency}.

With \(M\) bootstrap replications, the covariance of
\(\widehat{\boldsymbol\beta}\) is estimated by the sample covariance of the
bootstrap estimates. We use \(l_N\asymp N^{1/3}\), the standard order for
block-bootstrap variance estimation under short-range dependence
\citep{hall1995blocking}.

\section{Simulation studies}
\label{sec:simulation}

We evaluate the finite-sample performance of H-JADE in estimating
and testing bi-directional causal effects. The primary simulations
examine estimation accuracy, interval coverage and directional
testing in comparison with existing MR methods. We then compare
point estimation and computational cost with LHC-MR \citep{darrous2021simultaneous} and assess
sensitivity to local dependence among direct genetic effects.

\subsection{Primary simulations}
\label{subsec:sim_primary}
\label{subsec:sim_main_results}

We generated summary associations from the model in
Section~\ref{sec:setup}, with
\(
\mathbf B_0=
\bigl(\begin{smallmatrix}0&\beta_{12}\\ \beta_{21}&0\end{smallmatrix}\bigr)
\)
and \(|\beta_{12}\beta_{21}|<1\). The primitive direct effects were independent
across traits and variants and followed the two-component Gaussian mixture
\[
\alpha_{jk}
=
(1-U_{jk})Z_{jk}^{(\mathrm{small})}
+
U_{jk}Z_{jk}^{(\mathrm{large})},
\qquad
U_{jk}\sim\operatorname{Bernoulli}(\pi_{\mathrm{large}}),
\]
where
\(Z_{jk}^{(\mathrm{small})}\sim
\mathbb N(0,\sigma_{\mathrm{small},k}^2)\) and
\(Z_{jk}^{(\mathrm{large})}\sim
\mathbb N(0,\sigma_{\mathrm{large},k}^2)\).
Variants were partitioned into
independent LD blocks with within-block AR(1) correlation. Marginal effects were obtained after LD propagation and causal mixing, and heteroscedastic Gaussian estimation errors were generated from  effective sample
sizes and information scores. The errors shared the within-block correlation structure of the genotypes.

Table~\ref{tab:simulation_setups} gives the six primary designs. Setting 1 sets one directional effect to zero,
allowing us to assess how reverse causation affects inference for that null direction. Settings 2–4 have nonzero causal effects in both directions and vary the strength and heterogeneity of LD. Settings 5–6 weaken the genetic signal and Setting 6 additionally allows many small nonzero direct genetic effects. In the hotspot settings, the block-specific AR(1) parameter was sampled from
\(\{0.20,0.40,0.60,0.80\}\). Each design used $N \in\{2000,5000\}$ variants, GWAS sample sizes $n_1 = n_2 = 30000$ for the two traits, and 200 Monte Carlo replications.  H-JADE used
all variants and 200 moving-block-bootstrap replications. IVW
\citep{burgess2013summarized}, MR-Egger \citep{bowden2015egger}, the weighted
median \citep{bowden2016weightedmedian}, and MR-RAPS
\citep{zhao2020statistical} were fitted separately in each direction after
selection at \(p<10^{-5}\) and LD clumping at \(r^2<0.001\). MR-RAPS used
over-dispersion with Huber loss. Further implementation details and additional
numerical results are given in Supplementary
Section~\ref{supp:simulation_diagnostics}.

\begin{table}[!htbp]
\centering
\caption{Primary simulation settings.}
\label{tab:simulation_setups}
\footnotesize
\setlength{\tabcolsep}{2.3pt}
\resizebox{\textwidth}{!}{%
\begin{tabular}{clccccc}
\toprule
Setting & LD pattern & \(\beta_{12}\) & \(\beta_{21}\) &
\(\pi_{\mathrm{large}}\) &
\((\sigma_{\mathrm{small},1},\sigma_{\mathrm{large},1})\) &
\((\sigma_{\mathrm{small},2},\sigma_{\mathrm{large},2})\) \\
\midrule
1 & Homogeneous \((\rho=0.70)\) & 0.00  & 0.50 & 0.09 & (0.000, 0.060) & (0.000, 0.050) \\
2 & Homogeneous \((\rho=0.40)\) & -0.40 & 0.50 & 0.09 & (0.000, 0.060) & (0.000, 0.050) \\
3 & Homogeneous \((\rho=0.80)\) & -0.40 & 0.50 & 0.09 & (0.000, 0.060) & (0.000, 0.050) \\
4 & Hotspot & -0.40 & 0.50 & 0.09 & (0.000, 0.060) & (0.000, 0.050) \\
5 & Hotspot & -0.40 & 0.50 & 0.05 & (0.000, 0.030) & (0.000, 0.025) \\
6 & Hotspot & -0.40 & 0.50 & 0.05 & (0.003, 0.030) & (0.003, 0.025) \\
\bottomrule
\end{tabular}%
}
\end{table}

\paragraph{Estimation and interval coverage.}
Tables~\ref{tab:simulation_bias_sd} and~\ref{tab:simulation_cp_length}
report the empirical bias and Monte Carlo standard deviation
of the estimators, and the empirical coverage probability and 
average length of nominal 95\% confidence intervals,
respectively, for \(N=5000\). Each entry gives the first measure followed by the second in
parentheses. 
The corresponding results for \(N=2000\) are provided in
Supplementary Section~\ref{supp:primary_simulations}.

\begin{table}[!htbp]
\centering
\caption{Bias (standard deviation) of the causal effect estimators at \(N=5000\).}
\label{tab:simulation_bias_sd}
\footnotesize
\setlength{\tabcolsep}{2.4pt}
\renewcommand{\arraystretch}{1.08}
\begin{tabular*}{\textwidth}{@{\extracolsep{\fill}}clr rrrrr@{}}
\toprule
Setting & Target & True & H-JADE & IVW & MR-Egger & MR-RAPS & \shortstack{Weighted\\ median} \\
\midrule
1 & $\beta_{12}$ & 0.00 & -0.002 (0.048) & 0.429 (0.061) & -0.003 (0.146) & 0.364 (0.062) & 0.063 (0.032) \\
 & $\beta_{21}$ & 0.50 & 0.002 (0.031) & -0.015 (0.028) & -0.003 (0.061) & -0.010 (0.020) & -0.032 (0.020) \\
\addlinespace[0.15em]
2 & $\beta_{12}$ & -0.40 & 0.000 (0.030) & 0.466 (0.064) & -0.249 (0.175) & 0.364 (0.070) & 0.070 (0.023) \\
 & $\beta_{21}$ & 0.50 & 0.000 (0.021) & -0.135 (0.032) & 0.237 (0.100) & -0.067 (0.022) & -0.047 (0.018) \\
\addlinespace[0.15em]
3 & $\beta_{12}$ & -0.40 & -0.005 (0.088) & 0.536 (0.077) & 0.129 (0.199) & 0.470 (0.082) & 0.224 (0.073) \\
 & $\beta_{21}$ & 0.50 & 0.007 (0.056) & -0.193 (0.047) & 0.090 (0.120) & -0.143 (0.042) & -0.107 (0.042) \\
\addlinespace[0.15em]
4 & $\beta_{12}$ & -0.40 & -0.003 (0.042) & 0.497 (0.070) & -0.111 (0.184) & 0.408 (0.072) & 0.105 (0.037) \\
 & $\beta_{21}$ & 0.50 & 0.002 (0.027) & -0.155 (0.037) & 0.171 (0.112) & -0.088 (0.028) & -0.060 (0.024) \\
\addlinespace[0.15em]
5 & $\beta_{12}$ & -0.40 & -0.002 (0.077) & 0.349 (0.119) & -0.160 (0.543) & 0.260 (0.119) & 0.151 (0.072) \\
 & $\beta_{21}$ & 0.50 & 0.003 (0.054) & -0.115 (0.054) & -0.010 (0.237) & -0.079 (0.043) & -0.102 (0.050) \\
\addlinespace[0.15em]
6 & $\beta_{12}$ & -0.40 & 0.000 (0.075) & 0.374 (0.107) & -0.127 (0.545) & 0.293 (0.106) & 0.179 (0.074) \\
 & $\beta_{21}$ & 0.50 & 0.003 (0.055) & -0.134 (0.056) & 0.019 (0.241) & -0.097 (0.047) & -0.120 (0.052) \\
\bottomrule
\end{tabular*}
\end{table}

\begin{table}[!htbp]
\centering
\caption{Empirical coverage probability (average length) of  95\% confidence intervals at \(N=5000\).}
\label{tab:simulation_cp_length}
\footnotesize
\setlength{\tabcolsep}{2.4pt}
\renewcommand{\arraystretch}{1.08}
\begin{tabular*}{\textwidth}{@{\extracolsep{\fill}}clr rrrrr@{}}
\toprule
Setting & Target & True & H-JADE & IVW & MR-Egger & MR-RAPS & \shortstack{Weighted\\ median} \\
\midrule
1 & $\beta_{12}$ & 0.00 & 0.930 (0.160) & 0.000 (0.209) & 0.950 (0.533) & 0.000 (0.212) & 0.215 (0.077) \\
 & $\beta_{21}$ & 0.50 & 0.940 (0.109) & 0.855 (0.098) & 0.955 (0.242) & 0.955 (0.087) & 0.470 (0.064) \\
\addlinespace[0.15em]
2 & $\beta_{12}$ & -0.40 & 0.920 (0.104) & 0.000 (0.243) & 0.720 (0.695) & 0.000 (0.240) & 0.070 (0.078) \\
 & $\beta_{21}$ & 0.50 & 0.910 (0.075) & 0.005 (0.147) & 0.340 (0.385) & 0.245 (0.111) & 0.210 (0.065) \\
\addlinespace[0.15em]
3 & $\beta_{12}$ & -0.40 & 0.930 (0.295) & 0.000 (0.277) & 0.875 (0.746) & 0.000 (0.282) & 0.000 (0.114) \\
 & $\beta_{21}$ & 0.50 & 0.935 (0.204) & 0.015 (0.188) & 0.885 (0.462) & 0.065 (0.172) & 0.035 (0.088) \\
\addlinespace[0.15em]
4 & $\beta_{12}$ & -0.40 & 0.910 (0.141) & 0.000 (0.257) & 0.915 (0.724) & 0.000 (0.257) & 0.015 (0.090) \\
 & $\beta_{21}$ & 0.50 & 0.920 (0.097) & 0.000 (0.160) & 0.620 (0.407) & 0.160 (0.129) & 0.125 (0.071) \\
\addlinespace[0.15em]
5 & $\beta_{12}$ & -0.40 & 0.930 (0.284) & 0.110 (0.459) & 0.920 (2.031) & 0.275 (0.418) & 0.305 (0.239) \\
 & $\beta_{21}$ & 0.50 & 0.935 (0.202) & 0.365 (0.203) & 0.905 (0.815) & 0.645 (0.186) & 0.425 (0.185) \\
\addlinespace[0.15em]
6 & $\beta_{12}$ & -0.40 & 0.960 (0.284) & 0.025 (0.434) & 0.925 (1.905) & 0.135 (0.407) & 0.215 (0.236) \\
 & $\beta_{21}$ & 0.50 & 0.920 (0.210) & 0.285 (0.217) & 0.960 (0.858) & 0.535 (0.202) & 0.320 (0.186) \\
\bottomrule
\end{tabular*}
\end{table}

H-JADE estimates were approximately centred across the primary designs. At
\(N=5000\), the largest absolute bias was 0.007, and the Monte Carlo
standard deviation was smaller than at \(N=2000\) in every setting and
direction. In contrast, the direction-specific estimators showed substantial
bias in several feedback settings. For example, when \(\beta_{12}=-0.4\) in
Setting~3, IVW and MR-RAPS had positive mean estimates, while the H-JADE mean
was \(-0.405\). MR-Egger was more variable in several settings, particularly
under weaker genetic signals. 
H-JADE also maintained relatively stable
empirical coverage across the primary settings, whereas IVW, MR-RAPS
and the weighted median frequently showed substantial undercoverage.
MR-Egger generally produced considerably wider confidence intervals.

\paragraph{Directional testing.}
\label{par:sim_directional_testing}
We compared H-JADE with the directional focusing procedures of
\citet{LiYe2025}. Their method forms a direction-specific focused set using
strong exposure associations and small standardised outcome associations under
the directional null, and then applies IVW or median inference. We used focused
IVW (Fo-IVW) with focusing threshold \(\tau_f=1.2\) or 1.5 and focused median
(Fo-Median) with \(\tau_f=1.5\), following the released implementation; a
smaller \(\tau_f\) gives a more selective focused set. H-JADE rejected a
directional null when the corresponding
95\% confidence interval based on moving-block-bootstrap
standard errors excluded zero.

\begin{figure}[!p]
\centering
\includegraphics[width=0.92\textwidth,height=0.37\textheight,keepaspectratio]{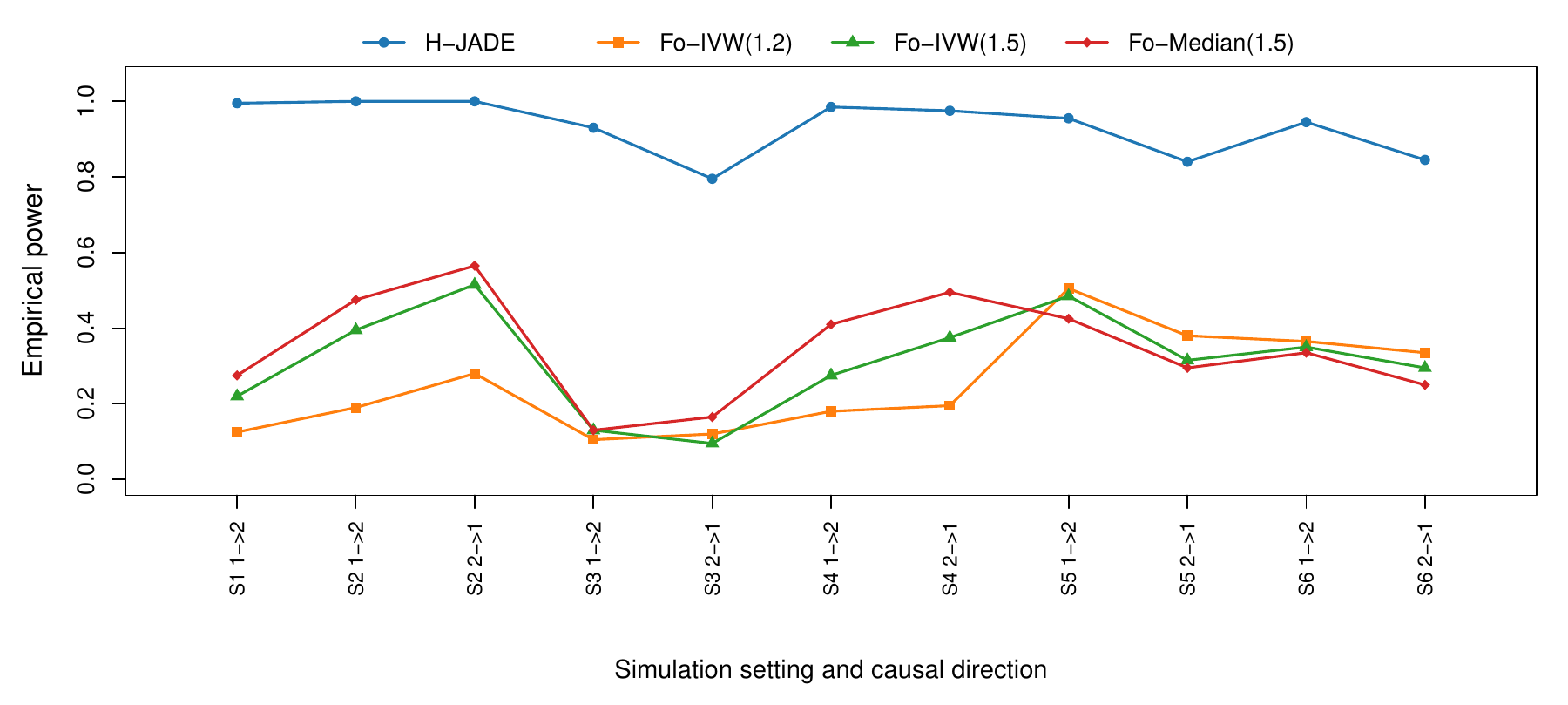}\par\vspace{-0.25em}
{\small (a) \(N=2000\)}\par\vspace{0.45em}
\includegraphics[width=0.92\textwidth,height=0.37\textheight,keepaspectratio]{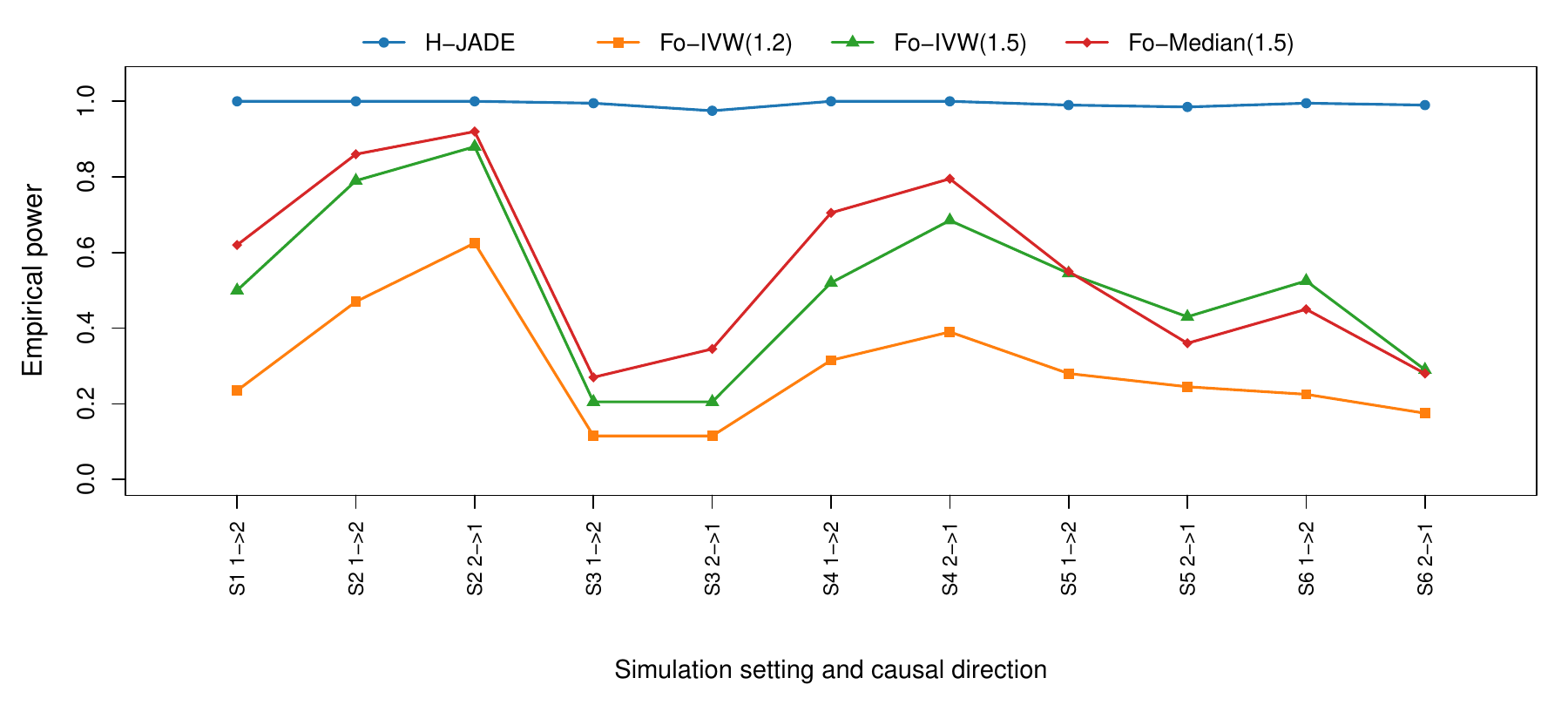}\par\vspace{-0.25em}
{\small (b) \(N=5000\)}
\caption{Empirical rejection rates for nonzero directional effects.
Setting~1 contributes only its nonzero \(1\to2\) direction.}
\label{fig:focusing_power}
\end{figure}

Setting~1 has a null \(2\to1\) effect. To examine the other orientation, we
also interchanged the zero and nonzero effects, setting
\((\beta_{12},\beta_{21})=(0.5,0)\). H-JADE rejection probabilities for the
two null directions were 0.090 and 0.070 at \(N=2000\), and 0.070 and 0.055
at \(N=5000\). The three focusing procedures ranged from 0.010 to 0.055 over
these four null cases. Figure~\ref{fig:focusing_power} reports the empirical
rejection rates for the nonzero setting--direction combinations.
H-JADE had the highest rejection rate in all 11 setting--direction combinations at both $N$. The complete results are given in Supplementary
Section~\ref{supp:focusing_comparison}.

\subsection{Comparison with LHC-MR}
\label{subsec:sim_lhcmr_comparison}
We also compared point estimation with LHC-MR
\citep{darrous2021simultaneous}, which jointly estimates the two causal
directions under a parametric model for the genetic architecture and uses multi-start likelihood optimisation for estimation. We considered two scenarios. In the first, the two primitive source processes were independent, making it compatible with the working structures of both
methods. In the second, we introduced the shared heritable component modelled by LHC-MR, inducing dependence between the two source processes and thereby violating the InDiE condition used by H-JADE.

Five independent data sets were generated in each scenario, with 5000 input
variants and \((\beta_{12},\beta_{21})=(-0.4,0.5)\). LHC-MR used 50 likelihood
starts and its preprocessing retained 500 variants. H-JADE was fitted both to
this same 500-variant panel and to all 5000 variants. Table~\ref{tab:lhcmr_main_comparison}
reports descriptive means, root mean squared errors (RMSEs) and point-estimation times. Across the five
data sets in the independent-source scenario, LHC-MR had smaller RMSE than H-JADE on the matched panel, while full-panel H-JADE had the smallest RMSE in both
directions. In the shared-heritable-component scenario, departures from the generating values were mainly in the \(\beta_{12}\) direction and were most pronounced for LHC-MR. Estimates of \(\beta_{21}\) remained close to the true value for all three fits. The computational difference was substantial: median LHC-MR point-estimation time
was about 7.9 hours per data set on our computing platform, compared with less
than 0.5 seconds for full-panel H-JADE. These timings cover point estimation
only; they exclude H-JADE moving-block-bootstrap inference and LHC-MR
block-jackknife inference. Individual estimates and further implementation
details are given in Supplementary Section~\ref{supp:lhcmr_comparison}.

\begin{table}[!htbp]
\centering
\caption{Point-estimation comparison with LHC-MR.}
\label{tab:lhcmr_main_comparison}
\footnotesize
\setlength{\tabcolsep}{3.7pt}
\begin{tabular}{llrccc}
\toprule
Scenario & Method & Variants &
\shortstack{\(\widehat\beta_{12}\)\\mean (RMSE)} &
\shortstack{\(\widehat\beta_{21}\)\\mean (RMSE)} & Runtime \\
\midrule
Independent sources & H-JADE, full & 5000 & -0.401 (0.040) & 0.510 (0.029) & 0.30 s \\
 & H-JADE, matched & 500 & -0.399 (0.189) & 0.482 (0.083) & 0.01 s \\
 & LHC-MR & 500 & -0.401 (0.137) & 0.529 (0.051) & 7.94 h \\
\addlinespace[0.2em]
Shared component & H-JADE, full & 5000 & -0.323 (0.086) & 0.524 (0.036) & 0.45 s \\
 & H-JADE, matched & 500 & -0.328 (0.178) & 0.511 (0.088) & 0.03 s \\
 & LHC-MR & 500 & -0.198 (0.345) & 0.514 (0.099) & 7.84 h \\
\bottomrule
\end{tabular}
\end{table}

\subsection{Sensitivity to local dependence}
\label{subsec:sim_source_dependence}

The primary simulations assume that the primitive direct effects are independent
across variants. To assess sensitivity to this condition, we introduced
short-range dependence within each trait-specific direct-effect sequence while
retaining independence between the two sequences. Using Setting~6, the direct genetic
effects within each LD block were generated from normalised MA(5) processes,
with \(\theta_k\) controlling the strength of local dependence for trait \(k\).
We considered
\((\theta_1,\theta_2) \in \{(0,0),(0.2,0.5),(0.5,0.5),(0.8,0.5)\}\).
The values \(\theta=0.5\) and 0.8 correspond to lag-one correlations of about
0.50 and 0.77, respectively, and \((\theta_1,\theta_2)=(0,0)\) reproduces Setting~6 replication by
replication.

Table~\ref{tab:simulation_local_dependence_summary} reports the results
at \(N=5000\). Empirical biases remained small across all dependence
settings. For \(\widehat\beta_{12}\), the Monte Carlo standard deviation
varied across settings, while the ratio of the average bootstrap
standard error to the Monte Carlo standard deviation declined from
0.968 to 0.855, with empirical coverage decreasing from 0.960 to
0.905. For \(\widehat\beta_{21}\), the variation in Monte Carlo standard
deviation and empirical coverage was smaller, although the
corresponding SE/SD ratio also declined from 0.974 to 0.936.

\begin{table}[!htbp]
\centering
\caption{MA(5) sensitivity results at \(N=5000\).}
\label{tab:simulation_local_dependence_summary}
\footnotesize
\setlength{\tabcolsep}{2.0pt}
\renewcommand{\arraystretch}{1.08}
\begin{tabular*}{\textwidth}{@{\extracolsep{\fill}}l*{10}{r}@{}}
\toprule
 & \multicolumn{5}{c}{\(\beta_{12}=-0.40\)} &
 \multicolumn{5}{c}{\(\beta_{21}=0.50\)} \\
\cmidrule(lr){2-6}\cmidrule(lr){7-11}
\(\theta_1,\theta_2\) & Bias & SD & Boot. SE & SE/SD & CP &
Bias & SD & Boot. SE & SE/SD & CP \\
\midrule
$(0,0)$ & 0.000 & 0.075 & 0.072 & 0.968 & 0.960 & 0.003 & 0.055 & 0.054 & 0.974 & 0.920 \\
$(0.2,0.5)$ & 0.004 & 0.058 & 0.053 & 0.915 & 0.925 & -0.001 & 0.059 & 0.056 & 0.959 & 0.900 \\
$(0.5,0.5)$ & 0.002 & 0.075 & 0.067 & 0.900 & 0.905 & 0.001 & 0.054 & 0.051 & 0.953 & 0.890 \\
$(0.8,0.5)$ & -0.001 & 0.102 & 0.087 & 0.855 & 0.905 & 0.002 & 0.049 & 0.046 & 0.936 & 0.900 \\
\bottomrule
\end{tabular*}
\end{table}

Under MA(5) dependence, the covariance of the LD-score-normalised
source vectors varies across variants. However, the two trait-specific
source processes remain independent, so their average covariance is
diagonal and whitening based on the average signal covariance retains
an orthogonal mixing structure. The MA construction introduces
additional dependence between neighbouring variants beyond that
induced by LD aggregation, altering the cross-variant covariance
of the corrected moment contributions. These features are consistent with
the small empirical biases and the changes in Monte Carlo standard
deviations and bootstrap standard errors observed in the simulations. Further
diagnostics are given in Supplementary
Section~\ref{supp:local_dependence}.

\section{Real-data analysis}\label{sec:real_data}

We analysed three cardiometabolic trait pairs: low-density lipoprotein
cholesterol (LDL-C)--coronary heart disease (CHD), apolipoprotein B
(ApoB)--CHD, and triglycerides (TG)--type 2 diabetes (T2D). 
GWAS summary statistics were obtained through the MRC IEU OpenGWAS
infrastructure \citep{Hemani2018,elsworth2020opengwas}. Dataset identifiers and original GWAS sources are listed
in Supplementary Table~\ref{tab:supp_realdata_gwas}. For each pair, the two GWAS were conducted in distinct studies using European-ancestry samples. All GWAS associations and standard errors were converted to the standardised-genotype scale. Lipid associations remain on standardised phenotype scales, whereas CHD and T2D associations remain on log-odds scales. Causal estimates involving binary disease exposures therefore require cautious interpretation \citep{burgess2018binary}.

After allele harmonisation, H-JADE used the common harmonised variant panel for the two traits together with a European-ancestry LD-score reference
\citep{buliksullivan2015atlas,bulik2015ldsc}. Standard errors were obtained
from 200 moving block bootstrap replications. The four
direction-specific estimators used in Section~\ref{sec:simulation}---IVW,
MR-Egger, MR-RAPS and the weighted median---were fitted separately after
exposure-specific selection and LD clumping. We also applied the Fo-IVW and
Fo-Median directional tests with the same tuning values as in
Section~\ref{subsec:sim_primary}. Further
implementation details and complete results are in Supplementary
Section~\ref{supp:real_data_information}.

\begin{figure}[!htbp]
\centering
\includegraphics[width=\textwidth]{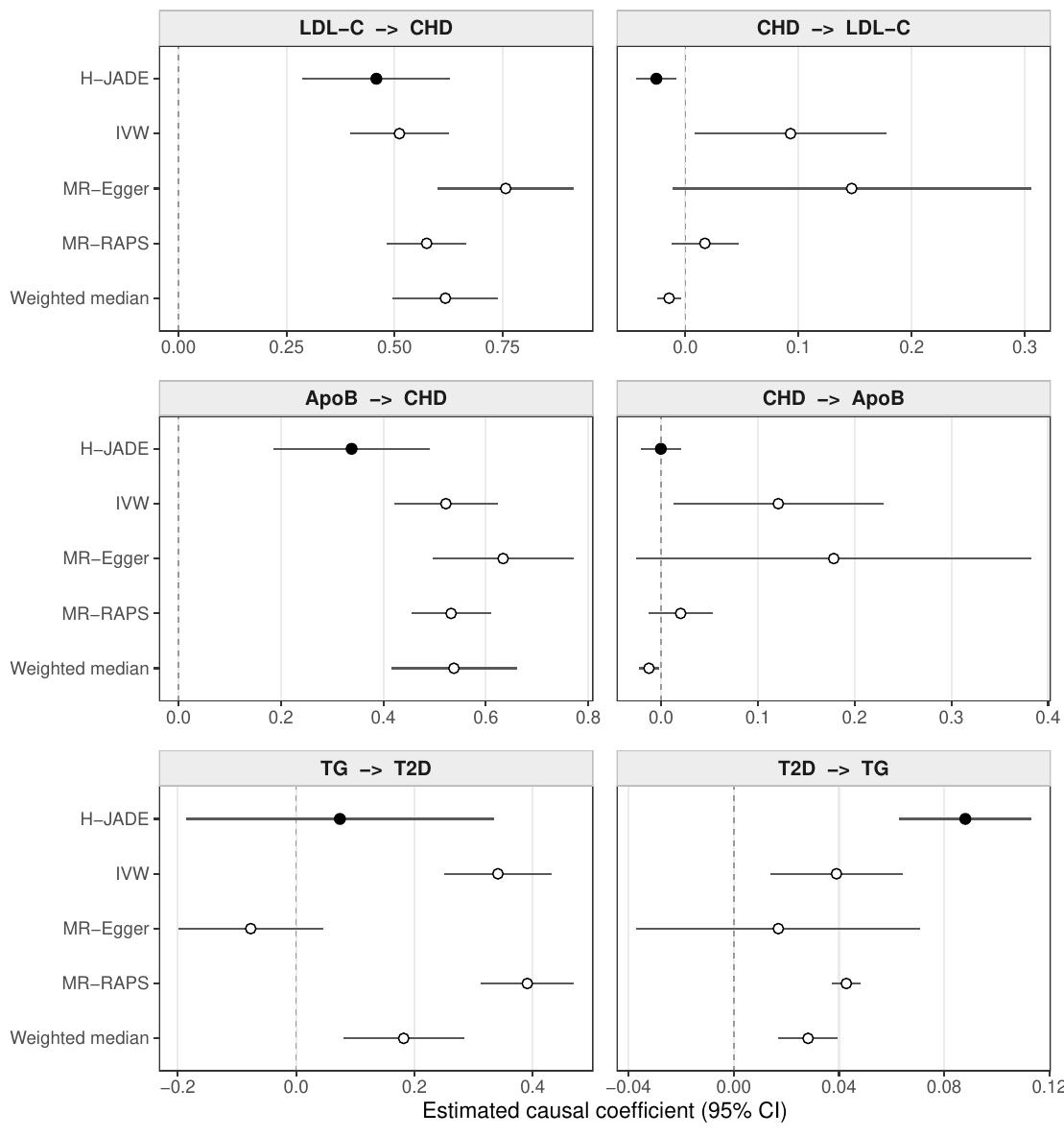}
\caption{Point estimates and 95\% confidence intervals for six causal
directions. H-JADE estimates the two coefficients jointly, whereas the
comparators use direction-specific SNP selection. Horizontal scales vary
across panels; estimates are on their respective GWAS association scales.}
\label{fig:real_data_results}
\end{figure}

Figure~\ref{fig:real_data_results} shows the point estimates from the five
methods and their 95\% confidence intervals.
Table~\ref{tab:realdata_focusing_comparison} reports the
\(p\)-values from H-JADE and the directional focusing tests
for each causal direction.

\begin{table}[!htbp]
\centering
\caption{Directional no-effect \(p\)-values in the real-data analyses.}
\label{tab:realdata_focusing_comparison}
\small
\setlength{\tabcolsep}{3.2pt}
\begin{threeparttable}
\begin{tabular*}{\textwidth}{@{\extracolsep{\fill}}lrrrr}
\toprule
Direction & H-JADE & \shortstack{Fo-IVW\\ $\tau_f=1.2$} &
\shortstack{Fo-IVW\\ $\tau_f=1.5$} &
\shortstack{Fo-Median\\ $\tau_f=1.5$} \\
\midrule
LDL-C $\rightarrow$ CHD & $1.61\times10^{-7}$ & 0.183 & 0.0352 & 0.249 \\
CHD $\rightarrow$ LDL-C & 0.00481 & 0.642 & 0.101 & 0.262 \\
ApoB $\rightarrow$ CHD & $1.52\times10^{-5}$ & 0.0166 & $7.98\times10^{-4}$ & 0.00100 \\
CHD $\rightarrow$ ApoB & 0.978 & 0.261 & 0.168 & 0.239 \\
TG $\rightarrow$ T2D & 0.577 & 0.0453 & $4.20\times10^{-5}$ & $<10^{-4}$ \\
T2D $\rightarrow$ TG & $5.87\times10^{-12}$ & 0.121 & 0.0927 & 0.0447 \\
\bottomrule
\end{tabular*}
\begin{tablenotes}
\footnotesize
\item H-JADE \(p\)-values are based on a two-sided normal
approximation with standard errors estimated by the moving block
bootstrap. Fo-Median \(p\)-values are based on 10,000 bootstrap
replications; \(<10^{-4}\) denotes no bootstrap exceedances.
\end{tablenotes}
\end{threeparttable}
\end{table}

For LDL-C--CHD, all five estimators supported a positive
LDL-C-to-CHD effect, consistent with established genetic
evidence \citep{richardson2020lipids}. The reverse direction
showed greater disagreement: H-JADE estimated a significant
negative effect ($-0.0257$, SE $0.0091$), consistent in sign
with the weighted median, whereas IVW gave a significant
positive estimate and none of the focusing tests rejected
the null. Previous reverse-MR evidence also suggested that
greater coronary disease liability may be associated with
lower LDL-C, potentially reflecting lipid-lowering treatment
\citep{smith2023reversemetabolic}.

For ApoB--CHD, all five estimators again supported a positive
forward effect, consistent with the established importance
of ApoB in coronary risk \citep{zuber2021apobcad}.
However, despite using the same CHD GWAS, the reverse
results differed from those for LDL-C: H-JADE estimated
an effect close to zero ($-0.0003$, SE $0.0107$), and none
of the focusing tests rejected the null, whereas IVW and
the weighted median yielded significant estimates with
opposite signs. These two applications highlight the
method sensitivity of reverse inference even for closely
related lipid traits.

For TG--T2D, the most notable disagreement arose in the
TG-to-T2D direction: IVW, MR-RAPS and the weighted median,
together with all three focusing tests, detected a positive
effect, whereas H-JADE did not reject the null ($0.074$,
SE $0.133$). The latter finding is consistent with an
earlier MR study reporting no evidence that genetically
elevated TG increases T2D risk
\citep{desilva2011triglycerides}.
In the reverse direction, H-JADE detected a positive
T2D-to-TG effect ($0.088$, SE $0.013$), supported by IVW,
MR-RAPS and the weighted median but only one of the
three focusing tests. This direction is also consistent
with genetic evidence linking diabetes liability and
glycaemic traits to elevated TG
\citep{smith2023reversemetabolic,zhu2022lipidglycemic}.

Across the three applications, H-JADE agreed with
direction-specific methods in several well-established
directions but yielded different conclusions for
CHD-to-LDL-C, CHD-to-ApoB and TG-to-T2D.
These contrasts illustrate how conclusions may change
when the two causal directions are modelled jointly
rather than analysed separately using direction-specific
instruments.

\section{Discussion and conclusion}
\label{sec:discussion}

This article introduces a novel method for bi-directional Mendelian randomization by linking it to independent component analysis:
LD-score-normalised GWAS associations can be expressed as noisy linear mixtures of
LD-aggregated direct genetic effects. Under the InDiE condition, informative
non-Gaussianity of the direct genetic effects identifies the total-effect matrix
up to the usual ICA ambiguities, while the no-self-feedback and stability
restrictions recover the two directional causal effects. Identification
therefore comes from the distributional structure of direct genetic effects rather than from
valid instrumental variables. 

A key challenge in using the InDiE condition for bi-directional MR is that, unlike unidirectional MR, selecting variants based on marginal associations can induce dependence between the two latent source components, thereby violating the independence assumption underlying ICA. Estimation from GWAS summary statistics additionally requires accounting for variant-specific estimation errors and LD. To address these, we use LD-score normalisation to avoid inverting the full LD matrix. The H-JADE algorithm developed here has computational complexity linear in the number of genetic variants. LD-induced local dependence is accounted for in the asymptotic covariance, and inference is based on a moving block bootstrap that respects chromosome boundaries. In the primary simulations, H-JADE produced accurate point estimates and generally satisfactory confidence interval coverage. It also achieved higher power for nonzero directional effects than the focusing procedures of \citet{LiYe2025}, with significantly shorter computation times than LHC-MR \citep{darrous2021simultaneous}. The real-data analyses show that joint estimation of the feedback system can lead to different conclusions from separate unidirectional MR analyses.

A natural extension is to feedback systems involving more than two traits.
The noisy-ICA representation extends directly, but structural recovery becomes more difficult: the two-trait problem reduces to choosing between
two orderings, whereas higher-dimensional systems admit many possible
permutations and feedback structures. Developing recovery rules
that resolve this ambiguity is an important direction for future work.

\section*{Data availability}

GWAS summary statistics used in the real-data analyses
were obtained from IEU OpenGWAS \citep{elsworth2020opengwas}.
The corresponding GWAS IDs are listed in
Supplementary Table~\ref{tab:supp_realdata_gwas}.
The R code implementing the proposed method and used for
the simulation studies and real-data analyses is available at
\url{https://github.com/molly3687/bidirectional-mr-hjade/tree/main/R}.

\section*{Acknowledgements}

LM was supported by an overseas visiting fellowship to the University of Cambridge by the China Scholarship Council.
The authors used ChatGPT to improve the writing and presentation of the manuscript and to assist with R code implementation and debugging, but AI was not used when the core methodological ideas and technical results were conceptualized. AI-assisted content and code were reviewed and revised by the authors, who take full responsibility for the final manuscript.

\nocite{Moser2015,Zhou2013,bowden2017_misconceptions_inside,darmois1953analyse,skitovich1953property}
\bibliographystyle{apalike}
\bibliography{refer}

\begin{thebibliography}{}

\bibitem[Boehm and Zhou, 2022]{BoehmZhou2022}
Boehm, F.~J. and Zhou, X. (2022).
\newblock Statistical methods for {Mendelian} randomization in genome-wide
  association studies: A review.
\newblock {\em Computational and Structural Biotechnology Journal},
  20:2338--2351.

\bibitem[Bowden, 2017]{bowden2017_misconceptions_inside}
Bowden, J. (2017).
\newblock Misconceptions on the use of {MR}-{Egger} regression and the
  evaluation of the {InSIDE} assumption.
\newblock {\em International Journal of Epidemiology}, 46(6):2097--2099.

\bibitem[Bowden et~al., 2015]{bowden2015egger}
Bowden, J., Davey~Smith, G., and Burgess, S. (2015).
\newblock Mendelian randomization with invalid instruments: effect estimation
  and bias detection through {E}gger regression.
\newblock {\em International Journal of Epidemiology}, 44(2):512--525.

\bibitem[Bowden et~al., 2016]{bowden2016weightedmedian}
Bowden, J., Davey~Smith, G., Haycock, P.~C., and Burgess, S. (2016).
\newblock Consistent estimation in {Mendelian} randomization with some invalid
  instruments using a weighted median estimator.
\newblock {\em Genetic Epidemiology}, 40(4):304--314.

\bibitem[Boyle et~al., 2017]{boyle2017expanded}
Boyle, E.~A., Li, Y.~I., and Pritchard, J.~K. (2017).
\newblock An expanded view of complex traits: From polygenic to omnigenic.
\newblock {\em Cell}, 169(7):1177--1186.

\bibitem[Brumpton et~al., 2020]{Brumpton2020}
Brumpton, B., Sanderson, E., Heilbron, K., et~al. (2020).
\newblock Avoiding dynastic, assortative mating, and population stratification
  biases in {Mendelian} randomization through within-family analyses.
\newblock {\em Nature Communications}, 11(1):3519.

\bibitem[Bulik-Sullivan et~al., 2015a]{buliksullivan2015atlas}
Bulik-Sullivan, B., Finucane, H.~K., Anttila, V., Gusev, A., Day, F.~R., Loh,
  P.-R., Duncan, L., Perry, J. R.~B., Patterson, N., Robinson, E.~B., Daly,
  M.~J., Price, A.~L., and Neale, B.~M. (2015a).
\newblock An atlas of genetic correlations across human diseases and traits.
\newblock {\em Nature Genetics}, 47(11):1236--1241.

\bibitem[Bulik-Sullivan et~al., 2015b]{bulik2015ldsc}
Bulik-Sullivan, B.~K., Loh, P.-R., Finucane, H.~K., Ripke, S., Yang, J.,
  Patterson, N., Daly, M.~J., Price, A.~L., and Neale, B.~M. (2015b).
\newblock {LD} score regression distinguishes confounding from polygenicity in
  genome-wide association studies.
\newblock {\em Nature Genetics}, 47(3):291--295.

\bibitem[Burgess et~al., 2013]{burgess2013summarized}
Burgess, S., Butterworth, A., and Thompson, S.~G. (2013).
\newblock Mendelian randomization analysis with multiple genetic variants using
  summarized data.
\newblock {\em Genetic Epidemiology}, 37(7):658--665.

\bibitem[Burgess and Labrecque, 2018]{burgess2018binary}
Burgess, S. and Labrecque, J.~A. (2018).
\newblock Mendelian randomization with a binary exposure variable:
  interpretation and presentation of causal estimates.
\newblock {\em European Journal of Epidemiology}, 33(10):947--952.

\bibitem[Burgess and Thompson, 2015]{BurgessThompson2015}
Burgess, S. and Thompson, S.~G. (2015).
\newblock {\em Mendelian Randomization: Methods for Using Genetic Variants in
  Causal Estimation}.
\newblock Chapman and Hall/CRC.

\bibitem[Cardoso and Souloumiac, 1993]{cardoso1993blind}
Cardoso, J.-F. and Souloumiac, A. (1993).
\newblock Blind beamforming for non-gaussian signals.
\newblock {\em IEE Proceedings F (Radar and Signal Processing)},
  140(6):362--370.

\bibitem[Comon, 1994]{Comon1994}
Comon, P. (1994).
\newblock Independent component analysis, a new concept?
\newblock {\em Signal Processing}, 36(3):287--314.

\bibitem[Darmois, 1953]{darmois1953analyse}
Darmois, G. (1953).
\newblock Analyse g{\'e}n{\'e}rale des liaisons stochastiques: {\'E}tude
  particuli{\`e}re de l'analyse factorielle lin{\'e}aire.
\newblock {\em Revue de l'Institut International de Statistique / Review of the
  International Statistical Institute}, 21(1/2):2--8.

\bibitem[Darrous et~al., 2021]{darrous2021simultaneous}
Darrous, L., Mounier, N., and Kutalik, Z. (2021).
\newblock Simultaneous estimation of bi-directional causal effects and
  heritable confounding from {GWAS} summary statistics.
\newblock {\em Nature Communications}, 12(1):7274.

\bibitem[Davey~Smith and Ebrahim, 2003]{DaveySmithEbrahim2003}
Davey~Smith, G. and Ebrahim, S. (2003).
\newblock ``mendelian randomization'': can genetic epidemiology contribute to
  understanding environmental determinants of disease?
\newblock {\em International Journal of Epidemiology}, 32(1):1--22.

\bibitem[Davey~Smith and Hemani, 2014]{DaveySmithHemani2014}
Davey~Smith, G. and Hemani, G. (2014).
\newblock Mendelian randomization: genetic anchors for causal inference in
  epidemiological studies.
\newblock {\em Human Molecular Genetics}, 23(R1):R89--R98.

\bibitem[de~Silva et~al., 2011]{desilva2011triglycerides}
de~Silva, N. M.~G., Freathy, R.~M., Palmer, T.~M., Donnelly, L.~A., Luan, J.,
  Gaunt, T., Langenberg, C., Weedon, M.~N., et~al. (2011).
\newblock {Mendelian} randomization studies do not support a role for raised
  circulating triglyceride levels influencing type 2 diabetes, glucose levels,
  or insulin resistance.
\newblock {\em Diabetes}, 60(3):1008--1018.

\bibitem[Elsworth et~al., 2020]{elsworth2020opengwas}
Elsworth, B., Lyon, M., Alexander, T., Liu, Y., Matthews, P., Hallett, J.,
  Bates, P., Palmer, T., Haberland, V., Davey~Smith, G., Zheng, J., Haycock,
  P., Gaunt, T.~R., and Hemani, G. (2020).
\newblock The {MRC IEU OpenGWAS} data infrastructure.
\newblock {\em bioRxiv}.

\bibitem[Gon{\c{c}}alves and White, 2002]{goncalveswhite2002}
Gon{\c{c}}alves, S. and White, H. (2002).
\newblock The bootstrap of the mean for dependent heterogeneous arrays.
\newblock {\em Econometric Theory}, 18(6):1367--1384.

\bibitem[Guo et~al., 2018]{Guo2018}
Guo, Z., Kang, H., Cai, T.~T., and Small, D.~S. (2018).
\newblock Confidence intervals for causal effects with invalid instruments by
  using two-stage hard thresholding with voting.
\newblock {\em Journal of the Royal Statistical Society: Series B (Statistical
  Methodology)}, 80(4):793--815.

\bibitem[Hall et~al., 1995]{hall1995blocking}
Hall, P., Horowitz, J.~L., and Jing, B.-Y. (1995).
\newblock On blocking rules for the bootstrap with dependent data.
\newblock {\em Biometrika}, 82(3):561--574.

\bibitem[Hemani et~al., 2017]{hemani2017steiger}
Hemani, G., Tilling, K., and Davey~Smith, G. (2017).
\newblock Orienting the causal relationship between imprecisely measured traits
  using {GWAS} summary data.
\newblock {\em PLoS Genetics}, 13(11):e1007081.

\bibitem[Hemani et~al., 2018]{Hemani2018}
Hemani, G., Zheng, J., Elsworth, B., Wade, K.~H., Haberland, V., et~al. (2018).
\newblock The {MR-Base} platform supports systematic causal inference across
  the human phenome.
\newblock {\em eLife}, 7:e34408.

\bibitem[Hyttinen et~al., 2012]{hyttinen2012learning}
Hyttinen, A., Eberhardt, F., and Hoyer, P.~O. (2012).
\newblock Learning linear cyclic causal models with latent variables.
\newblock {\em The Journal of Machine Learning Research}, 13(1):3387--3439.

\bibitem[Hyv{\"a}rinen et~al., 2001]{Hyvarinen2001}
Hyv{\"a}rinen, A., Karhunen, J., and Oja, E. (2001).
\newblock {\em Independent Component Analysis}.
\newblock John Wiley \& Sons, New York.

\bibitem[Hyv{\"a}rinen and Oja, 2000]{HyvarinenOja2000}
Hyv{\"a}rinen, A. and Oja, E. (2000).
\newblock Independent component analysis: algorithms and applications.
\newblock {\em Neural Networks}, 13(4–5):411--430.

\bibitem[Jiang et~al., 2023]{Jiang2023WinnerCurse}
Jiang, T., Gill, D., Butterworth, A.~S., and Burgess, S. (2023).
\newblock An empirical investigation into the impact of winner's curse on
  estimates from {Mendelian} randomization.
\newblock {\em International Journal of Epidemiology}, 52(4):1209--1219.

\bibitem[Kang et~al., 2016]{Kang2016}
Kang, H., Zhang, A., Cai, T.~T., and Small, D.~S. (2016).
\newblock Instrumental variables estimation with some invalid instruments and
  its application to {Mendelian} randomization.
\newblock {\em Journal of the American Statistical Association},
  111(513):132--144.

\bibitem[Lacerda et~al., 2008]{Lacerda2008}
Lacerda, G., Spirtes, P., Ramsey, J., and Hoyer, P.~O. (2008).
\newblock Discovering cyclic causal models by independent components analysis.
\newblock In {\em Proceedings of the Twenty-Fourth Conference on Uncertainty in
  Artificial Intelligence}, pages 366--374. AUAI Press.

\bibitem[Lahiri, 2003]{lahiri2003resampling}
Lahiri, S.~N. (2003).
\newblock {\em Resampling Methods for Dependent Data}.
\newblock Springer Series in Statistics. Springer, New York.

\bibitem[Lauritzen and Richardson, 2002]{lauritzen2002chain}
Lauritzen, S.~L. and Richardson, T.~S. (2002).
\newblock Chain graph models and their causal interpretations.
\newblock {\em Journal of the Royal Statistical Society: Series B (Statistical
  Methodology)}, 64(3):321--348.

\bibitem[Li and Ye, 2025]{LiYe2025}
Li, S. and Ye, T. (2025).
\newblock A focusing framework for testing bi-directional causal effects in
  {Mendelian} randomization.
\newblock {\em Journal of the Royal Statistical Society: Series B (Statistical
  Methodology)}, 87(2):529--548.

\bibitem[Li et~al., 2024]{li2024discovery}
Li, W., Duan, R., and Li, S. (2024).
\newblock Discovery and inference of possibly bi-directional causal
  relationships with invalid instrumental variables.
\newblock {\em arXiv preprint arXiv:2407.11646}.

\bibitem[Ma et~al., 2023]{MaWangWu2023}
Ma, X., Wang, J., and Wu, C. (2023).
\newblock Breaking the winner's curse in {Mendelian} randomization:
  Rerandomized inverse variance weighted estimator.
\newblock {\em The Annals of Statistics}, 51(1):211--232.

\bibitem[Morrison et~al., 2020]{morrison2020mendelian}
Morrison, J., Knoblauch, N., Marcus, J.~H., Stephens, M., and He, X. (2020).
\newblock {Mendelian} randomization accounting for correlated and uncorrelated
  pleiotropic effects using genome-wide summary statistics.
\newblock {\em Nature Genetics}, 52(7):740--747.

\bibitem[Moser et~al., 2015]{Moser2015}
Moser, G., Lee, S.~H., Hayes, B.~J., Goddard, M.~E., Wray, N.~R., and Visscher,
  P.~M. (2015).
\newblock Simultaneous discovery, estimation and prediction analysis of complex
  traits using a {Bayesian} mixture model.
\newblock {\em PLOS Genetics}, 11(4):e1004969.

\bibitem[Richardson et~al., 2020]{richardson2020lipids}
Richardson, T.~G., Sanderson, E., Palmer, T.~M., Ala-Korpela, M., Ference,
  B.~A., Davey~Smith, G., and Holmes, M.~V. (2020).
\newblock Evaluating the relationship between circulating lipoprotein lipids
  and apolipoproteins with risk of coronary heart disease: A multivariable
  {Mendelian} randomisation analysis.
\newblock {\em PLOS Medicine}, 17(3):e1003062.

\bibitem[Richmond and Davey~Smith, 2019]{RichmondDaveySmith2019}
Richmond, R.~C. and Davey~Smith, G. (2019).
\newblock Commentary: Orienting causal relationships between two phenotypes
  using bi-directional {Mendelian} randomization.
\newblock {\em International Journal of Epidemiology}, 48(3):907--911.

\bibitem[Richmond and Davey~Smith, 2022]{Richmond2022}
Richmond, R.~C. and Davey~Smith, G. (2022).
\newblock Mendelian randomization: Concepts and scope.
\newblock {\em Annual Review of Public Health}, 43:473--492.

\bibitem[Sanderson et~al., 2019]{sanderson2019mvmr}
Sanderson, E., Davey~Smith, G., Windmeijer, F., and Bowden, J. (2019).
\newblock An examination of multivariable mendelian randomization in the
  single-sample and two-sample summary data settings.
\newblock {\em International Journal of Epidemiology}, 48(3):713--727.

\bibitem[Sanderson et~al., 2022]{Sanderson2022}
Sanderson, E., Glymour, M.~M., Holmes, M.~V., Kang, H., Morrison, J.,
  Munaf{\`o}, M.~R., Palmer, T., Schooling, C.~M., Wallace, C., Zhao, Q., and
  Davey~Smith, G. (2022).
\newblock Mendelian randomization.
\newblock {\em Nature Reviews Methods Primers}, 22(1):1--21.

\bibitem[Skitovich, 1953]{skitovich1953property}
Skitovich, V.~P. (1953).
\newblock On a property of the normal distribution.
\newblock {\em Doklady Akademii Nauk SSSR}, 89(2):217--219.
\newblock In Russian.

\bibitem[Smith et~al., 2023]{smith2023reversemetabolic}
Smith, M.~L., Bull, C.~J., Holmes, M.~V., Davey~Smith, G., Sanderson, E.,
  Anderson, E.~L., and Bell, J.~A. (2023).
\newblock Distinct metabolic features of genetic liability to type 2 diabetes
  and coronary artery disease: A reverse {Mendelian} randomization study.
\newblock {\em eBioMedicine}, 90:104503.

\bibitem[Staley and Burgess, 2017]{StaleyBurgess2017}
Staley, J.~R. and Burgess, S. (2017).
\newblock Semiparametric methods for estimation of a nonlinear
  exposure--outcome relationship using instrumental variables with application
  to {Mendelian} randomization.
\newblock {\em Genetic Epidemiology}, 41(4):341--352.

\bibitem[Swerdlow et~al., 2016]{Swerdlow2016}
Swerdlow, D.~I., Kuchenbaecker, K.~B., Shah, S., Sofat, R., Holmes, M.~V.,
  White, J., Mindell, J.~S., Kivimaki, M., Brunner, E.~J., Whittaker, J.~C.,
  Casas, J.~P., and Hingorani, A.~D. (2016).
\newblock Selecting instruments for {Mendelian} randomization in the wake of
  genome-wide association studies.
\newblock {\em International Journal of Epidemiology}, 45(5):1600--1616.

\bibitem[Tudball et~al., 2022]{Tudball2022}
Tudball, M.~J., Davey~Smith, G., and Zhao, Q. (2022).
\newblock Almost exact {Mendelian} randomization.
\newblock {\em arXiv preprint arXiv:2208.14035}.
\newblock Version 2, 18 April 2023.

\bibitem[Visscher et~al., 2017]{Visscher2017}
Visscher, P.~M., Wray, N.~R., Zhang, Q., Sklar, P., McCarthy, M.~I., Brown,
  M.~A., and Yang, J. (2017).
\newblock 10 years of {GWAS} discovery: Biology, function, and translation.
\newblock {\em The American Journal of Human Genetics}, 101(1):5--22.

\bibitem[Wang et~al., 2021]{wang2021causal}
Wang, J., Zhao, Q., Bowden, J., Hemani, G., Davey~Smith, G., Small, D.~S., and
  Zhang, N.~R. (2021).
\newblock Causal inference for heritable phenotypic risk factors using
  heterogeneous genetic instruments.
\newblock {\em PLoS Genetics}, 17(6):e1009575.

\bibitem[Wang and Han, 2021]{WangHan2021}
Wang, K. and Han, S. (2021).
\newblock Effect of selection bias on two sample summary data based {Mendelian}
  randomization.
\newblock {\em Scientific Reports}, 11:7585.

\bibitem[Wu et~al., 2025]{wu2024causal}
Wu, Z., Lewis, E., Zhao, Q., and Wang, J. (2025).
\newblock Causal mediation analysis for time-varying heritable risk factors
  with {Mendelian} randomization.
\newblock {\em Nature Communications}, 16(1):6945.

\bibitem[Xie et~al., 2024]{Xie2024}
Xie, F., Yao, Z., Xie, L., Zeng, Y., and Geng, Z. (2024).
\newblock Identification and estimation of the bi-directional {MR} with some
  invalid instruments.
\newblock In {\em Advances in Neural Information Processing Systems},
  volume~37, pages 48256--48291.

\bibitem[Xue and Pan, 2020]{XuePan2020}
Xue, H. and Pan, W. (2020).
\newblock Inferring causal direction between two traits in the presence of
  horizontal pleiotropy with {GWAS} summary data.
\newblock {\em PLOS Genetics}, 16(11):e1009105.

\bibitem[Xue and Pan, 2022]{XuePan2022}
Xue, H. and Pan, W. (2022).
\newblock Robust inference of bi-directional causal relationships in presence
  of correlated pleiotropy with {GWAS} summary data.
\newblock {\em PLOS Genetics}, 18(5):e1010205.

\bibitem[Zhao et~al., 2019]{zhao2019powerful}
Zhao, Q., Chen, Y., Wang, J., and Small, D.~S. (2019).
\newblock Powerful three-sample genome-wide design and robust statistical
  inference in summary-data {Mendelian} randomization.
\newblock {\em International Journal of Epidemiology}, 48(5):1478--1492.

\bibitem[Zhao et~al., 2020]{zhao2020statistical}
Zhao, Q., Wang, J., Hemani, G., Bowden, J., and Small, D.~S. (2020).
\newblock Statistical inference in two-sample summary-data {Mendelian}
  randomization using robust adjusted profile score.
\newblock {\em The Annals of Statistics}, 48(3):1742--1769.

\bibitem[Zhou et~al., 2013]{Zhou2013}
Zhou, X., Carbonetto, P., and Stephens, M. (2013).
\newblock Polygenic modeling with {Bayesian} sparse linear mixed models.
\newblock {\em PLOS Genetics}, 9(2):e1003264.

\bibitem[Zhu et~al., 2022]{zhu2022lipidglycemic}
Zhu, Z., Wang, K., Hao, X., Chen, L., Liu, Z., and Wang, C. (2022).
\newblock Causal graph among serum lipids and glycemic traits: A {Mendelian}
  randomization study.
\newblock {\em Diabetes}, 71(8):1818--1826.

\bibitem[Zuber et~al., 2021]{zuber2021apobcad}
Zuber, V., Gill, D., Ala-Korpela, M., Langenberg, C., Butterworth, A., Bottolo,
  L., and Burgess, S. (2021).
\newblock High-throughput multivariable {Mendelian} randomization analysis
  prioritizes apolipoprotein {B} as key lipid risk factor for coronary artery
  disease.
\newblock {\em International Journal of Epidemiology}, 50(3):893--901.

\end{thebibliography}

\clearpage
\setcounter{section}{0}
\setcounter{subsection}{0}
\setcounter{figure}{0}
\setcounter{table}{0}
\setcounter{equation}{0}
\setcounter{theorem}{0}
\setcounter{proposition}{0}
\setcounter{lemma}{0}
\setcounter{assumption}{0}
\setcounter{corollary}{0}
\setcounter{definition}{0}
\setcounter{remark}{0}
\setcounter{example}{0}
\renewcommand{\thesection}{S\arabic{section}}
\renewcommand{\thesubsection}{S\arabic{section}.\arabic{subsection}}
\renewcommand{\theequation}{S\arabic{section}.\arabic{equation}}
\renewcommand{\thefigure}{S\arabic{figure}}
\renewcommand{\thetable}{S\arabic{table}}
\renewcommand{\thetheorem}{S\arabic{theorem}}
\renewcommand{\theproposition}{S\arabic{proposition}}
\renewcommand{\thelemma}{S\arabic{lemma}}
\renewcommand{\theassumption}{S\arabic{assumption}}
\renewcommand{\thecorollary}{S\arabic{corollary}}
\renewcommand{\theremark}{S\arabic{remark}}
\renewcommand{\theHsection}{supp.\arabic{section}}
\renewcommand{\theHsubsection}{supp.\arabic{section}.\arabic{subsection}}
\makeatletter
\@ifundefined{theHequation}{}{\renewcommand{\theHequation}{supp.\arabic{section}.\arabic{equation}}}
\@ifundefined{theHfigure}{}{\renewcommand{\theHfigure}{supp.\arabic{figure}}}
\@ifundefined{theHtable}{}{\renewcommand{\theHtable}{supp.\arabic{table}}}
\@ifundefined{theHtheorem}{}{\renewcommand{\theHtheorem}{supp.\arabic{theorem}}}
\@ifundefined{theHproposition}{}{\renewcommand{\theHproposition}{supp.\arabic{proposition}}}
\@ifundefined{theHlemma}{}{\renewcommand{\theHlemma}{supp.\arabic{lemma}}}
\@ifundefined{theHassumption}{}{\renewcommand{\theHassumption}{supp.\arabic{assumption}}}
\@ifundefined{theHcorollary}{}{\renewcommand{\theHcorollary}{supp.\arabic{corollary}}}
\makeatother
\makeatletter
\setlength{\@fptop}{0pt}
\setlength{\@fpsep}{14pt}
\setlength{\@fpbot}{0pt plus 1fil}
\makeatother
\pdfbookmark[0]{Supplementary Material}{supplementary-material}
\begin{center}
{\LARGE\bfseries Supplementary Material for\\[5pt]
\textit{Identification and Inference of Bi-directional Causal Effects in Mendelian Randomization without Instrument Selection}\par}
\vspace{1.4em}
{\large Liyuan Mo\textsuperscript{*}, Qingyuan Zhao\textsuperscript{\(\dagger,\ddagger\)}\par}
\end{center}
\vspace{1.5em}
\noindent This supplement provides technical details for the noisy independent component analysis formulation and asymptotic theory, complete proofs of the theoretical results, additional simulation results and diagnostics, and further information for the real-data applications.

\section{Technical details for the noisy ICA formulation}
\label{app:noisy_ica_details}

This section provides technical details for the noisy ICA representation used in
Section~\ref{sec:ica_formulation}. It compares the model with ordinary ICA,
expands the interpretation of the direct-effect assumption and the effect of
variant selection, establishes an ICA identification result, and derives the
noise-adjusted second- and fourth-order quantities used to estimate the
total-effect matrix \(\mathbf A\).

\subsection{Relation to ordinary ICA}
\label{app:relation_to_ordinary_ica}

For a single observation, ordinary ICA is usually written as
\(\mathbf x=\mathbf A\mathbf s\),
where the coordinates of \(\mathbf s\) are mutually independent and the
observations are often treated as independent and identically distributed. The
model used in this paper is different in three ways.

First, the observations are not individual-level phenotypes. They are variant-level
pairs of marginal GWAS association estimates after LD-score normalisation:
\begin{equation}
\label{eq:appendix_general_noisy_ica}
    \mathbf x_j
    =
    \mathbf A\mathbf s_j+\widetilde{\mathbf e}_j,
    \qquad
    j=1,\ldots,N.
\end{equation}
The mixing matrix \(\mathbf A=(\mathbf I_2-\mathbf B)^{-1}\) is the
total-effect matrix of the bi-directional trait system.

Second, the latent sources are not the original direct genetic effects
\(\boldsymbol\alpha_{j\cdot}\). They are LD-weighted effective sources,
\(\mathbf s_j = \ell_j^{-1/2}\sum_{r=1}^N R_{jr}\boldsymbol\alpha_{r\cdot}\).
Thus the sequence \(\{\mathbf s_j:1\leq j\leq N\}\) is generally dependent along
the genome, even when the direct effects
\(\{\boldsymbol\alpha_{j\cdot}:1\leq j\leq N\}\) are independent across variants.

Third, the additive error is heteroscedastic:
\(\operatorname{Cov}(\widetilde{\mathbf e}_j) = \ell_j^{-1}\boldsymbol\Sigma_j\).
Consequently, whitening based on the observed covariance
does not generally yield an orthogonal signal mixing matrix,
and conventional fourth-order sample cumulants do not
directly recover the corresponding signal cumulants. Both the whitening step and the fourth-order
cumulant step must therefore be corrected for the variant-specific error covariance.

The LD-score transformation is used to stabilise the scale of the latent source, not to make the variants independent. Indeed,

\[
    \operatorname{Var}(s_{jk}) = \ell_j^{-1}\sum_{r=1}^N R_{jr}^2\sigma_{\alpha,k}^2 = \sigma_{\alpha,k}^2,
\]
but nearby \(\mathbf s_j\)'s may still be dependent because their LD windows
overlap. The dependence affects large-sample variance and bootstrap inference,
but it does not by itself affect identification of the common mixing
matrix \(\mathbf A\). The following result gives the ICA part of the identification argument.

\begin{theorem}[ICA identifiability of the total-effect matrix]
\label{thm:ica_identification_A}
Consider the LD-score transformed model
\(\mathbf{x}_j = \mathbf{A}\mathbf{s}_j+\widetilde{\mathbf{e}}_j\).
Suppose Assumptions~\ref{assum:summary_error} and~\ref{assum:direct_sources}
hold and \(\mathbf A\) is invertible. Then \(\mathbf A\) is identifiable from
the distribution of \(\mathbf x_j\) up to column scaling and permutation.
\end{theorem}

\Needspace{6\baselineskip}
For later use, define the row-normalisation map for a matrix with nonzero diagonal entries by
\begin{equation}
\label{eq:row_normalisation_operator}
    \mathcal N(\widetilde{\mathbf M})
    =
    \{\operatorname{diag}(\widetilde{M}_{11},\widetilde{M}_{22})\}^{-1}\widetilde{\mathbf M}.
\end{equation}
If \(\widehat{\mathbf A}\) is an ICA mixing-matrix estimate, let
\(\mathbf P_\pi\) denote either of the two candidate row-permutation
matrices. The corresponding demixing candidate is
\begin{equation}
\label{eq:normalised_W_candidate}
    \widehat{\mathbf W}(\mathbf P_\pi)
    =
    \mathcal N\!\left(\mathbf P_\pi\widehat{\mathbf A}^{-1}\right),
\end{equation}
provided that both diagonal entries of
\(\mathbf P_\pi\widehat{\mathbf A}^{-1}\) are nonzero; otherwise
that ordering is inadmissible.

The following proposition gives the two-trait form of the stability-based
causal recovery argument for cyclic structural models \citep{Lacerda2008}.

\begin{proposition}[Causal recovery from the total-effect matrix]
\label{prop:causal_recovery_population}
Suppose \(|\beta_{12}\beta_{21}|<1\),
\(\mathbf A=(\mathbf I_2-\mathbf B)^{-1}\), and ICA identifies
\(\mathbf A\) up to column scaling and permutation. Then \(\mathbf B\) is
uniquely recovered by inverting the identified mixing matrix, normalising the
admissible row orderings to have unit diagonal, and retaining the stable
ordering.
\end{proposition}

Theorem~\ref{thm:ica_identification_A} therefore identifies
\(\mathbf A\) up to scaling and permutation. These remaining indeterminacies are
resolved by the cyclic MR structure in Section~\ref{sec:causal_recovery}.

\subsection{Source assumptions and variant selection}
\label{supp:source_selection_details}

Assumption~\ref{assum:direct_sources} is imposed on the primitive direct
genetic effects \(\boldsymbol\alpha_{j\cdot}\), before LD aggregation and causal
feedback, rather than on the observed marginal associations. Independence
across primitive variants is therefore compatible with dependence among nearby
rows of \(\mathbf R\boldsymbol\alpha\). The identical-distribution condition
posits a common marginal distribution for the variant-level direct genetic
effects and is compatible with polygenic architectures having many small
effects and a smaller fraction of larger effects
\citep{Zhou2013,Moser2015,boyle2017expanded}.

The cross-trait condition
\(\alpha_{j1}\perp\!\!\!\perp\alpha_{j2}\) rules out systematic alignment of
the two direct-effect architectures before causal propagation. It does not
require a variant to act directly on only one trait: both components may be
nonzero for the same variant. The condition is related to independence
restrictions that separate instrument strength from direct effects in robust
MR, but it is imposed on the latent direct genetic effects across the common
harmonised variant panel rather than on a selected instrument set
\citep{bowden2015egger,bowden2017_misconceptions_inside,wu2024causal}.
Non-Gaussianity supplies the distributional information required for ICA
identification \citep{Comon1994,HyvarinenOja2000} and is compatible with sparse,
heavy-tailed and mixture effect distributions
\citep{Zhou2013,Moser2015,boyle2017expanded}.

Proposition~\ref{prop:selection_breaks_independence} in the main paper shows
that selection on mixed marginal associations generally truncates the source
distribution over a non-product region and thereby changes its independence
structure. For noisy GWAS summary statistics, write
\(\mathbf x_j=\mathbf A\mathbf s_j+\widetilde{\mathbf e}_j\) and let
\(\mathcal E_j=\{\mathbf x_j\in\mathcal C\}\).
Define the selection probability conditional on the latent source by

\[
    q_j(\mathbf s)=\mathbb P\!\left(\mathbf A\mathbf s+\widetilde{\mathbf e}_j\in\mathcal C\mid\mathbf s_j=\mathbf s\right).
\]

The conditional law of \(\mathbf s_j\) given selection is proportional to
\(q_j(\mathbf s)\) times its original product law. Source independence is
therefore preserved only when this weight factorises, up to a multiplicative
constant, into separate functions of the two source coordinates. The
noise-free result is the special case
\(q_j(\mathbf s) =\mathbf 1\{\mathbf s\in\mathbf A^{-1}\mathcal C\}\).

Selection mechanisms independent of the marginal associations, or regions
that are separable in the latent source coordinates, preserve the product
structure. Direction-specific significance selection on mixed marginal
associations generally does not. This structural effect is distinct from
winner's curse, post-selection bias, loss of polygenic signal and sensitivity
to significance or LD-clumping thresholds
\citep{WangHan2021,MaWangWu2023,Jiang2023WinnerCurse,zhao2019powerful}.
The proof of Proposition~\ref{prop:selection_breaks_independence} is given in
Section~\ref{proof:selection_breaks_independence}.

\subsection{Noise-adjusted whitening}
\label{app:noise_adjusted_whitening}

For the transformed model
\(\mathbf x_j = \mathbf A\mathbf s_j+\widetilde{\mathbf e}_j\),
recall that \(\mathbf D_\alpha=\operatorname{Cov}(\mathbf s_j)\), with
\(\mathbf D_\alpha=\operatorname{diag}(\sigma_{\alpha,1}^2,\sigma_{\alpha,2}^2)\).
The signal covariance is
\(\mathbf R_s=\mathbf A\mathbf D_\alpha\mathbf A^\top\).
The observed covariance contains the additional contribution of the
summary-statistic error. Since this contribution varies with \(j\), we estimate
the signal covariance by
\[
    \widehat{\mathbf R}_s
    =
    \frac1N\sum_{j=1}^N
    \left\{
        \mathbf x_j^c(\mathbf x_j^c)^\top
        -        {\boldsymbol\Sigma}^{\,\mathrm{ld}}_j
    \right\},
    \qquad    {\boldsymbol\Sigma}^{\,\mathrm{ld}}_j
    =    \ell_j^{-1}{\boldsymbol\Sigma}_j .
\]
Let
\(\widehat{\mathbf H}=\widehat{\mathbf R}_s^{-1/2}\)
be the estimated whitening matrix. With
\(\mathbf H=\mathbf R_s^{-1/2}\),
we have
\(\mathbf O = \mathbf H\mathbf A\mathbf D_\alpha^{1/2} \in\mathcal O(2)\),
because

\[
    \mathbf O\mathbf O^\top = \mathbf H\mathbf A\mathbf D_\alpha\mathbf A^\top\mathbf H^\top = \mathbf I_2.
\]

Thus whitening reduces the estimation of \(\mathbf A\) to the estimation of an
orthogonal rotation \(\mathbf O\), followed by the inverse whitening map.

\subsection{Heteroscedastic-noise-adjusted fourth-order cumulants}
\label{app:corrected_cumulants}

Let
\(\mathbf z_j=\widehat{\mathbf H}\mathbf x_j^c\)
be the whitened observations. Under the theoretical whitening transformation,
\[
    \mathbf H\mathbf x_j
    =
    \mathbf O\mathbf t_j+\boldsymbol\zeta_j,
    \qquad
    \mathbf t_j=\mathbf D_\alpha^{-1/2}\mathbf s_j,
    \qquad
    \boldsymbol\zeta_j=\mathbf H\widetilde{\mathbf e}_j .
\]
The covariance of \(\boldsymbol\zeta_j\) is generally variant-specific:
\[
    \mathbf V_j
    =
    \mathbf H\boldsymbol\Sigma_j^{\mathrm{ld}}\mathbf H^\top,
    \qquad
    \boldsymbol\Sigma_j^{\mathrm{ld}}
    =
    \ell_j^{-1}\boldsymbol\Sigma_j .
\]
Accordingly, the covariance of the whitened observation is
\(\mathbf C_j = \mathbf I_2+\mathbf V_j\).
For any symmetric matrix \(\mathbf M\in\mathbb R^{2\times2}\), define the
heteroscedastic-noise-adjusted sample fourth-order cumulant matrix by
\begin{equation}
\label{eq:appendix_corrected_Q}
\begin{aligned}
    \widehat{\mathbf Q}(\mathbf M)
    =
    \frac1N\sum_{j=1}^N
    \Big[
        &\mathbf z_j\mathbf z_j^\top
        (\mathbf z_j^\top \mathbf M\mathbf z_j)
        -
        \widehat{\mathbf C}_j
        \operatorname{tr}(\mathbf M\widehat{\mathbf C}_j)
        -
        \widehat{\mathbf C}_j \mathbf M \widehat{\mathbf C}_j
        -
        \widehat{\mathbf C}_j \mathbf M^\top\widehat{\mathbf C}_j
    \Big],
\end{aligned}
\end{equation}
where
\[
    \widehat{\mathbf V}_j
    =
    \widehat{\mathbf H}    {\boldsymbol\Sigma}^{\,\mathrm{ld}}_j
    \widehat{\mathbf H}^\top,
    \qquad
    \widehat{\mathbf C}_j
    =
    \mathbf I_2+\widehat{\mathbf V}_j .
\]
Although \(\mathbf M\) is symmetric, the two terms
\(\widehat{\mathbf C}_j\mathbf M\widehat{\mathbf C}_j\) and
\(\widehat{\mathbf C}_j\mathbf M^\top\widehat{\mathbf C}_j\) are kept separately to
display the standard fourth-cumulant correction.

We now derive the cumulant identity underlying the common-diagonaliser
representation in \eqref{eq:population_joint_diagonalisation_main} of the main paper. The
population counterpart of \eqref{eq:appendix_corrected_Q} removes, for each
observation, the Gaussian covariance-pairing terms generated by the
heteroscedastic noise. Since \(\boldsymbol\zeta_j\) is Gaussian and
independent of \(\mathbf t_j\), its fourth-order cumulants vanish and all
mixed cumulants between \(\boldsymbol\zeta_j\) and \(\mathbf t_j\) are zero. Consequently, the corrected fourth-order cumulant operator equals that of the
signal component
\(\mathbf O\mathbf t_j\).

For $k=1,2$, write
$\bar\kappa_{k,N}=N^{-1}\sum_{j=1}^N\operatorname{cum}_4(t_{jk})$.
By component-wise independence of \(t_{j1}\) and \(t_{j2}\), the corrected
fourth-order cumulant matrix satisfies
\begin{equation}
\label{eq:appendix_population_Q}
    \mathbf Q_N(\mathbf M)
    =
    \mathbf O
    \operatorname{diag}
    \left\{
        (\mathbf O^\top \mathbf M\mathbf O)_{11}\bar\kappa_{1,N},
        (\mathbf O^\top \mathbf M\mathbf O)_{22}\bar\kappa_{2,N}
    \right\}
    \mathbf O^\top .
\end{equation}
The LD attenuation formula in
\eqref{eq:ld_attenuated_fourth_contrast} gives
$\bar\kappa_{k,N}=a_N\kappa_{\alpha,k}/\sigma_{\alpha,k}^4$.
Consequently, pre- and postmultiplying
\eqref{eq:appendix_population_Q} by $\mathbf O^\top$ and $\mathbf O$
yields \eqref{eq:population_joint_diagonalisation_main} of the main paper.
In particular, the same $\mathbf O$ jointly diagonalises these matrices
for every symmetric $\mathbf M$.
For estimation, choose a finite collection
\(\mathcal M=\{\mathbf M_1,\ldots,\mathbf M_L\}\)
spanning the space of symmetric \(2\times2\) matrices. Compute
\(\widehat{\mathbf Q}_\ell = \widehat{\mathbf Q}(\mathbf M_\ell), \quad \ell=1,\ldots,L\),
and estimate the orthogonal rotation by
\[
    \widehat{\mathbf V}
    =
    \arg\min_{\mathbf V\in\mathcal O(2)}
    \sum_{\ell=1}^L
    \left\|
        \operatorname{off}
        \left(
            \mathbf V^\top\widehat{\mathbf Q}_\ell\mathbf V
        \right)
    \right\|_F^2 .
\]
The estimated mixing matrix is then
\(\widehat{\mathbf A} = \widehat{\mathbf H}^{-1}\widehat{\mathbf V}\).
The orientation of \(\widehat{\mathbf A}\) remains subject to column scaling and
column permutation, as in ordinary ICA. These ambiguities are handled using the
causal structure in Section~\ref{sec:causal_recovery}.

\subsection{Fourth-order contrast and weak dependence}
\label{app:fourth_order_contrast}

Theorem~\ref{thm:ica_identification_A} establishes ICA identification under
component-wise independence and the usual non-Gaussianity condition. The H-JADE estimator, however, uses fourth-order cumulants. We therefore
need fourth-order contrast for the cumulant-based implementation.

A sufficient condition is that \(\bar\kappa_{k,N}\to\kappa_k\) for \(k=1,2\) as \(N\to\infty\), where \(\kappa_1\) and \(\kappa_2\) are not both zero. A formal version of this condition is given in Assumption~\ref{assum:primitive_fourth_contrast}. Under this condition,
the family of matrices in \eqref{eq:appendix_population_Q} contains enough
fourth-order information to identify the orthogonal rotation \(\mathbf O\) up to
the usual signed permutation. This condition is the fourth-order analogue of the
non-Gaussianity requirement in ordinary ICA.

Weak dependence along the genome does not change the diagonal form in
\eqref{eq:appendix_population_Q}. It affects the stochastic convergence of the
estimated matrices \(\widehat{\mathbf Q}(\mathbf M)\) and the covariance of the resulting
estimator. 

\section{Additional details for the asymptotic theory}
\label{supp:theory_details}

This section gives the technical details behind the asymptotic results in
Section~\ref{sec:theory} of the main paper. We first describe the corrected
moment vector and its long-run covariance under local LD, then record the
fourth-order contrast after LD aggregation, and finally give the
chromosome-respecting moving block bootstrap construction. Complete proofs
follow in Section~\ref{supp:proofs}. Unless otherwise specified,
asymptotic statements in this section and the proofs are for
\(N\to\infty\), with the number of traits fixed.

\subsection{Moment conditions and local LD}
\label{supp:moment_dependence_details}

Assumption~\ref{assum:moment_mixing} is stated for the primitive direct effects
because they are the independent genetic inputs in the structural model. For
\(p=8+\eta\), write
\(w_{jr}=R_{jr}/\ell_j^{1/2}\), so that \(\sum_{r=1}^Nw_{jr}^2=1\).
Standard moment inequalities for weighted sums of independent, mean-zero
vectors then give
\(\sup_j\mathbb E\|\mathbf s_j\|^{8+\eta}<\infty\).
The uniform operator-norm bound on
\(\boldsymbol\Sigma_j^{\mathrm{ld}}\) similarly bounds every Gaussian error
moment needed in the second- and fourth-order corrections. Hence
\(\mathbf x_j=\mathbf A\mathbf s_j+\widetilde{\mathbf e}_j\) has uniformly
bounded \(8+\eta\) moments.

The raw sample contributions entering the corrected moment vector are
polynomials of degree at most four in \(\mathbf x_j\), with coefficients depending on
\(\boldsymbol\Sigma_j^{\mathrm{ld}}\). They have uniformly bounded
\(2+\delta\) moments with \(\delta=\eta/4\), and the standard covariance
inequality for strongly mixing sequences therefore uses
\(\frac{\delta}{2+\delta}=\frac{\eta}{8+\eta}\),
which explains the exponent in \eqref{eq:mixing_summability}. The assumption is
a high-level representation of local LD rather than an independence condition.
A useful measure of overlap between two LD-weighted source vectors is

\[
    \omega(h) = \sup_j \frac{\sum_{r=1}^N|R_{jr}R_{j+h,r}|} {\ell_j^{1/2}\ell_{j+h}^{1/2}}.
\]
Decay of \(\omega(h)\) with genomic distance motivates the mixing condition. In
the idealized case \(R_{jr}=0\) whenever \(|j-r|>m\), the transformed sequence
is \(2m\)-dependent, apart from chromosome boundaries, and the summability
condition holds automatically.

For the proofs below, let \(\widehat{\boldsymbol\mu}_N\) denote the finite-dimensional vector of corrected second- and fourth-order sample moments used by H-JADE, let \(\boldsymbol\mu_N\) be its population counterpart, and let \(\mathbf m_{j,N}\) denote the corresponding centred first-order contribution from variant \(j\). The long-run covariance is
\begin{equation}
\label{eq:long_run_covariance}
    \boldsymbol\Omega_N
    =
    \frac1N
    \sum_{i=1}^N\sum_{j=1}^N
    \operatorname{Cov}(\mathbf m_{i,N},\mathbf m_{j,N})
    \longrightarrow
    \boldsymbol\Omega.
\end{equation}
It retains covariance terms between nearby variants and reduces to a sum of marginal variances only under independence.

\begin{lemma}[Subarray CLT and square-tail control]
\label{lem:mixing_subarray}
Let \(\{u_{j,N}:1\leq j\leq N\}\) be a mean-zero scalar triangular array.
Suppose that, for some \(\delta>0\),
\[
    \sup_{N\geq1}\max_{j\leq N}
    \mathbb E|u_{j,N}|^{2+\delta}<\infty,
    \qquad
    \sum_{h=1}^{\infty}
    \alpha_{\mathrm{mix}}(h)^{\delta/(2+\delta)}<\infty,
\]
where the strong-mixing coefficients of every row are bounded by
\(\alpha_{\mathrm{mix}}(h)\). Let \(I_N\) be any contiguous interval of
length \(m_N\to\infty\) as \(N\to\infty\). If

\[
    \frac1{m_N}\operatorname{Var} \left(\sum_{j\in I_N}u_{j,N}\right) \longrightarrow v,
\]

then

\[
    \frac1{\sqrt{m_N}}\sum_{j\in I_N}u_{j,N} \xrightarrow{d}\mathbb N(0,v).
\]

Moreover, for every deterministic sequence \(m_N\to\infty\) with
\(m_N\leq N\),
\begin{equation}
\label{eq:mixing_subarray_UI}
    \lim_{M\to\infty}\limsup_{N\to\infty}
    \sup_{\substack{I\subseteq[N]\text{ contiguous}\\|I|=m_N}}
    \mathbb E\!\left[
        \frac{(\sum_{j\in I}u_{j,N})^2}{m_N}
        \mathbf1\!\left\{
          \frac{(\sum_{j\in I}u_{j,N})^2}{m_N}>M
        \right\}
    \right]
    =0.
\end{equation}
\end{lemma}

\begin{lemma}[Corrected-moment central limit theorem]
\label{lem:moment_clt}
Under Assumptions~\ref{assum:summary_error}, \ref{assum:direct_sources}, and~\ref{assum:moment_mixing},
\[
    \sqrt N\left(
        \widehat{\boldsymbol\mu}_N-\boldsymbol\mu_N
    \right)
    \xrightarrow{d}
    \mathbb N(\mathbf0,\boldsymbol\Omega).
\]
\end{lemma}

\subsection{Fourth-order contrast under LD aggregation}
\label{supp:fourth_contrast_details}

Let \(t_{jk}=s_{jk}/\sigma_{\alpha,k}\). Independence of the primitive direct
effects across variants gives
\[
    \operatorname{cum}_4(t_{jk})
    =
    \frac{\kappa_{\alpha,k}}{\sigma_{\alpha,k}^4}
    \frac{\sum_{r=1}^N R_{jr}^4}
         {\left(\sum_{r=1}^N R_{jr}^2\right)^2}.
\]
Accordingly, let \(\mathbf Q_{j,N}(\mathbf M)\) denote the theoretical
noise-adjusted fourth-order cumulant contribution of variant \(j\). Then
\[
    \mathbf Q_{j,N}(\mathbf M)
    =
    \mathbf O
    \operatorname{diag}\!\left\{
        (\mathbf O^\top \mathbf M\mathbf O)_{11}\operatorname{cum}_4(t_{j1}),
        (\mathbf O^\top \mathbf M\mathbf O)_{22}\operatorname{cum}_4(t_{j2})
    \right\}
    \mathbf O^\top
    =
    a_{j,N}\mathbf Q_{0,N}(\mathbf M),
\]
where
\[
    \mathbf Q_{0,N}(\mathbf M)
    =
    \mathbf O
    \operatorname{diag}\!\left\{
        (\mathbf O^\top \mathbf M\mathbf O)_{11}
        \frac{\kappa_{\alpha,1}}{\sigma_{\alpha,1}^4},
        (\mathbf O^\top \mathbf M\mathbf O)_{22}
        \frac{\kappa_{\alpha,2}}{\sigma_{\alpha,2}^4}
    \right\}
    \mathbf O^\top .
\]
Averaging over variants yields
\[
    \bar\kappa_{k,N}=\frac1N\sum_{j=1}^N\operatorname{cum}_4(t_{jk})
    =
    \frac{\kappa_{\alpha,k}}{\sigma_{\alpha,k}^4}a_N
    \longrightarrow
    \frac{\kappa_{\alpha,k}}{\sigma_{\alpha,k}^4}a.
\]
Thus, when \(a>0\), LD aggregation attenuates but does not asymptotically
eliminate any primitive fourth-order contrast for which
\(\kappa_{\alpha,k}\neq0\).
As shown in Section~\ref{app:corrected_cumulants}, the noise-adjusted fourth-order cumulant matrices
\(\mathbf Q_N(\mathbf M)\) in \eqref{eq:appendix_population_Q} admit representations of
the form
\[
\mathbf Q_N(\mathbf M)
=
\mathbf O \operatorname{diag}
    \left\{
        (\mathbf O^\top \mathbf M\mathbf O)_{11}\bar\kappa_{1,N},
        (\mathbf O^\top \mathbf M\mathbf O)_{22}\bar\kappa_{2,N}
    \right\} \mathbf O^\top,
\]
and all the matrices
\(\mathbf Q_N(\mathbf M)\) can be jointly diagonalized by the same
orthogonal matrix \(\mathbf O\). Under
Assumption~\ref{assum:primitive_fourth_contrast}, at most one source direction
has a vanishing fourth-order contrast. The joint diagonalization therefore
identifies \(\mathbf O\) uniquely up to signed permutation.

Assumption~\ref{assum:primitive_fourth_contrast} therefore combines a primitive
fourth-cumulant condition with a requirement that LD aggregation not wash out
the contrast asymptotically. This is a condition for the fourth-order H-JADE
implementation, not for general independent-component identification. A non-Gaussian distribution may have zero
fourth cumulant and remain identifiable through another contrast, while
providing no identifying information to a method based only on fourth-order
cumulants.

\begin{proposition}[Local asymptotic distribution of the H-JADE mixing estimate]
\label{prop:A_asymptotic}
Under the conditions of Theorem~\ref{thm:B_asymptotic}, fix a local
signed-permutation branch of the joint diagonaliser. Let
\(\widehat{\mathbf A}^{\mathrm{loc}}\) be the corresponding locally
oriented representative of the H-JADE mixing estimate. There exist a
permutation matrix \(\mathbf P_0\) and a nonsingular diagonal matrix
\(\boldsymbol\Lambda_0\) such that, with
\(\mathbf A_0=\mathbf A\mathbf P_0\boldsymbol\Lambda_0\),
\[
    \widehat{\mathbf A}^{\mathrm{loc}}\xrightarrow{p}\mathbf A_0,
    \qquad
    \sqrt N\,\operatorname{vec}
    (\widehat{\mathbf A}^{\mathrm{loc}}-\mathbf A_0)
    \xrightarrow{d}
    \mathbb N(\mathbf0,\boldsymbol\Sigma_A),
\]
where \(\boldsymbol\Sigma_A=\mathbf J_A\boldsymbol\Omega\mathbf J_A^\top\)
and \(\mathbf J_A\) is the Jacobian of the locally oriented H-JADE map.
\end{proposition}

\subsection{Chromosome-respecting moving block bootstrap}
\label{supp:bootstrap_details}

Let
\(\mathcal D_N =\{(\mathbf x_j,\boldsymbol\Sigma_j^{\mathrm{ld}}):j=1,\ldots,N\}\),
with variants ordered by chromosome and physical position. The transformed
association and its variant-specific noise covariance are resampled together,
so every bootstrap observation retains the covariance used in the
heteroscedastic Gaussian-noise correction.

Suppose chromosome \(c\) contributes \(N_c\) variants. For a global block
length \(l_N\), set \(l_{N,c}=\min(l_N,N_c)\). The overlapping blocks on
chromosome \(c\) are
\[
    \mathcal B_{c,r}
    =\{(\mathbf x_{c,r+k},\boldsymbol\Sigma_{c,r+k}^{\mathrm{ld}}):
       0\leq k\leq l_{N,c}-1\},
    \qquad
    r=1,\ldots,N_c-l_{N,c}+1.
\]
Blocks are sampled independently with replacement within each chromosome until
exactly \(N_c\) bootstrap observations have been generated; the last sampled
block is truncated if necessary. The chromosome-specific samples are then
concatenated. Hence every bootstrap data set contains exactly \(N\)
observation--covariance pairs and preserves the original number of variants on
each chromosome. Under Proposition~\ref{prop:mbb_cov_consistency},
\(N_c/N\to\pi_c>0\) and \(l_N=o(N^{1/2})\), so
\(l_{N,c}=l_N\) for all chromosomes for sufficiently large \(N\).

Each bootstrap sample repeats centring, covariance correction, whitening,
corrected cumulant construction, joint diagonalisation, structural
normalisation and stability-based ordering. The implementation uses
\(l_N=\operatorname{round}(3N^{1/3})\)
and \(M=200\) bootstrap replications in the reported analyses.

The proposition concerns the ideal conditional covariance. For the full causal
matrix, write
\[
    \widehat{\boldsymbol\Sigma}_{B,\mathrm{MBB}}
    =\operatorname{Var}^{*}\!\left[
        \sqrt N\,\operatorname{vec}(\widehat{\mathbf B}^{*})
      \right].
\]
If \(\mathbf E_\beta\) extracts the two off-diagonal entries from
\(\operatorname{vec}(\mathbf B)\), then
\(\boldsymbol\Sigma_\beta=\mathbf E_\beta
\boldsymbol\Sigma_B\mathbf E_\beta^\top\), and the covariance in
Proposition~\ref{prop:mbb_cov_consistency} is the corresponding projection of
\(\widehat{\boldsymbol\Sigma}_{B,\mathrm{MBB}}\). With \(M\) Monte Carlo
replications, let
\(\widehat{\mathbf b}^{*(m)}=\operatorname{vec}(\widehat{\mathbf B}^{*(m)})\)
and
\(\overline{\mathbf b}^{*}=M^{-1}\sum_{m=1}^M\widehat{\mathbf b}^{*(m)}\).
The sample covariance
\begin{equation}
\label{eq:finite_m_bootstrap_cov}
    \widehat{\mathbf V}_{B,M}
    =\frac1{M-1}
    \sum_{m=1}^M
    (\widehat{\mathbf b}^{*(m)}-\overline{\mathbf b}^{*})
    (\widehat{\mathbf b}^{*(m)}-\overline{\mathbf b}^{*})^\top
\end{equation}
is the Monte Carlo approximation to the conditional covariance of
\(\operatorname{vec}(\widehat{\mathbf B}^{*})\); equivalently,
\(N\widehat{\mathbf V}_{B,M}\) approximates
\(\widehat{\boldsymbol\Sigma}_{B,\mathrm{MBB}}\), which is the ideal
bootstrap covariance studied in Proposition~\ref{prop:mbb_cov_consistency}.

The centred first-order moment array has a uniformly bounded
\(2+\eta/4\) moment and satisfies the same mixing summability condition as in
Assumption~\ref{assum:moment_mixing}. The proof of
Proposition~\ref{prop:mbb_cov_consistency} shows directly that these
conditions give the required conditional covariance and second-moment tail
control.

The remaining issue is variation in the expected fourth-order cumulant
contributions across variants. The expected noise-adjusted second-order moment
is constant across variants, whereas
\[
    \mathbf Q_{j,N}(\mathbf M)
    =a_{j,N}\mathbf Q_{0,N}(\mathbf M),
    \qquad
    a_{j,N}
    =\frac{\sum_{r=1}^N R_{jr}^4}
           {(\sum_{r=1}^N R_{jr}^2)^2}.
\]
Because each bootstrap sample has exactly \(N\) observations, bootstrap
reweighting replaces \(a_N=N^{-1}\sum_ja_{j,N}\) by a weighted average of
the same positive \(a_{j,N}\)'s. Thus, the fourth-order cumulant matrices
are all rescaled by one common positive factor. Joint
diagonalisation and causal recovery are exactly invariant to this common
rescaling, including the change in weights caused by truncating the final
block. The proof consequently expands only the centred bootstrap fluctuation
around this scale-invariant path of theoretical moments.

The ideal bootstrap map uses a uniformly bounded extension of the locally
identified H-JADE and causal-recovery map outside a fixed regular
neighbourhood. It agrees with the stated estimator on that neighbourhood. The
proof shows that the complement has conditional probability \(o_p(N^{-1})\)
and makes an asymptotically negligible contribution to the root-\(N\) second
moment.

\section{Proofs of the theoretical results}
\label{supp:proofs}
This section contains the proofs of the theoretical results in the main paper and this supplement.

\subsection{Proof of Theorem~\ref{thm:ica_identification_A}}
\label{proof:ica_identification_A}

\begin{proof}
The argument concerns identifiability from the observed distribution. It does not require the
sequence
\(\{\mathbf s_j\}_{j=1}^{N}\) to be independent or identically distributed
across \(j\). Dependence between effective source vectors corresponding to
nearby variants is therefore irrelevant for the present result. The proof uses
only the marginal distribution at a fixed index \(j\), the non-Gaussianity and within-index independence of the two source components, and the fact that the same mixing matrix \(\mathbf A\) is shared across all variants.

Fix \(j\in[N]\). Recall that
\(\mathbf x_j = \mathbf A\mathbf s_j + \widetilde{\mathbf e}_j\),
where
\(\widetilde{\mathbf e}_j \sim \mathbb N \left( \mathbf 0, \boldsymbol\Sigma_j^{\mathrm{ld}} \right)\)
is independent of \(\mathbf s_j\). Since
\(\boldsymbol\Sigma_j^{\mathrm{ld}}\) is known as part of the
summary-statistic error model, the Gaussian noise distribution is known. The
characteristic function of \(\mathbf x_j\) is therefore
\begin{equation}
\label{eq:ica_identification_characteristic}
    \varphi_{\mathbf x_j}(\mathbf t)
    =
    \varphi_{\mathbf A\mathbf s_j}(\mathbf t)
    \exp
    \left\{
        -\frac12
        \mathbf t^\top
        \boldsymbol\Sigma_j^{\mathrm{ld}}
        \mathbf t
    \right\},
    \qquad
    \mathbf t\in\mathbb R^2.
\end{equation}
The Gaussian characteristic function is strictly positive for every
\(\mathbf t\). Consequently, the characteristic function of the signal part
is uniquely determined by the distribution of \(\mathbf x_j\):
\begin{equation}
\label{eq:ica_identification_deconvolution}
    \varphi_{\mathbf A\mathbf s_j}(\mathbf t)
    =
    \varphi_{\mathbf x_j}(\mathbf t)
    \exp
    \left\{
        \frac12
        \mathbf t^\top
        \boldsymbol\Sigma_j^{\mathrm{ld}}
        \mathbf t
    \right\}.
\end{equation}
Thus the distribution of
\(\mathbf A\mathbf s_j\) is identified by Gaussian deconvolution. In
particular, the additive Gaussian noise introduces no additional ambiguity in
the identification of the signal distribution.

We next verify that the effective source vector
\(\mathbf s_j=(s_{j1},s_{j2})^\top\) satisfies the standard ICA
non-Gaussianity condition. From the LD-score transformation,
\(s_{jk}=\ell_j^{-1/2}\sum_{r=1}^N R_{jr}\alpha_{rk}\) for \(k=1,2\).
The two components \(s_{j1}\) and \(s_{j2}\) are mutually independent because
they are measurable functions of the two mutually independent families
\(\{\alpha_{r1}:r\in[N]\} \quad\text{and}\quad \{\alpha_{r2}:r\in[N]\}\),
respectively.

Moreover, at most one of \(s_{j1}\) and \(s_{j2}\) is Gaussian. To see this,
suppose that \(s_{jk}\) is Gaussian for some \(k\). It is a finite sum of
independent random variables. By Cramér's decomposition theorem, every
non-degenerate summand
\(\ell_j^{-1/2}R_{jr}\alpha_{rk}\)
with \(R_{jr}\neq0\) must then be Gaussian. Since \(R_{jj}=1\), the coefficient
of \(\alpha_{jk}\) in this sum is nonzero. Hence \(\alpha_{jk}\) must be
Gaussian. Assumption~\ref{assum:direct_sources} allows at most one of
the two primitive component distributions to be Gaussian. It follows that at
most one component of \(\mathbf s_j\) is Gaussian.

Now suppose that the same marginal distribution of \(\mathbf x_j\) admits
another representation in the same model class:
\begin{equation}
\label{eq:ica_identification_alternative}
    \mathbf x_j
    \stackrel{d}{=}
    \mathbf A^\star\mathbf s_j^\star
    +
    \widetilde{\mathbf e}_j^\star,
\end{equation}
where \(\mathbf A^\star\) is nonsingular,
\(\widetilde{\mathbf e}_j^\star \sim \mathbb N \left( \mathbf 0, \boldsymbol\Sigma_j^{\mathrm{ld}} \right)\)
is independent of \(\mathbf s_j^\star\), and the two components of
\(\mathbf s_j^\star\) are mutually independent, non-degenerate, and contain
at most one Gaussian component.

Applying the same Gaussian deconvolution as in
\eqref{eq:ica_identification_deconvolution} to the two representations gives
\(\mathbf A\mathbf s_j \stackrel{d}{=} \mathbf A^\star\mathbf s_j^\star\).
Since \(\mathbf A\) is nonsingular under the stability condition, define
\(\mathbf C = \mathbf A^{-1}\mathbf A^\star\).
Then
\begin{equation}
\label{eq:ica_identification_source_equivalence}
    \mathbf s_j
    \stackrel{d}{=}
    \mathbf C\mathbf s_j^\star.
\end{equation}
Because the components of \(\mathbf s_j\) are independent,
the two components of
\(\mathbf C\mathbf s_j^\star\) are also independent.

Write
\[
    \mathbf C
    =
    \begin{pmatrix}
        C_{11} & C_{12}\\
        C_{21} & C_{22}
    \end{pmatrix},
\]
and define the two linear forms

\[
    L_1 = C_{11}s_{j1}^\star + C_{12}s_{j2}^\star, \qquad L_2 = C_{21}s_{j1}^\star + C_{22}s_{j2}^\star.
\]

By \eqref{eq:ica_identification_source_equivalence},
\(L_1\) and \(L_2\) are independent. The Darmois--Skitovich theorem
\citep{darmois1953analyse,skitovich1953property} therefore implies that any
source component \(s_{jk}^\star\) whose coefficients are nonzero in both
\(L_1\) and \(L_2\) must be Gaussian.

First suppose that both components of \(\mathbf s_j^\star\) are non-Gaussian.
Then neither source component can appear in both linear forms. Consequently,
each column of \(\mathbf C\) contains at most one nonzero entry. Since
\(\mathbf C\) is nonsingular, every column and every row must contain exactly
one nonzero entry.

It remains to consider the case in which exactly one source component is
Gaussian. Without loss of generality, suppose that
\(s_{j2}^\star\) is Gaussian and \(s_{j1}^\star\) is non-Gaussian. The
Darmois--Skitovich theorem \citep{darmois1953analyse,skitovich1953property} implies
\(C_{11}C_{21}=0\),
because the non-Gaussian component \(s_{j1}^\star\) cannot enter both
\(L_1\) and \(L_2\).

Suppose, for example, that
\(C_{11}\neq0, \quad C_{21}=0\).
Nonsingularity of \(\mathbf C\) then implies \(C_{22}\neq0\). Since
\(L_1\) and \(L_2\) are independent and have finite second moments,
\[
\begin{aligned}
    0
    =
    \operatorname{Cov}(L_1,L_2)
    =
    C_{12}C_{22}
    \operatorname{Var}(s_{j2}^\star),
\end{aligned}
\]
where the cross-covariance involving \(s_{j1}^\star\) vanishes by source
independence. Since \(C_{22}\neq0\) and
\(\operatorname{Var}(s_{j2}^\star)>0\), it follows that
\(C_{12}=0\). Hence \(\mathbf C\) has exactly one nonzero entry in every row
and column. The case
\(C_{11}=0, \quad C_{21}\neq0\)
is symmetric. Therefore the same conclusion holds when one source component
is Gaussian.

We have shown in all cases that \(\mathbf C\) is a generalized permutation
matrix. Hence there exist a permutation matrix \(\mathbf P\) and a
nonsingular diagonal matrix \(\mathbf D\) such that
\(\mathbf C = \mathbf P\mathbf D\).
It follows that
\(\mathbf A^\star = \mathbf A\mathbf C = \mathbf A\mathbf P\mathbf D\).

Therefore the mixing matrix \(\mathbf A\) is identified from the marginal
distribution of the LD-score-transformed summary statistics up to column
permutation and nonsingular column scaling.

The argument was carried out at a fixed variant index \(j\). It does not require
the sequence \(\{\mathbf s_j\}_{j=1}^{N}\) to be independent or identically
distributed across variants. Cross-variant dependence induced by LD affects
estimation and inference, but not identification of the common mixing matrix.
\end{proof}

\subsection{Proof of Proposition~\ref{prop:selection_breaks_independence}}
\label{proof:selection_breaks_independence}

\begin{proof}
Let
\(\mathcal D = \mathbf A^{-1}\mathcal C\)
denote the selected region in the source coordinates, and write
\(p_j = \mathbb P(\mathcal E_j) = \mu_j(\mathcal D)\).
By assumption, \(0<p_j<1\).

Let \(\nu_j\) denote the conditional distribution of \(\mathbf s_j\) given
\(\mathcal E_j\). For every Borel set
\(H\subset\mathbb R^2\),

\[
    \nu_j(H) = \mathbb P(\mathbf s_j\in H\mid\mathcal E_j) = \frac{\mu_j(H\cap\mathcal D)}{p_j}.
\]

Consequently, \(\nu_j\) is absolutely continuous with respect to \(\mu_j\), with
Radon--Nikodym derivative
\[
    \frac{\mathrm d\nu_j}{\mathrm d\mu_j}(u,v)
    =
    \frac{\mathbf 1_{\mathcal D}(u,v)}{p_j},
    \qquad
    \mu_j\text{-almost everywhere}.
\]

Suppose, to the contrary, that \(s_{j1}\) and \(s_{j2}\) remain independent
under the conditional distribution given \(\mathcal E_j\). Let
\(\nu_{j1}\) and \(\nu_{j2}\) denote the two marginal distributions induced by
\(\nu_j\). Conditional independence implies
\(\nu_j = \nu_{j1}\otimes\nu_{j2}\).

Moreover,
\(\nu_{j1}\ll\mu_{j1}, \quad \nu_{j2}\ll\mu_{j2}\).
Indeed, if \(G\subset\mathbb R\) satisfies \(\mu_{j1}(G)=0\), then
\(\mu_j(G\times\mathbb R) = \mu_{j1}(G)\mu_{j2}(\mathbb R) = 0\),
and hence
\(\nu_{j1}(G) = \nu_j(G\times\mathbb R) = 0\).
The argument for \(\nu_{j2}\ll\mu_{j2}\) is identical.

Let
\(h_{j1}=\frac{\mathrm d\nu_{j1}}{\mathrm d\mu_{j1}}\) and \(h_{j2}=\frac{\mathrm d\nu_{j2}}{\mathrm d\mu_{j2}}\).
Choosing nonnegative Borel-measurable versions of these derivatives, the product
measure identity gives

\[
    \frac{ \mathrm d(\nu_{j1}\otimes\nu_{j2}) }{ \mathrm d(\mu_{j1}\otimes\mu_{j2}) }(u,v) = h_{j1}(u)h_{j2}(v)
\]

for \(\mu_j\)-almost every \((u,v)\). Since
\(\nu_j=\nu_{j1}\otimes\nu_{j2}\) and
\(\mu_j=\mu_{j1}\otimes\mu_{j2}\), it follows that
\(\mathbf 1_{\mathcal D}(u,v)/p_j=h_{j1}(u)h_{j2}(v)\)
for \(\mu_j\)-almost every \((u,v)\).

Define
\(B_1 = \{u\in\mathbb R:h_{j1}(u)>0\}, \quad B_2 = \{v\in\mathbb R:h_{j2}(v)>0\}\).
Because \(p_j>0\), the preceding equality implies
\[
    \mathbf 1_{\mathcal D}(u,v)>0
    \quad\Longleftrightarrow\quad
    h_{j1}(u)h_{j2}(v)>0
    \quad\Longleftrightarrow\quad
    (u,v)\in B_1\times B_2
\]
for \(\mu_j\)-almost every \((u,v)\). Therefore,
\(\mu_j\!\left( \mathcal D\mathbin{\triangle}(B_1\times B_2) \right) = 0\),
where \(\triangle\) denotes symmetric difference. Equivalently,
\(\mathbf A^{-1}\mathcal C = B_1\times B_2\)
up to a \(\mu_j\)-null set.

This contradicts the assumed non-rectangularity of
\(\mathbf A^{-1}\mathcal C\). Hence \(s_{j1}\) and \(s_{j2}\) cannot remain
independent after conditioning on \(\mathcal E_j\), and therefore
\(s_{j1} \not\!\perp\!\!\!\perp s_{j2} \mid\mathcal E_j\).
\end{proof}

\subsection{Proof of Proposition~\ref{prop:causal_recovery_population}}
\begin{proof}
Let
\[
    \mathbf W
    =
    \mathbf A^{-1}
    =
    \mathbf I_2-\mathbf B
    =
    \begin{pmatrix}
        1 & -\beta_{12}\\
        -\beta_{21} & 1
    \end{pmatrix}.
\]
Since
\(\mathbf A^{\mathrm{ICA}} = \mathbf A\mathbf P\mathbf D\),
the corresponding ICA demixing matrix is

\[
    \mathbf W^{\mathrm{ICA}} := (\mathbf A^{\mathrm{ICA}})^{-1} = \mathbf D^{-1}\mathbf P^\top\mathbf W.
\]

Thus \(\mathbf W^{\mathrm{ICA}}\) differs from the structural demixing matrix
\(\mathbf W\) only by row scaling and row permutation.

Let \(\mathcal P_2\) denote the set of \(2\times2\) permutation matrices.
Here \(\mathbf P\) is the column permutation in the ICA representation,
whereas \(\mathbf P_\pi\) indexes a candidate row ordering of the
demixing matrix. For each \(\mathbf P_\pi\in\mathcal P_2\), define
\(\widetilde{\mathbf W}(\mathbf P_\pi) = \mathbf P_\pi\mathbf W^{\mathrm{ICA}}\).
Whenever the diagonal entries of
\(\widetilde{\mathbf W}(\mathbf P_\pi)\) are nonzero, define the normalised
demixing candidate
\(\mathbf W(\mathbf P_\pi) = \mathcal N\{ \widetilde{\mathbf W}(\mathbf P_\pi) \}\),
where \(\mathcal N(\cdot)\) is the row-normalisation map in
\eqref{eq:row_normalisation_operator}, and let
\(\mathbf B(\mathbf P_\pi) = \mathbf I_2-\mathbf W(\mathbf P_\pi)\)
be the corresponding causal candidate.

Consider first the row permutation \(\mathbf P_\pi=\mathbf P\). Then
\(\widetilde{\mathbf W}(\mathbf P) = \mathbf P\mathbf D^{-1}\mathbf P^\top\mathbf W\).
The matrix
\(\boldsymbol\Lambda := \mathbf P\mathbf D^{-1}\mathbf P^\top\)
is nonsingular and diagonal. Because the diagonal entries of \(\mathbf W\) are
equal to one, the normalisation is well defined, and
\(\mathbf W(\mathbf P) = \mathcal N(\boldsymbol\Lambda\mathbf W) = \mathbf W\).
Consequently,
\(\mathbf B(\mathbf P) = \mathbf I_2-\mathbf W = \mathbf B\).
This candidate satisfies the stability condition \(|\beta_{12}\beta_{21}|<1\).

It remains to examine the only other row permutation. Let
\[
    \mathbf J
    =
    \begin{pmatrix}
        0 & 1\\
        1 & 0
    \end{pmatrix}
\]
denote the swap matrix. Since \(\mathcal P_2\) contains exactly two elements,
the alternative row permutation is
\(\mathbf P_\pi=\mathbf J\mathbf P\).
For this permutation,
\[
    \widetilde{\mathbf W}(\mathbf J\mathbf P)
    =
    \mathbf J\boldsymbol\Lambda\mathbf W
    =
    (\mathbf J\boldsymbol\Lambda\mathbf J)
    (\mathbf J\mathbf W).
\]
Because \(\mathbf J\boldsymbol\Lambda\mathbf J\) is again a nonsingular
diagonal matrix, row normalisation removes this factor whenever the
normalisation is well defined. Hence
\(\mathbf W(\mathbf J\mathbf P) = \mathcal N(\mathbf J\mathbf W)\).
Now
\[
    \mathbf J\mathbf W
    =
    \begin{pmatrix}
        -\beta_{21} & 1\\
        1 & -\beta_{12}
    \end{pmatrix}.
\]

If \(\beta_{12}\beta_{21}=0\), then at least one diagonal entry of
\(\mathbf J\mathbf W\) is zero. The row normalisation
\(\mathcal N(\mathbf J\mathbf W)\) is therefore not well defined, so the
row-swapped candidate is inadmissible.

Suppose instead that \(\beta_{12}\beta_{21}\neq0\). Row normalisation then gives
\[
    \mathbf W(\mathbf J\mathbf P)
    =
    \begin{pmatrix}
        1 & -1/\beta_{21}\\
        -1/\beta_{12} & 1
    \end{pmatrix},
\]
and therefore
\[
    \mathbf B(\mathbf J\mathbf P)
    =
    \mathbf I_2-\mathbf W(\mathbf J\mathbf P)
    =
    \begin{pmatrix}
        0 & 1/\beta_{21}\\
        1/\beta_{12} & 0
    \end{pmatrix}.
\]
The product of the off-diagonal entries of this candidate satisfies

\[
    \left|\frac{1}{\beta_{21}}\frac{1}{\beta_{12}}\right| =\frac{1}{|\beta_{12}\beta_{21}|}>1.
\]
Thus the swapped candidate violates the stability condition.

Thus, if \(\beta_{12}\beta_{21}=0\), the swapped candidate cannot be
row-normalised, while if \(\beta_{12}\beta_{21}\neq0\), it yields an unstable
causal matrix. In either case, the correctly ordered candidate is the unique
admissible stable candidate. Since this candidate satisfies
\(\mathbf B(\mathbf P)=\mathbf B\),
the causal matrix \(\mathbf B\) is uniquely recovered.
\end{proof}

\subsection{Proof of Theorem~\ref{thm:B_identification}}

\begin{proof}
Theorem~\ref{thm:ica_identification_A} identifies the total-effect matrix
\(\mathbf A\) up to the usual ICA scaling and permutation ambiguities.
Proposition~\ref{prop:causal_recovery_population} resolves these ambiguities under the
no-self-feedback constraint and stability condition, and hence uniquely
recovers \(\mathbf B\). The conclusion follows.
\end{proof}

\subsection{Proof of Lemma~\ref{lem:mixing_subarray}}

\begin{proof}
Put
\(p=2+\delta\), \(q=\delta/(2+\delta)\), \(a_0=1\), and \(a_h=\alpha_{\mathrm{mix}}(h)^q\) for \(h\geq1\).
Davydov's inequality gives, uniformly in the row and the indices,
\begin{equation}
\label{eq:subarray_covariance_bound}
    |\operatorname{Cov}(u_{i,N},u_{j,N})|
    \leq C a_{|i-j|},
    \qquad
    \sum_{h=0}^{\infty}a_h<\infty.
\end{equation}
In particular, the variance of a sum over any set of \(m\) indices is at
most \(Cm\).

We first establish an approximation by bounded variables that is uniform over
all contiguous intervals. For \(A>0\), define
\[
    u_{j,N}^{(A)}
    =u_{j,N}\mathbf1\{|u_{j,N}|\leq A\}
     -\mathbb E[u_{j,N}\mathbf1\{|u_{j,N}|\leq A\}],
    \qquad
    r_{j,N}^{(A)}=u_{j,N}-u_{j,N}^{(A)}.
\]
Uniform boundedness of the \(p\)-th moments implies
\(\tau_A :=\sup_{N,j}\|r_{j,N}^{(A)}\|_2\to0\) as \(A\to\infty\).
For lags not exceeding an arbitrary integer \(H\), Cauchy--Schwarz bounds
the covariance of two remainder variables by \(\tau_A^2\); for larger lags,
Davydov's inequality and the uniform \(p\)-moment bound give \(Ca_h\).
Consequently,
\begin{equation}
\label{eq:subarray_truncation_L2}
    \lim_{A\to\infty}
    \sup_{N}\sup_{I\subseteq[N]}
    \frac1{|I|}\mathbb E\!\left(
        \sum_{j\in I}r_{j,N}^{(A)}
    \right)^2
    =0,
\end{equation}
where the second supremum is over nonempty contiguous intervals. Indeed, the
displayed supremum is bounded by
\(C\{(H+1)\tau_A^2+\sum_{h>H}a_h\}\).
For each fixed \(H\), let \(A\to\infty\) to remove the first term;
then let \(H\to\infty\), using summability of \(a_h\).

We next prove the CLT for the bounded array
\(u_{j,N}^{(A)}\), with \(A\) fixed. Since \(a_h\) is nonincreasing and
summable, \(h a_h\to0\) as \(h\to\infty\), and hence
\(\alpha_{\mathrm{mix}}(h)=o(h^{-1/q})\).
Choose constants \(b,g\) such that
\(\frac{q}{1+q}<b<\frac12, \quad q(1-b)<g<b\),
and, for an interval of length \(m\), use big blocks of length
\(p_m=\lfloor m^b\rfloor\) separated by gaps of length
\(r_m=\lfloor m^g\rfloor\). The total number of observations in the gaps
and the terminal remainder is \(o(m)\). By
\eqref{eq:subarray_covariance_bound}, their sum is therefore \(o_{L^2}(m^{1/2})\).

Let \(B_{1,m},\ldots,B_{k_m,m}\) be the sums over the big blocks. A
telescoping application of the strong-mixing covariance inequality for
bounded functions gives, for each fixed \(t\), as \(m\to\infty\),
\begin{equation}
\label{eq:subarray_independent_blocks}
    \left|
      \mathbb E\exp\!\left(\frac{it}{\sqrt m}\sum_{r=1}^{k_m}B_{r,m}\right)
      -\prod_{r=1}^{k_m}
       \mathbb E\exp\!\left(\frac{it}{\sqrt m}B_{r,m}\right)
    \right|
    \leq C k_m\alpha_{\mathrm{mix}}(r_m)
    \longrightarrow0,
\end{equation}
because \(k_m=O(m^{1-b})\) and
\(1-b-g/q<0\). The sum of the covariances between distinct big blocks is
bounded in absolute value by
\(Cm\sum_{h\geq r_m}a_h=o(m)\).
It follows that the sum of the variances of independent copies of the big
blocks differs by \(o(m)\) from the variance of the complete bounded-array
sum. Finally, \(\max_{r\leq k_m}|B_{r,m}|/\sqrt m\leq 2Ap_m/\sqrt m\to0\) as \(m\to\infty\).
Thus the Lindeberg--Feller theorem for the independent block copies, together
with \eqref{eq:subarray_independent_blocks}, proves that along every sequence
on which the normalized variance of the bounded-array sum converges to
\(v_A\), that sum divided by \(\sqrt m\) converges to
\(\mathbb N(0,v_A)\). The conclusion also holds for \(v_A=0\), by
Chebyshev's inequality.

Now consider intervals \(I_N\) satisfying the variance convergence in the
statement. From any subsequence, use a diagonal argument to select a further
subsequence on which the normalized variances of the truncated sums converge
for every integer \(A\). The bounded-array result applies for each fixed
\(A\), while \eqref{eq:subarray_truncation_L2} shows that the truncated and
untruncated normalized sums are uniformly close in \(L^2\), and that their
limiting variances converge to \(v\) as \(A\to\infty\). Hence the untruncated
sum converges to \(\mathbb N(0,v)\) along the selected further subsequence.
The subsequence principle proves the first assertion.

It remains to prove \eqref{eq:mixing_subarray_UI}. If it failed, there would
exist \(M_k\to\infty\), rows \(N_k\), and intervals \(I_k\) of length
\(m_{N_k}\) for which the displayed tail expectations stay bounded away
from zero. By \eqref{eq:subarray_covariance_bound}, the normalized variances
are bounded, so a further subsequence has a limit \(v\). The first assertion
then gives, as \(k\to\infty\),
\[
    X_k:=m_{N_k}^{-1/2}\sum_{j\in I_k}u_{j,N_k}
    \xrightarrow{d}\mathbb N(0,v),
    \qquad
    \mathbb E(X_k^2)\longrightarrow v.
\]
Convergence in distribution together with convergence of the second moments
implies uniform integrability of \(\{X_k^2\}\), contradicting the choice of
\(M_k\) and \(I_k\). This proves the second assertion.
\end{proof}

\subsection{Proof of Lemma~\ref{lem:moment_clt}}

\begin{proof}
Throughout the proof, the dependence of
\(\mathbf x_j\), \(\mathbf s_j\), and
\(\boldsymbol\Sigma_j^{\mathrm{ld}}\) on \(N\) is suppressed in the notation.
Let
\(\delta=\frac{\eta}{4}\).
Then
\(2+\delta=2+\eta/4\) and \(\delta/(2+\delta)=\eta/(8+\eta)\).

We first express the sample H-JADE moments as a function of a finite-dimensional vector of
raw sample moments. Because \(\mathcal M\) is finite and the dimension of
\(\mathbf x_j\) is fixed at two, there exists a finite-dimensional measurable
map
\(\boldsymbol\psi: \mathbb R^2\times\mathbb R^{2\times2} \longrightarrow \mathbb R^d\)
such that
\(\mathbf r_{j,N} = \boldsymbol\psi \left( \mathbf x_j, \boldsymbol\Sigma_j^{\mathrm{ld}} \right)\)
collects all raw quantities required to construct the entries of
\(\widehat{\boldsymbol\mu}_N\). In particular, it is sufficient for
\(\mathbf r_{j,N}\) to contain the monomials in the entries of \(\mathbf x_j\)
of orders one through four, together with the entries of
\(\boldsymbol\Sigma_j^{\mathrm{ld}}\) and their second-order products. Define
\[
    \widehat{\boldsymbol\theta}_N
    =
    \frac1N\sum_{j=1}^N\mathbf r_{j,N},
    \qquad
    \boldsymbol\theta_N
    =
    \frac1N\sum_{j=1}^N
    \mathbb E(\mathbf r_{j,N}).
\]

Centering, noise-covariance subtraction, inverse-square-root whitening, and
the corrected cumulant construction are finite-dimensional smooth operations
whenever the signal covariance is positive definite. Consequently, there is a
map \(\boldsymbol\Phi\), continuously twice differentiable on a neighborhood of
\(\boldsymbol\theta_N\), such that
\(\widehat{\boldsymbol\mu}_N=\boldsymbol\Phi(\widehat{\boldsymbol\theta}_N)\) and \(\boldsymbol\mu_N=\boldsymbol\Phi(\boldsymbol\theta_N)\).
Indeed, the centring identities express all sample central moments of order
at most four as polynomials in the corresponding raw moments. Moreover, the
signal covariance is
\(\mathbf R_s = \mathbf A\mathbf D_\alpha\mathbf A^\top\).
Since \(\mathbf A\) is nonsingular and
\(\mathbf D_\alpha\) is positive definite under
Assumption~\ref{assum:direct_sources}, \(\mathbf R_s\) is positive definite.
The inverse-square-root map is therefore smooth in a neighborhood of
\(\mathbf R_s\). Because \(\mathcal M\) is finite, the derivatives of
\(\boldsymbol\Phi\) are uniformly bounded on a sufficiently small neighborhood
of \(\boldsymbol\theta_N\).

We next establish the required moment bounds. Set
\(p=8+\eta\).
For
\(w_{jr} = \frac{R_{jr}}{\ell_j^{1/2}}\),
we have
\(\sum_{r=1}^N w_{jr}^2=1\).
Since \(p>2\), it also follows that
\(\sum_{r=1}^N |w_{jr}|^p \leq \sum_{r=1}^N w_{jr}^2 = 1\).
Applying the standard moment inequality for weighted sums of independent
mean-zero random vectors to
\(\mathbf s_j=\sum_{r=1}^N w_{jr}\boldsymbol\alpha_{r\cdot}\)
and using Assumption~\ref{assum:moment_mixing}(i) gives

\[
    \sup_{N\geq1} \max_{j\in[N]} \mathbb E\|\mathbf s_j\|^{8+\eta} < \infty.
\]

Under Assumption~\ref{assum:summary_error}, the transformed error
\(\widetilde{\mathbf e}_j\) is Gaussian with covariance
\(\boldsymbol\Sigma_j^{\mathrm{ld}}\). The uniform boundedness of
\(\boldsymbol\Sigma_j^{\mathrm{ld}}\) therefore implies

\[
    \sup_{N\geq1} \max_{j\in[N]} \mathbb E \|\widetilde{\mathbf e}_j\|^{8+\eta} < \infty.
\]

Since
\(\mathbf x_j = \mathbf A\mathbf s_j+\widetilde{\mathbf e}_j\),
we consequently obtain
\begin{equation}
\label{eq:proof_uniform_x_moment}
    \sup_{N\geq1}
    \max_{j\in[N]}
    \mathbb E\|\mathbf x_j\|^{8+\eta}
    <
    \infty.
\end{equation}

Every coordinate of \(\mathbf r_{j,N}\) is a polynomial of degree at most four
in \(\mathbf x_j\), with coefficients involving bounded functions of
\(\boldsymbol\Sigma_j^{\mathrm{ld}}\). Because
\(4(2+\delta) = 4\left(2+\frac{\eta}{4}\right) = 8+\eta\),
equation~\eqref{eq:proof_uniform_x_moment} implies
\begin{equation}
\label{eq:proof_uniform_raw_moment}
    \sup_{N\geq1}
    \max_{j\in[N]}
    \mathbb E
    \|\mathbf r_{j,N}\|^{2+\delta}
    <
    \infty.
\end{equation}

The array
\(\left\{ \mathbf r_{j,N}:j\in[N] \right\}\)
is strongly mixing with coefficients bounded by
\(\alpha_{\mathrm{mix}}(h)\), because \(\mathbf r_{j,N}\) is a measurable
function of
\((\mathbf x_j,\boldsymbol\Sigma_j^{\mathrm{ld}})\). For each coordinate
\(a\in[d]\), set
\(Y_{j,N}^{(a)} = r_{j,N}^{(a)} - \mathbb E\bigl(r_{j,N}^{(a)}\bigr)\).
Davydov's covariance inequality and
\eqref{eq:proof_uniform_raw_moment} give a constant \(C<\infty\), independent
of \(i,j\), and \(N\), such that
\[
    \left|
        \operatorname{Cov}
        \left(
            Y_{i,N}^{(a)},
            Y_{j,N}^{(a)}
        \right)
    \right|
    \leq
    C\,
    \alpha_{\mathrm{mix}}(|i-j|)
    ^{\delta/(2+\delta)}
    =
    C\,
    \alpha_{\mathrm{mix}}(|i-j|)
    ^{\eta/(8+\eta)}.
\]
It follows that
\[
\begin{aligned}
    \operatorname{Var}
    \left(
        \frac1N\sum_{j=1}^NY_{j,N}^{(a)}
    \right)
    &\leq
    \frac{C}{N^2}
    \sum_{i=1}^N\sum_{j=1}^N
    \alpha_{\mathrm{mix}}(|i-j|)
    ^{\eta/(8+\eta)}
    \\
    &\leq
    \frac{C}{N}
    \left\{
        1+
        2\sum_{h=1}^{N-1}
        \alpha_{\mathrm{mix}}(h)
        ^{\eta/(8+\eta)}
    \right\}.
\end{aligned}
\]
By \eqref{eq:mixing_summability}, the expression in braces is uniformly
bounded. Hence
\(\widehat{\boldsymbol\theta}_N - \boldsymbol\theta_N = O_p(N^{-1/2})\).

Let
\(\mathbf G_N = D\boldsymbol\Phi \left( \boldsymbol\theta_N \right)\)
be the Jacobian of \(\boldsymbol\Phi\) at
\(\boldsymbol\theta_N\). A second-order Taylor expansion gives
\begin{equation}
\label{eq:proof_moment_linearization}
\begin{aligned}
    \widehat{\boldsymbol\mu}_N
    -
    \boldsymbol\mu_N
    &=
    \mathbf G_N
    \left(
        \widehat{\boldsymbol\theta}_N
        -
        \boldsymbol\theta_N
    \right)
    +
    \mathbf R_N,
\end{aligned}
\end{equation}
where
\[
    \|\mathbf R_N\|
    =
    O_p
    \left(
        \|
        \widehat{\boldsymbol\theta}_N
        -
        \boldsymbol\theta_N
        \|^2
    \right)
    =
    O_p(N^{-1})
    =
    o_p(N^{-1/2}).
\]

The centred first-order contribution of variant \(j\) is

\[
    \mathbf m_{j,N} = \mathbf G_N \left\{ \mathbf r_{j,N} - \mathbb E(\mathbf r_{j,N}) \right\}.
\]

Then \(\mathbb E(\mathbf m_{j,N})=\mathbf0\) and
\[
    \mathbf G_N
    \left(
        \widehat{\boldsymbol\theta}_N
        -
        \boldsymbol\theta_N
    \right)
    =
    \frac1N
    \sum_{j=1}^N
        \mathbf m_{j,N}.
\]
Thus \eqref{eq:proof_moment_linearization} becomes
\begin{equation}
\label{eq:proof_mu_asymptotic_linear}
    \widehat{\boldsymbol\mu}_N
    -
    \boldsymbol\mu_N
    =
    \frac1N
    \sum_{j=1}^N
        \mathbf m_{j,N}
    +
    o_p(N^{-1/2}).
\end{equation}

Since the matrices \(\mathbf G_N\) are uniformly bounded,
\eqref{eq:proof_uniform_raw_moment} also gives uniformly bounded
\((2+\delta)\)-th moments for \(\mathbf m_{j,N}\).
The same covariance argument as above consequently gives
\(\frac1N \sum_{j=1}^N \mathbf m_{j,N} = O_p(N^{-1/2})\).
Together with \eqref{eq:proof_mu_asymptotic_linear}, this proves
\(\widehat{\boldsymbol\mu}_N - \boldsymbol\mu_N = o_p(1)\).

\Needspace{7\baselineskip}
It remains to establish asymptotic normality. Fix
\(\mathbf t\in\mathbb R^q\), and define
\(U_{j,N}(\mathbf t) = \mathbf t^\top \mathbf m_{j,N}\).
The scalar triangular array
\(\{U_{j,N}(\mathbf t):j\in[N]\}\) is strongly mixing with coefficients bounded
by \(\alpha_{\mathrm{mix}}(h)\), and

\[
    \sup_{N\geq1} \max_{j\in[N]} \mathbb E |U_{j,N}(\mathbf t)|^{2+\delta} < \infty.
\]

Moreover, by \eqref{eq:long_run_covariance},
\[
\begin{aligned}
    \frac1N
    \operatorname{Var}
    \left\{
        \sum_{j=1}^N U_{j,N}(\mathbf t)
    \right\}
    &=
    \mathbf t^\top
    \boldsymbol\Omega_N
    \mathbf t
    \longrightarrow
    \mathbf t^\top
    \boldsymbol\Omega
    \mathbf t.
\end{aligned}
\]
Lemma~\ref{lem:mixing_subarray}, applied to the full interval
\(I_N=[N]\), therefore gives
\[
    \frac1{\sqrt N}
    \sum_{j=1}^NU_{j,N}(\mathbf t)
    \xrightarrow{d}
    \mathbb N
    \left(
        0,
        \mathbf t^\top
        \boldsymbol\Omega
        \mathbf t
    \right).
\]
Therefore, for every fixed \(\mathbf t\in\mathbb R^q\),
\[
    \mathbf t^\top
    \frac1{\sqrt N}
    \sum_{j=1}^N
    \mathbf m_{j,N}
    \xrightarrow{d}
    \mathbb N
    \left(
        0,
        \mathbf t^\top
        \boldsymbol\Omega
        \mathbf t
    \right).
\]
The Cram\'er--Wold device now yields
\[
    \frac1{\sqrt N}
    \sum_{j=1}^N
    \mathbf m_{j,N}
    \xrightarrow{d}
    \mathbb N
    \left(
        \mathbf 0,
        \boldsymbol\Omega
    \right).
\]
Finally, multiplying \eqref{eq:proof_mu_asymptotic_linear} by \(\sqrt N\) and
applying Slutsky's theorem gives
\[
    \sqrt N
    \left(
        \widehat{\boldsymbol\mu}_N
        -
        \boldsymbol\mu_N
    \right)
    \xrightarrow{d}
    \mathbb N
    \left(
        \mathbf 0,
        \boldsymbol\Omega
    \right).
\]
This completes the proof.
\end{proof}

\subsection{Proof of Proposition~\ref{prop:A_asymptotic}}

\begin{proof}
Let
\(\mathbf D_\alpha=\operatorname{diag}(\sigma_{\alpha,1}^2,\sigma_{\alpha,2}^2)\).
By Assumption~\ref{assum:direct_sources},
\(\mathbf D_\alpha\) is positive definite. Besides,
\(\mathbf A = (\mathbf I_2-\mathbf B)^{-1}\)
is nonsingular. Hence the signal covariance
\(\mathbf R_s = \mathbf A\mathbf D_\alpha\mathbf A^\top\)
is positive definite. Define
\(\mathbf H=\mathbf R_s^{-1/2}\) and \(\mathbf O=\mathbf H\mathbf A\mathbf D_\alpha^{1/2}\).
Then
\[
\begin{aligned}
    \mathbf O\mathbf O^\top
    &=
    \mathbf H\mathbf A
    \mathbf D_\alpha
    \mathbf A^\top\mathbf H^\top
    \\
    &=
    \mathbf H\mathbf R_s\mathbf H^\top
    =
    \mathbf I_2,
\end{aligned}
\]
so that
\(\mathbf O\in\mathcal O(2)\).

Writing
\(\mathbf t_j = \mathbf D_\alpha^{-1/2}\mathbf s_j\),
applying the theoretical whitening matrix gives
\[
    \mathbf H\mathbf x_j
    =
    \mathbf O\mathbf t_j+\boldsymbol\zeta_j,
    \qquad
    \boldsymbol\zeta_j
    =
    \mathbf H\widetilde{\mathbf e}_j.
\]
The components of \(\mathbf t_j\) are mutually independent and have unit
variance. The covariance matrix of \(\boldsymbol\zeta_j\) is
\(\mathbf V_j = \mathbf H \boldsymbol\Sigma_j^{\mathrm{ld}} \mathbf H^\top\).
For \(k=1,2\), define

\[
    \bar\kappa_{k,N} = \frac1N \sum_{j=1}^N \operatorname{cum}_4(t_{jk}).
\]
Because
\[
    t_{jk} = \frac{1}{ \sigma_{\alpha,k}\ell_j^{1/2} } \sum_{r=1}^N R_{jr}\alpha_{rk},
\]
the independence of the primitive effects across \(r\) gives
\[
    \operatorname{cum}_4(t_{jk})
    =
    \frac{\kappa_{\alpha,k}}{
        \sigma_{\alpha,k}^4
    }
    \frac{
        \sum_{r=1}^N R_{jr}^4
    }{
        \left(
            \sum_{r=1}^N R_{jr}^2
        \right)^2
    }.
\]
Consequently,
\[
    \bar\kappa_{k,N}
    =
    \frac{\kappa_{\alpha,k}}{
        \sigma_{\alpha,k}^4
    }a_N
    \longrightarrow
    \kappa_k
    :=
    \frac{\kappa_{\alpha,k}}{
        \sigma_{\alpha,k}^4
    }a.
\]
By Assumption~\ref{assum:primitive_fourth_contrast}, at most one of
\(\kappa_1\) and \(\kappa_2\) is zero.

For every symmetric \(\mathbf M\in\mathbb R^{2\times2}\), the population counterpart
of the corrected cumulant matrix in
\eqref{eq:appendix_corrected_Q} is, by
\eqref{eq:appendix_population_Q},
\[
    \mathbf Q_N(\mathbf M)
    =
    \mathbf O
    \operatorname{diag}
    \left\{
        (\mathbf O^\top \mathbf M\mathbf O)_{11}
        \bar\kappa_{1,N},
        \,
        (\mathbf O^\top \mathbf M\mathbf O)_{22}
        \bar\kappa_{2,N}
    \right\}
    \mathbf O^\top.
\]
The subscript \(N\) is retained here because the averaged effective-source
cumulants depend on \(N\).

Let
\(\mathcal M = \{\mathbf M_1,\ldots,\mathbf M_L\}\)
span the space of symmetric \(2\times2\) matrices, and define
\(\mathbf Q_{\ell,N} = \mathbf Q_N(\mathbf M_\ell), \quad \ell=1,\ldots,L\).
Then
\(\mathbf O^\top \mathbf Q_{\ell,N} \mathbf O = \mathbf D_{\ell,N}\),
where
\(\mathbf D_{\ell,N} = \operatorname{diag} \left( d_{\ell1,N}, d_{\ell2,N} \right)\)
and
\(d_{\ell k,N} = \bar\kappa_{k,N} (\mathbf O^\top \mathbf M_\ell\mathbf O)_{kk}\).

Define the joint-diagonalization separation quantity
\[
    \Delta_N = \sum_{\ell=1}^L \left( d_{\ell1,N}-d_{\ell2,N} \right)^2.
\]
We claim that
\begin{equation}
\label{eq:proof_A_separation}
    \liminf_{N\to\infty}\Delta_N>0.
\end{equation}
To verify this, let
\(d_{\ell k} = \kappa_k (\mathbf O^\top \mathbf M_\ell\mathbf O)_{kk}\)
and
\(\Delta = \sum_{\ell=1}^L (d_{\ell1}-d_{\ell2})^2\).
Since
\(\bar\kappa_{k,N}\to\kappa_k\), we have
\(\Delta_N\to\Delta\).

Suppose that \(\Delta=0\). Then, for every \(\ell\),
\[
    \kappa_1
    (\mathbf O^\top \mathbf M_\ell\mathbf O)_{11}
    =
    \kappa_2
    (\mathbf O^\top \mathbf M_\ell\mathbf O)_{22}.
\]
Because \(\mathcal M\) spans the symmetric \(2\times2\) matrices, the same
identity holds for every symmetric matrix \(\mathbf M\). Taking
\(\mathbf M = \mathbf O\mathbf e_1\mathbf e_1^\top\mathbf O^\top\)
gives
\(\kappa_1=0\),
whereas taking
\(\mathbf M = \mathbf O\mathbf e_2\mathbf e_2^\top\mathbf O^\top\)
gives
\(\kappa_2=0\).
This contradicts Assumption~\ref{assum:primitive_fourth_contrast}. Therefore
\(\Delta>0\), proving \eqref{eq:proof_A_separation}.

We next derive the stochastic behaviour of the whitening step. The entries of
\(\widehat{\mathbf R}_s\) form part of the estimated moment vector
\(\widehat{\boldsymbol\mu}_N\). Lemma~\ref{lem:moment_clt} therefore implies
\(\widehat{\mathbf R}_s-\mathbf R_s = O_p(N^{-1/2})\).
Since \(\mathbf R_s\) is positive definite,
\(\mathbb P \left\{ \lambda_{\min} (\widehat{\mathbf R}_s)>0 \right\} \longrightarrow1\).

Both matrix maps
\(\mathbf R \longmapsto \mathbf R^{-1/2} \)and \( \mathbf R \longmapsto \mathbf R^{1/2}\)
are continuously Fr\'echet differentiable on the cone of positive-definite
matrices. Hence
\(\widehat{\mathbf H}-\mathbf H = O_p(N^{-1/2})\)
and
\(\widehat{\mathbf H}^{-1} - \mathbf H^{-1} = O_p(N^{-1/2})\).
More precisely, since \(\widehat{\mathbf R}_s\) is a coordinate of
\(\widehat{\boldsymbol\mu}_N\), there exists a matrix
\(\mathbf J_{H^{-1}}\) such that
\begin{equation}
\label{eq:proof_H_inverse_linearization}
    \operatorname{vec}
    \left(
        \widehat{\mathbf H}^{-1}
        -
        \mathbf H^{-1}
    \right)
    =
    \mathbf J_{H^{-1}}
    \left(
        \widehat{\boldsymbol\mu}_N
        -
        \boldsymbol\mu_N
    \right)
    +
    o_p(N^{-1/2}).
\end{equation}
Consequently,
\(\sqrt N\, \operatorname{vec} \left( \widehat{\mathbf H}^{-1} - \mathbf H^{-1} \right)\)
is asymptotically normal.

We now consider the orthogonal joint-diagonalization step. Define the
population and empirical criteria
\[
    \mathcal L_N(\mathbf V)
    =
    \sum_{\ell=1}^L
    \left\|
        \operatorname{off}
        \left(
            \mathbf V^\top
            \mathbf Q_{\ell,N}
            \mathbf V
        \right)
    \right\|_F^2
\]
and
\[
    \widehat{\mathcal L}_N(\mathbf V)
    =
    \sum_{\ell=1}^L
    \left\|
        \operatorname{off}
        \left(
            \mathbf V^\top
            \widehat{\mathbf Q}_\ell
            \mathbf V
        \right)
    \right\|_F^2.
\]
Because the entries of the corrected cumulant matrices are included in
\(\widehat{\boldsymbol\mu}_N\), Lemma~\ref{lem:moment_clt} gives

\[
    \max_{1\leq\ell\leq L} \left\| \widehat{\mathbf Q}_\ell - \mathbf Q_{\ell,N} \right\|_F = O_p(N^{-1/2}).
\]

Since \(L\) is fixed and \(\mathcal O(2)\) is compact,
\[
    \sup_{\mathbf V\in\mathcal O(2)}
    \left|
        \widehat{\mathcal L}_N(\mathbf V)
        -
        \mathcal L_N(\mathbf V)
    \right|
    =
    o_p(1).
\]
By \eqref{eq:proof_A_separation}, the population minimisers of
\(\mathcal L_N\), for all sufficiently large \(N\), are precisely
\(\mathbf O\mathbf S\),
where \(\mathbf S\) ranges over the \(2\times2\) signed permutation matrices.
Thus the orthogonal rotation is identified up to the usual sign and
permutation indeterminacies.

Fix one local signed-permutation branch and choose its representative
\(\mathbf V_0 = \mathbf O\mathbf S_0\).
Since the signed-permutation orbit is finite, uniform convergence and
\eqref{eq:proof_A_separation} allow a signed-permutation matrix
\(\mathbf S_N\) to be chosen so that the locally oriented estimator

\[
    \widehat{\mathbf V}^{\mathrm{loc}} :=\widehat{\mathbf V}\mathbf S_N \xrightarrow{p} \mathbf V_0.
\]

The subsequent local expansion concerns this representative, not an
arbitrary orientation returned by the joint-diagonalisation algorithm.

We next verify the local nonsingularity needed for the
root-\(N\) expansion. Let
\[
    \mathbf R(\vartheta)
    =
    \begin{pmatrix}
        \cos\vartheta & -\sin\vartheta\\
        \sin\vartheta & \cos\vartheta
    \end{pmatrix}
\]
and use the local parameterization
\(\mathbf V(\vartheta) = \mathbf V_0\mathbf R(\vartheta)\).
After the fixed signed permutation has been absorbed into the diagonal entries,
write

\[
    \mathbf V_0^\top\mathbf Q_{\ell,N}\mathbf V_0=\operatorname{diag}(d_{\ell1,N}^{(0)},d_{\ell2,N}^{(0)}).
\]
The corresponding theoretical criterion is
\[
\begin{aligned}
    \mathcal L_N\{
        \mathbf V(\vartheta)
    \}
    &=
    2
    \sum_{\ell=1}^L
    \left(
        d_{\ell1,N}^{(0)}
        -
        d_{\ell2,N}^{(0)}
    \right)^2
    \sin^2\vartheta\cos^2\vartheta
    \\
    &=
    2\Delta_N
    \sin^2\vartheta\cos^2\vartheta.
\end{aligned}
\]
Therefore,
\[
    \left.
    \frac{\partial}{
        \partial\vartheta
    }
    \mathcal L_N\{
        \mathbf V(\vartheta)
    \}
    \right|_{\vartheta=0}
    =
    0
\]
and
\begin{equation}
\label{eq:proof_A_jd_hessian}
    \left.
    \frac{\partial^2}{
        \partial\vartheta^2
    }
    \mathcal L_N\{
        \mathbf V(\vartheta)
    \}
    \right|_{\vartheta=0}
    =
    4\Delta_N.
\end{equation}
By \eqref{eq:proof_A_separation}, the quantity in
\eqref{eq:proof_A_jd_hessian} is bounded away from zero for all sufficiently
large \(N\).

The first-order condition for the sample criterion at the locally oriented minimiser is
\[
    \left.
    \frac{\partial}{
        \partial\vartheta
    }
    \widehat{\mathcal L}_N
    \{
        \mathbf V_0\mathbf R(\vartheta)
    \}
    \right|_{\vartheta=\widehat\vartheta_N}
    =
    0.
\]
The score is a continuously differentiable function of the entries of the
corrected cumulant matrices. The nonsingularity in
\eqref{eq:proof_A_jd_hessian} therefore permits application of the implicit
function theorem. It follows that the locally oriented joint diagonalizer is a
continuously differentiable function of the estimated moment vector in a
neighborhood of \(\boldsymbol\mu_N\). Consequently, for some matrix
\(\mathbf J_V\),
\begin{equation}
\label{eq:proof_V_linearization}
    \operatorname{vec}
    \left(
        \widehat{\mathbf V}^{\mathrm{loc}}-\mathbf V_0
    \right)
    =
    \mathbf J_V
    \left(
        \widehat{\boldsymbol\mu}_N
        -
        \boldsymbol\mu_N
    \right)
    +
    o_p(N^{-1/2}).
\end{equation}
In particular,
\(\widehat{\mathbf V}^{\mathrm{loc}}-\mathbf V_0 = O_p(N^{-1/2})\),
and Lemma~\ref{lem:moment_clt} gives the asymptotic normality of the locally
oriented \(\widehat{\mathbf V}^{\mathrm{loc}}\). This is the two-dimensional local
joint-diagonalization perturbation argument underlying the general results for
fourth-order ICA estimators.

We now combine the whitening and rotation steps. By
\eqref{eq:Ahat_section3} and the definition of the local orientation,
\(\widehat{\mathbf A}^{\mathrm{loc}} = \widehat{\mathbf H}^{-1} \widehat{\mathbf V}^{\mathrm{loc}}\).
Its target under the chosen local orientation is
\[
\begin{aligned}
    \mathbf A_0
    :=
    \mathbf H^{-1}\mathbf V_0=
    \mathbf H^{-1}
    \mathbf O\mathbf S_0=
    \mathbf A
    \mathbf D_\alpha^{1/2}
    \mathbf S_0.
\end{aligned}
\]
Write the signed permutation matrix as
\(\mathbf S_0 = \mathbf P_0\mathbf D_0\),
where \(\mathbf P_0\) is a permutation matrix and
\(\mathbf D_0\) is a diagonal matrix with diagonal entries in
\(\{-1,1\}\). Then
\[
\begin{aligned}
    \mathbf D_\alpha^{1/2}
    \mathbf S_0
    =
    \mathbf D_\alpha^{1/2}
    \mathbf P_0\mathbf D_0
    =
    \mathbf P_0
    \left(
        \mathbf P_0^\top
        \mathbf D_\alpha^{1/2}
        \mathbf P_0
    \right)
    \mathbf D_0.
\end{aligned}
\]
Hence, with
\[
    \boldsymbol\Lambda_0 = \left( \mathbf P_0^\top \mathbf D_\alpha^{1/2} \mathbf P_0 \right) \mathbf D_0,
\]
the matrix \(\boldsymbol\Lambda_0\) is nonsingular and diagonal, and
\(\mathbf A_0 = \mathbf A\mathbf P_0 \boldsymbol\Lambda_0\).

Since
\[
    \widehat{\mathbf H}^{-1}
    \xrightarrow{p}
    \mathbf H^{-1}
    \qquad\text{and}\qquad
    \widehat{\mathbf V}^{\mathrm{loc}}
    \xrightarrow{p}
    \mathbf V_0,
\]
we obtain

\[
    \widehat{\mathbf A}^{\mathrm{loc}} \xrightarrow{p} \mathbf A\mathbf P_0 \boldsymbol\Lambda_0.
\]

For the root-\(N\) expansion, write
\[
\begin{aligned}
    \widehat{\mathbf A}^{\mathrm{loc}}-\mathbf A_0
    ={}
    \left(
        \widehat{\mathbf H}^{-1}
        -
        \mathbf H^{-1}
    \right)
    \mathbf V_0
    +
    \mathbf H^{-1}
    \left(
        \widehat{\mathbf V}^{\mathrm{loc}}
        -
        \mathbf V_0
    \right)    
    +
    \left(
        \widehat{\mathbf H}^{-1}
        -
        \mathbf H^{-1}
    \right)
    \left(
        \widehat{\mathbf V}^{\mathrm{loc}}
        -
        \mathbf V_0
    \right).
\end{aligned}
\]
The last term is \(O_p(N^{-1})\) because both factors are
\(O_p(N^{-1/2})\). Therefore,
\[
\begin{aligned}
    \operatorname{vec}
    \left(
        \widehat{\mathbf A}^{\mathrm{loc}}-\mathbf A_0
    \right)
    =
    \left(
        \mathbf V_0^\top\otimes\mathbf I_2
    \right)
    \operatorname{vec}
    \left(
        \widehat{\mathbf H}^{-1}
        -
        \mathbf H^{-1}
    \right)
    +
    \left(
        \mathbf I_2\otimes\mathbf H^{-1}
    \right)
    \operatorname{vec}
    \left(
        \widehat{\mathbf V}^{\mathrm{loc}}
        -
        \mathbf V_0
    \right)
    +
    o_p(N^{-1/2}).
\end{aligned}
\]
Substituting
\eqref{eq:proof_H_inverse_linearization} and
\eqref{eq:proof_V_linearization} yields
\[
    \operatorname{vec}
    \left(
        \widehat{\mathbf A}^{\mathrm{loc}}-\mathbf A_0
    \right)
    =
    \mathbf J_A
    \left(
        \widehat{\boldsymbol\mu}_N
        -
        \boldsymbol\mu_N
    \right)
    +
    o_p(N^{-1/2}),
\]
where
\begin{equation}
\label{eq:proof_J_A}
\begin{aligned}
    \mathbf J_A
    =
    \left(
        \mathbf V_0^\top\otimes\mathbf I_2
    \right)
    \mathbf J_{H^{-1}}
    +
    \left(
        \mathbf I_2\otimes\mathbf H^{-1}
    \right)
    \mathbf J_V.
\end{aligned}
\end{equation}
Thus \(\mathbf J_A\) is precisely the Jacobian of the locally oriented H-JADE
map from the population moment vector to
\(\mathbf A\mathbf P_0\boldsymbol\Lambda_0\).

Finally, by Lemma~\ref{lem:moment_clt} and
\eqref{eq:long_run_covariance},
\[
    \sqrt N
    \left(
        \widehat{\boldsymbol\mu}_N
        -
        \boldsymbol\mu_N
    \right)
    \xrightarrow{d}
    \mathbb N
    \left(
        \mathbf 0,
        \boldsymbol\Omega
    \right).
\]
Multiplying the preceding asymptotic linear expansion by \(\sqrt N\) and
applying Slutsky's theorem gives
\[
    \sqrt N\,
    \operatorname{vec}
    \left(
        \widehat{\mathbf A}^{\mathrm{loc}}
        -
        \mathbf A\mathbf P_0
        \boldsymbol\Lambda_0
    \right)
    \xrightarrow{d}
    \mathbb N
    \left(
        \mathbf 0,
        \mathbf J_A
        \boldsymbol\Omega
        \mathbf J_A^\top
    \right).
\]
Therefore,
\(\boldsymbol\Sigma_A = \mathbf J_A \boldsymbol\Omega \mathbf J_A^\top\),
which completes the proof.
\end{proof}

\subsection{Proof of Theorem~\ref{thm:B_asymptotic}}

\begin{proof}
By Proposition~\ref{prop:A_asymptotic}, on a fixed local
signed-permutation branch, there exist a permutation matrix
\(\mathbf P_0\) and a nonsingular diagonal matrix
\(\boldsymbol\Lambda_0\) such that, with
\(\mathbf A_0 = \mathbf A\mathbf P_0\boldsymbol\Lambda_0\),
we have
\begin{equation}
\label{eq:proof_B_A_limit}
    \widehat{\mathbf A}^{\mathrm{loc}}
    \xrightarrow{p}
    \mathbf A_0
\end{equation}
and
\begin{equation}
\label{eq:proof_B_A_clt}
    \sqrt N\,
    \operatorname{vec}
    \left(
        \widehat{\mathbf A}^{\mathrm{loc}}-\mathbf A_0
    \right)
    \xrightarrow{d}
    \mathbb N
    \left(
        \mathbf 0,
        \boldsymbol\Sigma_A
    \right).
\end{equation}

The causal-effect estimator does not depend on this choice of local
orientation. Indeed, any two ICA representatives differ by right
multiplication by a signed-permutation matrix \(\mathbf S\). For each
row permutation \(\mathbf P_\pi\), there are another row permutation
\(\mathbf P_{\pi'}\) and a nonsingular diagonal matrix
\(\mathbf D_\pi\) such that
\(\mathbf P_\pi\mathbf S^{-1} =\mathbf D_\pi\mathbf P_{\pi'}\).
As \(\mathcal N(\mathbf D_\pi\mathbf M)=\mathcal N(\mathbf M)\)
whenever the relevant diagonal entries are nonzero, applying the
recovery rule to \(\widehat{\mathbf A}\mathbf S\) gives exactly the same
set of admissible causal candidates as applying it to the original
\(\widehat{\mathbf A}\). The stability-selected causal-effect estimate
is consequently invariant to signed permutations. We may therefore
analyse the locally oriented representative
\(\widehat{\mathbf A}^{\mathrm{loc}}\) without changing the
estimator \(\widehat{\mathbf B}\).

Recall that
\[
    \mathbf W
    =
    \mathbf A^{-1}
    =
    \mathbf I_2-\mathbf B
    =
    \begin{pmatrix}
        1 & -\beta_{12}\\
        -\beta_{21} & 1
    \end{pmatrix}.
\]
Since
\(\mathbf A_0^{-1} = \boldsymbol\Lambda_0^{-1} \mathbf P_0^\top\mathbf W\),
the correct row permutation of \(\mathbf A_0^{-1}\) is
\(\mathbf P_0\). Indeed,

\[
    \mathbf P_0\mathbf A_0^{-1} = \mathbf P_0 \boldsymbol\Lambda_0^{-1} \mathbf P_0^\top\mathbf W.
\]

Define
\(\boldsymbol\Delta_0 = \mathbf P_0 \boldsymbol\Lambda_0^{-1} \mathbf P_0^\top\).
Because \(\boldsymbol\Lambda_0\) is nonsingular and diagonal,
\(\boldsymbol\Delta_0\) is also nonsingular and diagonal. Therefore,
\(\mathbf P_0\mathbf A_0^{-1} = \boldsymbol\Delta_0\mathbf W\).
Since the diagonal entries of \(\mathbf W\) are equal to one, the diagonal
entries of \(\boldsymbol\Delta_0\mathbf W\) are nonzero, and the row
normalisation in \eqref{eq:row_normalisation_operator} gives
\[
    \mathcal N
    \left(
        \mathbf P_0\mathbf A_0^{-1}
    \right)
    =
    \mathcal N
    \left(
        \boldsymbol\Delta_0\mathbf W
    \right)
    =
    \mathbf W.
\]
Consequently, the candidate causal matrix associated with
\(\mathbf P_0\) is

\[
    \mathbf I_2- \mathcal N \left( \mathbf P_0\mathbf A_0^{-1} \right) = \mathbf I_2-\mathbf W = \mathbf B.
\]
We first show that the stability-based procedure selects this row permutation
with probability tending to one. Let
\[
    \mathbf J
    =
    \begin{pmatrix}
        0&1\\
        1&0
    \end{pmatrix}
\]
be the swap matrix. Since \(\mathcal P_2\) contains only two elements, the
alternative row permutation is
\(\mathbf J\mathbf P_0\).
For the correctly ordered candidate, continuity of matrix inversion and
\eqref{eq:proof_B_A_limit} imply

\[
    \mathbf P_0(\widehat{\mathbf A}^{\mathrm{loc}})^{-1} \xrightarrow{p} \boldsymbol\Delta_0\mathbf W.
\]
The diagonal entries of the limit are nonzero. Hence the correct candidate is
well defined with probability tending to one, and
\[
\begin{aligned}
    \widehat{\mathbf B}(\mathbf P_0)
    &=
    \mathbf I_2-
    \mathcal N
    \left(
        \mathbf P_0(\widehat{\mathbf A}^{\mathrm{loc}})^{-1}
    \right)
    \\
    &\xrightarrow{p}
    \mathbf I_2-
    \mathcal N
    \left(
        \boldsymbol\Delta_0\mathbf W
    \right)
    =
    \mathbf B.
\end{aligned}
\]
Continuity of the off-diagonal product gives
\begin{equation}
\label{eq:proof_B_correct_product}
    \bigl|\widehat B_{12}(\mathbf P_0)
    \widehat B_{21}(\mathbf P_0)\bigr|
    \xrightarrow{p}|\beta_{12}\beta_{21}|<1.
\end{equation}

Consider now the alternative row permutation. We have
\[
\begin{aligned}
    \mathbf J\mathbf P_0\mathbf A_0^{-1}
    =
    \mathbf J\boldsymbol\Delta_0\mathbf W
    =
    \left(
        \mathbf J\boldsymbol\Delta_0\mathbf J
    \right)
    \mathbf J\mathbf W.
\end{aligned}
\]
The matrix
\(\mathbf J\boldsymbol\Delta_0\mathbf J\) is nonsingular and diagonal, so
row normalisation removes it whenever the normalisation is well defined.
Moreover,
\[
    \mathbf J\mathbf W
    =
    \begin{pmatrix}
        -\beta_{21}&1\\
        1&-\beta_{12}
    \end{pmatrix}.
\]

Suppose first that
\(\beta_{12}\beta_{21}\neq0\).
Then both diagonal entries of \(\mathbf J\mathbf W\) are nonzero, and the
candidate causal matrix under the swapped ordering is
\[
\begin{aligned}
    \mathbf B^{\mathrm{swap}}
    &=
    \mathbf I_2-
    \mathcal N(\mathbf J\mathbf W)
    \\
    &=
    \begin{pmatrix}
        0&1/\beta_{21}\\
        1/\beta_{12}&0
    \end{pmatrix}.
\end{aligned}
\]
Since the swapped candidate is well defined with probability tending
to one, its off-diagonal product satisfies
\[
    \bigl|\widehat B_{12}(\mathbf J\mathbf P_0)
    \widehat B_{21}(\mathbf J\mathbf P_0)\bigr|
    \xrightarrow{p}
    \frac{1}{|\beta_{12}\beta_{21}|}>1.
\]
Together with \eqref{eq:proof_B_correct_product}, this shows that the
correct ordering is the unique stable ordering with probability tending to one.

Suppose instead that
\(\beta_{12}\beta_{21}=0\).
At least one diagonal entry of \(\mathbf J\mathbf W\) is then zero, whereas
both off-diagonal entries are equal to one. To make the finite-sample behaviour
explicit, set
\(\widehat{\mathbf M}_N = \mathbf J\mathbf P_0 (\widehat{\mathbf A}^{\mathrm{loc}})^{-1}\).
Then
\(\widehat{\mathbf M}_N \xrightarrow{p} \mathbf J\boldsymbol\Delta_0\mathbf W\).
Write
\(\boldsymbol\Delta_0=\operatorname{diag}(\delta_1,\delta_2)\), with \(\delta_1\delta_2\ne0\).
The limiting matrix is
\[
    \mathbf J\boldsymbol\Delta_0\mathbf W
    =
    \begin{pmatrix}
        -\delta_2\beta_{21}&\delta_2\\
        \delta_1&-\delta_1\beta_{12}
    \end{pmatrix}.
\]
Whenever the swapped candidate is well defined, its off-diagonal
product satisfies
\[
    \bigl|\widehat B_{12}(\mathbf J\mathbf P_0)
    \widehat B_{21}(\mathbf J\mathbf P_0)\bigr|
    =\left|
        \frac{\widehat M_{N,12}\widehat M_{N,21}}
             {\widehat M_{N,11}\widehat M_{N,22}}
      \right|.
\]
The numerator converges in probability to \(\delta_1\delta_2\neq0\),
whereas the denominator converges in probability to
\(\delta_1\delta_2\beta_{12}\beta_{21}=0\). Consequently, for any
fixed \(K>0\), the probability that the swapped candidate is both well
defined and has an off-diagonal product of absolute value at most \(K\)
tends to zero.
If one of the relevant sample diagonal entries is exactly zero, the swapped
candidate is excluded from the admissible set by
\eqref{eq:normalised_W_candidate}. Thus, in either event, it cannot be selected
over the correctly ordered candidate with asymptotic probability bounded away
from zero.

Combining the two cases, the row ordering
\(\widehat{\mathbf P}^{\mathrm{loc}}\) selected from the locally oriented
mixing matrix satisfies
\begin{equation}
\label{eq:proof_B_selection_consistency}
    \mathbb P
    \left(
        \widehat{\mathbf P}^{\mathrm{loc}}
        =
        \mathbf P_0
    \right)
    \longrightarrow1.
\end{equation}

We next study the causal recovery map locally around \(\mathbf A_0\). Define

\[
    \mathcal G_{\mathbf P_0}(\widetilde{\mathbf M})=\mathbf I_2-\mathcal N(\mathbf P_0\widetilde{\mathbf M}^{-1}).
\]

Since \(\mathbf A_0\) is nonsingular and the diagonal entries of
\(\mathbf P_0\mathbf A_0^{-1} = \boldsymbol\Delta_0\mathbf W\)
are nonzero, \(\mathcal G_{\mathbf P_0}\) is continuously differentiable on a
neighborhood of \(\mathbf A_0\). Furthermore,
\(\mathcal G_{\mathbf P_0}(\mathbf A_0) = \mathbf B\).
On the event
\(\{\widehat{\mathbf P}^{\mathrm{loc}}=\mathbf P_0\}\),
\(\widehat{\mathbf B} = \mathcal G_{\mathbf P_0} \left( \widehat{\mathbf A}^{\mathrm{loc}} \right)\).
Together with
\eqref{eq:proof_B_selection_consistency} and
\eqref{eq:proof_B_A_limit}, this proves
\(\widehat{\mathbf B} \xrightarrow{p} \mathbf B\).

The derivative of the causal recovery map is as follows. For a
matrix \(\mathbf K\in\mathbb R^{2\times2}\), define
\(\mathscr D(\mathbf K) = \operatorname{diag}(K_{11},K_{22})\).
The derivative of the row-normalisation map at
\(\mathbf M_0 = \boldsymbol\Delta_0\mathbf W\)
in the direction \(\mathbf K\) is
\begin{equation}
\label{eq:proof_B_normalisation_derivative}
    D\mathcal N(\mathbf M_0)[\mathbf K]
    =
    \boldsymbol\Delta_0^{-1}\mathbf K
    -
    \boldsymbol\Delta_0^{-1}
    \mathscr D(\mathbf K)\mathbf W.
\end{equation}
Indeed, writing
\[
    \mathcal N(\widetilde{\mathbf M}) = \{\mathscr D(\widetilde{\mathbf M})\}^{-1}\widetilde{\mathbf M}
\]
and differentiating the inverse of the diagonal matrix
\(\mathscr D(\widetilde{\mathbf M})\) gives
\eqref{eq:proof_B_normalisation_derivative}.

For a matrix perturbation \(\mathbf F\), the derivative of matrix
inversion at \(\mathbf A_0\) maps \(\mathbf F\) to
\(-\mathbf A_0^{-1}\mathbf F\mathbf A_0^{-1}\).
Define
\(\mathbf K(\mathbf F) = \mathbf P_0 \mathbf A_0^{-1} \mathbf F \mathbf A_0^{-1}\).
The negative sign in the derivative of the inverse cancels with
the negative sign in the definition of \(\mathcal G_{\mathbf P_0}\). Hence
\begin{equation}
\label{eq:proof_B_recovery_derivative}
\begin{aligned}
    D\mathcal G_{\mathbf P_0}(\mathbf A_0)
    [\mathbf F]
    =
    \boldsymbol\Delta_0^{-1}
    \mathbf K(\mathbf F)
    -
    \boldsymbol\Delta_0^{-1}
    \mathscr D
    \left\{
        \mathbf K(\mathbf F)
    \right\}
    \mathbf W.
\end{aligned}
\end{equation}
Let \(\mathbf J_B\) denote the matrix representation of the linear map in
\eqref{eq:proof_B_recovery_derivative}, so that
\[
    \operatorname{vec}
    \left\{
        D\mathcal G_{\mathbf P_0}(\mathbf A_0)
        [\mathbf F]
    \right\}
    =
    \mathbf J_B\operatorname{vec}(\mathbf F).
\]
This is the Jacobian of the causal recovery map evaluated at the correctly
scaled and permuted mixing matrix
\(\mathbf A_0=\mathbf A\mathbf P_0\boldsymbol\Lambda_0\).

The multivariate delta method applied to
\eqref{eq:proof_B_A_clt} now gives
\[
\begin{aligned}
    \sqrt N\,
    \operatorname{vec}
    \left[
        \mathcal G_{\mathbf P_0}
        \left(
            \widehat{\mathbf A}^{\mathrm{loc}}
        \right)
        -
        \mathcal G_{\mathbf P_0}
        (\mathbf A_0)
    \right]
    \xrightarrow{d}
    \mathbb N
    \left(
        \mathbf 0,
        \mathbf J_B
        \boldsymbol\Sigma_A
        \mathbf J_B^\top
    \right).
\end{aligned}
\]
By \eqref{eq:proof_B_selection_consistency} and the signed-permutation
invariance established above,
\(\widehat{\mathbf B} = \mathcal G_{\mathbf P_0} \left( \widehat{\mathbf A}^{\mathrm{loc}} \right)\)
with probability tending to one. Therefore replacing the locally fixed
candidate by the stability-selected estimator does not affect its limiting
distribution. It follows that
\[
    \sqrt N\,
    \operatorname{vec}
    \left(
        \widehat{\mathbf B}-\mathbf B
    \right)
    \xrightarrow{d}
    \mathbb N
    \left(
        \mathbf 0,
        \boldsymbol\Sigma_B
    \right),
\]
where
\(\boldsymbol\Sigma_B = \mathbf J_B \boldsymbol\Sigma_A \mathbf J_B^\top\).
Let \(\mathbf E_\beta\) extract the entries
\((B_{12},B_{21})^\top\) from \(\operatorname{vec}(\mathbf B)\). Since
\(\boldsymbol\Sigma_A=\mathbf J_A\boldsymbol\Omega\mathbf J_A^\top\),
the delta method gives
\(\boldsymbol\Sigma_\beta=\mathbf J_\beta\boldsymbol\Omega\mathbf J_\beta^\top\), where \(\mathbf J_\beta=\mathbf E_\beta\mathbf J_B\mathbf J_A\),
which is the covariance representation stated in
Theorem~\ref{thm:B_asymptotic}. This completes the proof.
\end{proof}

\subsection{Proof of Proposition~\ref{prop:mbb_cov_consistency}}
\label{proof:mbb_cov_consistency}

\begin{proof}
Write
\(\widehat{\mathbf b}=\operatorname{vec}(\widehat{\mathbf B})\), \(\widehat{\mathbf b}^{*}=\operatorname{vec}(\widehat{\mathbf B}^{*})\), and \(\mathbf b=\operatorname{vec}(\mathbf B)\).
Recall from the proof of Lemma~\ref{lem:moment_clt} that
\(\mathbf r_{j,N}=\boldsymbol\psi(\mathbf x_j,\boldsymbol\Sigma_j^{\mathrm{ld}})\), \(\mathbf q_{j,N}=\mathbb E(\mathbf r_{j,N})\), and \(\mathbf v_{j,N}=\mathbf r_{j,N}-\mathbf q_{j,N}\),
and
\[
    \widehat{\boldsymbol\theta}_N=N^{-1}\sum_{j=1}^N\mathbf r_{j,N}
\]
and 
\[
    \boldsymbol\theta_N=N^{-1}\sum_{j=1}^N\mathbf q_{j,N}.
\]

The corrected H-JADE moment vector is
\(\widehat{\boldsymbol\mu}_N
=\boldsymbol\Phi(\widehat{\boldsymbol\theta}_N)\), with population value
\(\boldsymbol\mu_N=\boldsymbol\Phi(\boldsymbol\theta_N)\). Let
\(\mathbf G_N=D\boldsymbol\Phi(\boldsymbol\theta_N)\) and \(\mathbf m_{j,N}=\mathbf G_N\mathbf v_{j,N}\).
Then \(\mathbb E(\mathbf m_{j,N})=\mathbf0\), and
\eqref{eq:long_run_covariance} is the long-run covariance of
\(\{\mathbf m_{j,N}\}\).

Let \(K_{j,N}^{*}\) be the bootstrap multiplicity of variant \(j\) and put
\(\omega_{j,N}^{*}=K_{j,N}^{*}/N\). Since the bootstrap preserves every
chromosome size,
\(\sum_{j=1}^N\omega_{j,N}^{*}=1\), and resampling
\((\mathbf x_j,\boldsymbol\Sigma_j^{\mathrm{ld}})\) as one unit gives
\begin{equation}
\label{eq:mbb_raw_decomposition}
    \widehat{\boldsymbol\theta}_N^{*}
    =\boldsymbol\theta_N^{\dagger *}+\boldsymbol\xi_N^{*},
    \qquad
    \boldsymbol\theta_N^{\dagger *}
    =\sum_{j=1}^N\omega_{j,N}^{*}\mathbf q_{j,N},
    \qquad
    \boldsymbol\xi_N^{*}
    =\sum_{j=1}^N\omega_{j,N}^{*}\mathbf v_{j,N}.
\end{equation}

For nonnegative weights \(w_1,\ldots,w_N\) summing to one, set
\(\boldsymbol\theta_N(\mathbf w)=\sum_jw_j\mathbf q_{j,N}\). The
expected noise-adjusted second-order moment equals
$
    \mathbb E(\mathbf x_j\mathbf x_j^\top)-\boldsymbol\Sigma_j^{\mathrm{ld}}=\mathbf A\mathbf D_\alpha\mathbf A^\top
$
for every \(j\), whereas Supplementary
Section~\ref{supp:fourth_contrast_details} gives
\[
    \mathbf Q_{j,N}(\mathbf M)
    =a_{j,N}\mathbf Q_{0,N}(\mathbf M),
    \qquad
    a_{j,N}
    =\frac{\sum_{r=1}^N R_{jr}^4}
           {(\sum_{r=1}^N R_{jr}^2)^2}.
\]
Writing \(\bar a_N(\mathbf w)=\sum_jw_ja_{j,N}\) and
\(a_N=N^{-1}\sum_ja_{j,N}\), we therefore have
\begin{equation}
\label{eq:mbb_weighted_population_path}
    \boldsymbol\Phi\{\boldsymbol\theta_N(\mathbf w)\}
    =\boldsymbol\mu_N
     +\{\bar a_N(\mathbf w)-a_N\}\mathbf d_N,
\end{equation}
where the second-order coordinates of \(\mathbf d_N\) vanish and its
fourth-order coordinates are the vectorised matrices
\(\{\mathbf Q_{0,N}(\mathbf M):\mathbf M\in\mathcal M\}\). In particular,
if
\(\bar a_N^*=\sum_{j=1}^N\omega_{j,N}^*a_{j,N}\) and \(\boldsymbol\mu_N^{\dagger *}=\boldsymbol\Phi(\boldsymbol\theta_N^{\dagger *})\),
then \(\bar a_N^*>0\), because \(R_{jj}=1\), and
\begin{equation}
\label{eq:mbb_exact_cumulant_rescaling}
    \mathbf Q_N^{\dagger *}(\mathbf M)
    =\frac{\bar a_N^*}{a_N}\mathbf Q_N(\mathbf M),
    \qquad \mathbf M\in\mathcal M.
\end{equation}
Let \(\mathcal T_N\) be the locally oriented map from the corrected moments
to \(\operatorname{vec}(\mathbf B)\), using the ordering selected by the stability condition at the theoretical
target. Joint diagonalisation and causal recovery are invariant to the
common positive rescaling in
\eqref{eq:mbb_exact_cumulant_rescaling}; hence
\begin{equation}
\label{eq:mbb_exact_scale_invariance}
    \mathcal T_N(\boldsymbol\mu_N^{\dagger *})=\mathbf b
\end{equation}
whenever \(\boldsymbol\theta_N^{\dagger *}\) lies in the regular
neighbourhood of \(\boldsymbol\theta_N\).

The uniform moment bound in
\eqref{eq:proof_uniform_raw_moment} implies that
\(\sup_{N,j}\|\mathbf q_{j,N}\|<\infty\). The conditional expectation of
\(\boldsymbol\theta_N^{\dagger *}\) differs from
\(\boldsymbol\theta_N\) only through the MBB edge weights, by
\(O(l_N/N)\), while Hoeffding's inequality applied to the independent
resampled blocks gives, for every \(\varepsilon>0\),
\begin{equation}
\label{eq:mbb_deterministic_mean_concentration}
    \boldsymbol\theta_N^{\dagger *}-\boldsymbol\theta_N
    =O_{p^*}\!\left(\sqrt{\frac{l_N}{N}}+\frac{l_N}{N}\right)
    =o_{p^*}(1),
    \qquad
    N\mathbb P^*\!\left(
      \|\boldsymbol\theta_N^{\dagger *}-\boldsymbol\theta_N\|>
      \varepsilon
    \right)\xrightarrow{p}0.
\end{equation}
Indeed, after the edge bias is smaller than \(\varepsilon/2\), the
conditional tail probability is bounded by
\(C\exp(-cN/l_N)\), and \(l_N=o(N^{1/2})\).

We next derive the covariance and second-moment bounds for the centred
bootstrap component. Let \(u_{j,N}\) be any scalar projection of either
\(\mathbf v_{j,N}\) or \(\mathbf m_{j,N}\), and put
\(\delta=\eta/4\), \(q=\delta/(2+\delta)\), \(a_0=1\), and \(a_h=\alpha_{\mathrm{mix}}(h)^q\) for \(h\geq1\).
Davydov's inequality and
\eqref{eq:proof_uniform_raw_moment} yield
\begin{equation}
\label{eq:mbb_covariance_envelope}
    |\operatorname{Cov}(u_{i,N},u_{j,N})|
    \leq Ca_{|i-j|},
    \qquad
    \sum_{h=0}^{\infty}a_h<\infty.
\end{equation}
Thus every sum over \(m\) consecutive observations has variance at most
\(Cm\). Lemma~\ref{lem:mixing_subarray}, applied with \(m_N=l_N\), also
gives
\begin{equation}
\label{eq:mbb_block_square_UI}
    \lim_{M\to\infty}\limsup_{N\to\infty}
    \sup_{I:\,|I|=l_N}
    \mathbb E\!\left[
      \frac{(\sum_{j\in I}u_{j,N})^2}{l_N}
      \mathbf1\!\left\{
        \frac{(\sum_{j\in I}u_{j,N})^2}{l_N}>M
      \right\}
    \right]=0.
\end{equation}

Consider one chromosome of length \(n\), and write \(l=l_N\). For
\(i=1,\ldots,n-l+1\), define

\[
    B_{i,l}=\sum_{j=i}^{i+l-1}u_j, \qquad W_{i,l}=\frac{B_{i,l}^2}{l}.
\]

For fixed \(M\), let \(W_{i,l}^{(M)}=W_{i,l}\wedge M\), and partition the
block starting points according to their residues modulo \(l\). Blocks in
each residue class are disjoint, and the strong-mixing covariance inequality
gives
\[
    \left|
      \operatorname{Cov}\{W_{i,l}^{(M)},W_{i+hl,l}^{(M)}\}
    \right|
    \leq4M^2\alpha_{\mathrm{mix}}((h-1)l+1).
\]
Since \(\sum_h\alpha_{\mathrm{mix}}(h)<\infty\), the variance of every
residue-class average is \(O(M^2l/n)\). Hence
\[
    \frac1{n-l+1}\sum_{i=1}^{n-l+1}W_{i,l}^{(M)}
    -\frac1{n-l+1}\sum_{i=1}^{n-l+1}\mathbb EW_{i,l}^{(M)}
    \xrightarrow{p}0.
\]
Equation~\eqref{eq:mbb_block_square_UI} removes the truncation and yields
\begin{equation}
\label{eq:mbb_block_square_lln}
    \frac1{n-l+1}\sum_{i=1}^{n-l+1}W_{i,l}
    -\frac1{n-l+1}\sum_{i=1}^{n-l+1}\mathbb EW_{i,l}
    \xrightarrow{p}0,
\end{equation}
as well as, for every \(\varepsilon>0\),
\begin{equation}
\label{eq:mbb_empirical_block_UI}
    \lim_{M\to\infty}\limsup_{N\to\infty}
    \mathbb P\!\left\{
      \frac1{n-l+1}\sum_{i=1}^{n-l+1}
      W_{i,l}\mathbf1(W_{i,l}>M)>\varepsilon
    \right\}=0.
\end{equation}
Counting the overlapping blocks that contain a pair at lag \(h\) and using
\eqref{eq:mbb_covariance_envelope}, we obtain
\begin{align}
\left|
  \frac1{n-l+1}\sum_{i=1}^{n-l+1}\mathbb EW_{i,l}
  -\frac1n\operatorname{Var}\!\left(\sum_{j=1}^nu_j\right)
 \right|
\leq C\left\{
  \sum_{h\geq l}a_h
  +\sum_{h=1}^{l-1}\frac{h}{l}a_h
  +\frac{l}{n}\sum_{h=0}^{\infty}a_h
 \right\}
=o(1).
\label{eq:mbb_bartlett_bias}
\end{align}

Let \(k=\lceil n/l\rceil\) and
\(r=n-(k-1)l\in\{1,\ldots,l\}\). The bootstrap sum on this chromosome is
\(S_n^*=\sum_{b=1}^{k-1}B_{I_b,l}+B_{I_k,r}\),
where the \(I_b\)'s are conditionally independent and uniform on
\(\{1,\ldots,n-l+1\}\). The variance bound following
\eqref{eq:mbb_covariance_envelope} makes the final truncated block
\(o_p(1)\) after division by \(n\), and therefore
\begin{align}
    \frac1n\operatorname{Var}^*(S_n^*)
    & =\frac{(k-1)l}{n}\left[
        \frac1{n-l+1}\sum_{i=1}^{n-l+1}W_{i,l}
        -\frac1l\left\{
          \frac1{n-l+1}\sum_{i=1}^{n-l+1}B_{i,l}
        \right\}^2
      \right]+o_p(1).
\label{eq:mbb_conditional_variance_formula}
\end{align}
The average inside braces divided by \(l\) is a weighted mean of the
\(u_j\)'s with maximal weight \(O(1/n)\). Hence
\eqref{eq:mbb_covariance_envelope} gives

\[
    \frac1l\left\{ \frac1{n-l+1}\sum_{i=1}^{n-l+1}B_{i,l} \right\}^2 =O_p(l/n)=o_p(1).
\]

Since \((k-1)l/n\to1\), equations
\eqref{eq:mbb_block_square_lln} and
\eqref{eq:mbb_bartlett_bias} imply
\begin{equation}
\label{eq:mbb_one_chromosome_variance}
    \frac1n\operatorname{Var}^*(S_n^*)
    -\frac1n\operatorname{Var}\!\left(\sum_{j=1}^nu_j\right)
    \xrightarrow{p}0.
\end{equation}

The chromosomes are resampled independently. The covariance terms removed
between two adjacent chromosome segments are negligible because, across the
fixed number of chromosome boundaries, at most \(Ch\) pairs have
concatenated-index distance \(h\). Thus
\[
    \frac1N\left|
      \sum_{c\ne c'}
      \operatorname{Cov}\!\left(
        \sum_{j\in c}u_{j,N},\sum_{j\in c'}u_{j,N}
      \right)
    \right|
    \leq\frac{C}{N}\sum_{h=1}^Nh a_h=o(1).
\]
Summing \eqref{eq:mbb_one_chromosome_variance} over the fixed number of
chromosomes and applying polarisation to fixed linear projections gives, for
either vector array \(\{\mathbf u_{j,N}\}=\{\mathbf v_{j,N}\}\) or
\(\{\mathbf m_{j,N}\}\),
\begin{equation}
\label{eq:mbb_centred_covariance_approximation}
    \operatorname{Var}^*\!\left(
      \frac1{\sqrt N}\sum_{j=1}^NK_{j,N}^*\mathbf u_{j,N}
    \right)
    -\frac1N\operatorname{Var}\!\left(
      \sum_{j=1}^N\mathbf u_{j,N}
    \right)
    \xrightarrow{p}\mathbf0.
\end{equation}

Equation~\eqref{eq:mbb_empirical_block_UI} also gives the conditional
Lindeberg condition
\[
    \frac1n\sum_{b=1}^{k-1}
    \mathbb E^*\!\left[
      B_{I_b,l}^2
      \mathbf1\{|B_{I_b,l}|>\varepsilon\sqrt n\}
    \right]
    \xrightarrow{p}0.
\]
The final truncated block has conditional second moment
\(O_p(r/n)=o_p(1)\) after normalization, and the conditional mean of a full
block satisfies
\((n-l+1)^{-1}\sum_iB_{i,l}/\sqrt n=o_p(1)\). From every subsequence,
select a further subsequence on which the normalized within-chromosome
variance converges. The conditional Lindeberg--Feller theorem and
\eqref{eq:mbb_one_chromosome_variance} then give convergence to a centred
normal law and convergence of the corresponding conditional second moments.
The squared centred and normalized chromosome sums are therefore
conditionally uniformly integrable in probability. This property is
preserved when the fixed number of chromosome contributions are added.
Moreover,
\[
    \mathbb E^*\!\left(
      \frac1{\sqrt N}\sum_{j=1}^NK_{j,N}^*u_{j,N}
    \right)
    =\frac1{\sqrt N}\sum_{j=1}^Nu_{j,N}+o_p(1)=O_p(1),
\]
where the first equality follows from the MBB edge weights and the second
from \eqref{eq:mbb_covariance_envelope}. Thus the uncentred bootstrap sums
also have conditionally uniformly integrable squares. 

Define
\[
\begin{aligned}
\mathbf Y_N^*
=\sqrt N\boldsymbol\xi_N^*
=\frac1{\sqrt N}\sum_{j=1}^N K_{j,N}^*\mathbf v_{j,N},\quad
\mathbf Z_N^*
=\mathbf G_N\mathbf Y_N^*
=\frac1{\sqrt N}\sum_{j=1}^N K_{j,N}^*\mathbf m_{j,N}.
\end{aligned}
\]
Applying the preceding argument coordinatewise, together with
\eqref{eq:mbb_centred_covariance_approximation} and
\eqref{eq:long_run_covariance}, yields, for every
$\varepsilon>0$,
\begin{equation}
\label{eq:mbb_centred_UI_covariance}
\begin{aligned}
\lim_{M\to\infty}\limsup_{N\to\infty}
\mathbb P\!\left\{
\mathbb E^*\!\left[
\|\mathbf Y_N^*\|^2
\mathbf1\{\|\mathbf Y_N^*\|>M\}
\right]>\varepsilon
\right\}
=0, \quad \text{and} \quad
\operatorname{Var}^*(\mathbf Z_N^*)
\xrightarrow{p}\boldsymbol\Omega.
\end{aligned}
\end{equation}

Let \(\mathbf F_N=\mathcal T_N\circ\boldsymbol\Phi\), using the bounded
extension specified in Supplementary Section~\ref{supp:bootstrap_details}.
The uniform separation bound in \eqref{eq:proof_A_separation} and the
strict stability condition imply that, for all sufficiently large \(N\),
there is a fixed \(\rho>0\) such that
\(\mathbf F_N\) is twice continuously differentiable on the
\(3\rho\)-neighbourhood of \(\boldsymbol\theta_N\), its first two
derivatives are uniformly bounded there, and the stability rule selects the
correct local branch. Define
\[
    \mathcal R_N^*=\{\|\boldsymbol\theta_N^{\dagger *}-\boldsymbol\theta_N\|\leq\rho,\ \|\boldsymbol\xi_N^*\|\leq\rho\}.
\]
Equations \eqref{eq:mbb_deterministic_mean_concentration} and
\eqref{eq:mbb_centred_UI_covariance} give
\begin{equation}
\label{eq:mbb_regular_event_rate}
    N\mathbb P^*\{(\mathcal R_N^*)^c\}\xrightarrow{p}0,
\end{equation}
because
\[
    N\mathbb P^*(\|\boldsymbol\xi_N^*\|>\rho)
    \leq\rho^{-2}\mathbb E^*\!\left[
      \|\mathbf Y_N^*\|^2
      \mathbf1\{\|\mathbf Y_N^*\|>\rho\sqrt N\}
    \right]
    \xrightarrow{p}0.
\]
On \(\mathcal R_N^*\), the line segment between
\(\boldsymbol\theta_N^{\dagger *}\) and
\(\widehat{\boldsymbol\theta}_N^*\) remains in the regular neighbourhood;
therefore \eqref{eq:mbb_exact_scale_invariance} and Taylor's theorem give
\begin{equation}
\label{eq:mbb_random_derivative_expansion}
    \sqrt N(\widehat{\mathbf b}^*-\mathbf b)
    =\mathbf L_N(\boldsymbol\theta_N^{\dagger *})\mathbf Y_N^*
     +\mathbf R_{B,N}^*,
    \qquad
    \mathbf L_N(\boldsymbol\theta)=D\mathbf F_N(\boldsymbol\theta),
\end{equation}
with
\begin{equation}
\label{eq:mbb_taylor_bound}
    \|\mathbf R_{B,N}^*\|
    \leq\frac{C}{\sqrt N}\|\mathbf Y_N^*\|^2.
\end{equation}
For fixed \(M<\infty\),
\begin{align*}
\frac1N\mathbb E^*\!\left[
  \|\mathbf Y_N^*\|^4\mathbf1_{\mathcal R_N^*}
\right] \leq
  \frac{M^2}{N}\mathbb E^*\|\mathbf Y_N^*\|^2
  +\rho^2\mathbb E^*\!\left[
       \|\mathbf Y_N^*\|^2
       \mathbf1\{\|\mathbf Y_N^*\|>M\}
    \right].
\end{align*}
Letting first \(N\to\infty\) and then \(M\to\infty\) in
\eqref{eq:mbb_centred_UI_covariance} gives
\begin{equation}
\label{eq:mbb_taylor_L2}
    \mathbb E^*\!\left[
      \|\mathbf R_{B,N}^*\|^2\mathbf1_{\mathcal R_N^*}
    \right]\xrightarrow{p}0.
\end{equation}

Set \(\mathbf L_N=D\mathbf F_N(\boldsymbol\theta_N)\). Uniform boundedness
of the second derivative gives, on \(\mathcal R_N^*\),
\[
    \|\mathbf L_N(\boldsymbol\theta_N^{\dagger *})-\mathbf L_N\|
    \leq
    C\|\boldsymbol\theta_N^{\dagger *}-\boldsymbol\theta_N\|.
\]
The right-hand side is bounded and converges to zero in conditional
probability by \eqref{eq:mbb_deterministic_mean_concentration}. Conditional
uniform integrability in \eqref{eq:mbb_centred_UI_covariance} therefore gives
\begin{equation}
\label{eq:mbb_random_jacobian_L2}
    \mathbb E^*\!\left[
      \big\|
        \{\mathbf L_N(\boldsymbol\theta_N^{\dagger *})-\mathbf L_N\}
        \mathbf Y_N^*
      \big\|^2
      \mathbf1_{\mathcal R_N^*}
    \right]
    \xrightarrow{p}0.
\end{equation}
Combining \eqref{eq:mbb_random_derivative_expansion},
\eqref{eq:mbb_taylor_L2}, and \eqref{eq:mbb_random_jacobian_L2}, and then
using boundedness of the extended map,
\eqref{eq:mbb_regular_event_rate}, and conditional uniform integrability on
\((\mathcal R_N^*)^c\), yields
\begin{equation}
\label{eq:mbb_global_L2_linearization}
    \mathbb E^*\!\left[
      \left\|
        \sqrt N(\widehat{\mathbf b}^*-\mathbf b)
        -\mathbf L_N\mathbf Y_N^*
      \right\|^2
    \right]
    \xrightarrow{p}0.
\end{equation}
Indeed, the two terms outside \(\mathcal R_N^*\) satisfy
\[
    \mathbb E^*\!\left[
      N\|\widehat{\mathbf b}^*-\mathbf b\|^2
      \mathbf1_{(\mathcal R_N^*)^c}
    \right]
    \leq CN\mathbb P^*\{(\mathcal R_N^*)^c\}=o_p(1)
\]
and

\[
    \mathbb E^*\!\left[ \|\mathbf L_N\mathbf Y_N^*\|^2 \mathbf1_{(\mathcal R_N^*)^c} \right]=o_p(1).
\]
By the chain rule,
\(\mathbf L_N=\mathbf J_N\mathbf G_N\), with \(\mathbf J_N=D\mathcal T_N(\boldsymbol\mu_N)\).
The smooth-map argument in the proof of
Theorem~\ref{thm:B_asymptotic} gives \(\mathbf J_N\to\mathbf J\), where
\(\boldsymbol\Sigma_B=\mathbf J\boldsymbol\Omega\mathbf J^\top\). Since
\(\mathbf J_N\) is nonrandom under the conditional bootstrap law,
\eqref{eq:mbb_global_L2_linearization} and
\eqref{eq:mbb_centred_UI_covariance} imply
\begin{align}
    \operatorname{Var}^*\!\left[
      \sqrt N(\widehat{\mathbf b}^*-\mathbf b)
    \right]
    &=\mathbf J_N
      \operatorname{Var}^*(\mathbf Z_N^*)
      \mathbf J_N^\top+o_p(1)
\notag\\
    &\xrightarrow{p}
      \mathbf J\boldsymbol\Omega\mathbf J^\top
      =\boldsymbol\Sigma_B.
\label{eq:mbb_B_covariance_limit}
\end{align}
Subtracting \(\widehat{\mathbf b}\), which is fixed under the conditional
law, leaves the covariance unchanged. Therefore
\[
    \widehat{\boldsymbol\Sigma}_{B,\mathrm{MBB}}
    =\operatorname{Var}^*\!\left[
      \sqrt N(\widehat{\mathbf b}^*-\widehat{\mathbf b})
    \right]
    \xrightarrow{p}\boldsymbol\Sigma_B.
\]
Projecting onto the two off-diagonal entries proves

\[
    \widehat{\boldsymbol\Sigma}_{\beta,\mathrm{MBB}} \xrightarrow{p}\boldsymbol\Sigma_\beta.
\]

\end{proof}

\section{Additional simulation results and diagnostics}
\label{supp:simulation_diagnostics}

\makeatletter
\setlength{\@fptop}{0pt}
\makeatother

This section gives implementation details and additional numerical results for
Section~\ref{sec:simulation}.

\subsection{Primary simulations}
\label{supp:primary_simulations}

The summary-data standard errors were generated on the standardised-genotype scale as
\(1/\sqrt{n_{\mathrm{eff},jk}q_j}\), where $n_{\mathrm{eff},jk}$ is the variant- and trait-specific effective sample size and $q_j$ is the information score. H-JADE used 200 moving-block-bootstrap
replications. The four direction-specific comparators
used the same \(p<10^{-5}\) selection threshold and \(r^2<0.001\) clumping
rule; MR-RAPS used over-dispersion with Huber loss.

Bias, SD, CP and average confidence-interval length are calculated over 200 Monte
Carlo replications. All methods returned a point estimate and interval in every
primary replication. Table~\ref{tab:hjade_se_vs_sd} compares the Monte Carlo
standard deviations with the average H-JADE bootstrap standard errors for both $N$. The two measures are generally close, supporting the use of the moving block bootstrap for standard error estimation.

\begin{table}[!htbp]
\centering
\caption{Monte Carlo standard deviations and average H-JADE bootstrap standard errors.}
\label{tab:hjade_se_vs_sd}
\footnotesize
\setlength{\tabcolsep}{5pt}
\begin{tabular}{ccrrrr}
\toprule
& & \multicolumn{2}{c}{\(N=2000\)} & \multicolumn{2}{c}{\(N=5000\)} \\
\cmidrule(lr){3-4}\cmidrule(lr){5-6}
Setting & Target & SD & Boot. SE & SD & Boot. SE \\
\midrule
1 & $\beta_{12}$ & 0.075 & 0.067 & 0.048 & 0.041 \\
 & $\beta_{21}$ & 0.046 & 0.046 & 0.031 & 0.028 \\
\addlinespace[0.15em]
2 & $\beta_{12}$ & 0.047 & 0.042 & 0.030 & 0.026 \\
 & $\beta_{21}$ & 0.033 & 0.030 & 0.021 & 0.019 \\
\addlinespace[0.15em]
3 & $\beta_{12}$ & 0.142 & 0.145 & 0.088 & 0.075 \\
 & $\beta_{21}$ & 0.090 & 0.106 & 0.056 & 0.052 \\
\addlinespace[0.15em]
4 & $\beta_{12}$ & 0.068 & 0.064 & 0.042 & 0.036 \\
 & $\beta_{21}$ & 0.048 & 0.044 & 0.027 & 0.025 \\
\addlinespace[0.15em]
5 & $\beta_{12}$ & 0.120 & 0.139 & 0.077 & 0.072 \\
 & $\beta_{21}$ & 0.095 & 0.096 & 0.054 & 0.052 \\
\addlinespace[0.15em]
6 & $\beta_{12}$ & 0.118 & 0.137 & 0.075 & 0.072 \\
 & $\beta_{21}$ & 0.102 & 0.103 & 0.055 & 0.054 \\
\bottomrule
\end{tabular}
\end{table}

Tables~\ref{tab:supp_primary_bias_sd_n2000} and~\ref{tab:supp_primary_cp_length_n2000} report the corresponding estimation and interval results at \(N=2000\). The results show broadly similar patterns
to those at $N=5000$, with greater estimation variability
and generally longer confidence intervals at \(N=2000\).

\begin{table}[!htbp]
\centering
\caption{Bias (standard deviation) of the causal effect estimators at \(N=2000\).}
\label{tab:supp_primary_bias_sd_n2000}
\footnotesize
\setlength{\tabcolsep}{2.4pt}
\renewcommand{\arraystretch}{1.08}
\begin{tabular*}{\textwidth}{@{\extracolsep{\fill}}clr rrrrr@{}}
\toprule
Setting & Target & True & H-JADE & IVW & MR-Egger & MR-RAPS & \shortstack{Weighted\\ median} \\
\midrule
1 & $\beta_{12}$ & 0.00 & 0.001 (0.075) & 0.434 (0.098) & -0.004 (0.213) & 0.371 (0.100) & 0.068 (0.054) \\
 & $\beta_{21}$ & 0.50 & 0.001 (0.046) & -0.014 (0.041) & 0.002 (0.092) & -0.010 (0.031) & -0.033 (0.033) \\
\addlinespace[0.15em]
2 & $\beta_{12}$ & -0.40 & 0.003 (0.047) & 0.473 (0.114) & -0.237 (0.300) & 0.376 (0.114) & 0.074 (0.040) \\
 & $\beta_{21}$ & 0.50 & 0.002 (0.033) & -0.136 (0.057) & 0.250 (0.163) & -0.069 (0.038) & -0.047 (0.031) \\
\addlinespace[0.15em]
3 & $\beta_{12}$ & -0.40 & -0.009 (0.142) & 0.543 (0.125) & 0.134 (0.316) & 0.481 (0.129) & 0.229 (0.112) \\
 & $\beta_{21}$ & 0.50 & 0.008 (0.090) & -0.191 (0.070) & 0.098 (0.195) & -0.144 (0.062) & -0.111 (0.062) \\
\addlinespace[0.15em]
4 & $\beta_{12}$ & -0.40 & 0.001 (0.068) & 0.509 (0.115) & -0.107 (0.293) & 0.418 (0.119) & 0.115 (0.070) \\
 & $\beta_{21}$ & 0.50 & 0.000 (0.048) & -0.156 (0.056) & 0.163 (0.178) & -0.090 (0.044) & -0.063 (0.039) \\
\addlinespace[0.15em]
5 & $\beta_{12}$ & -0.40 & 0.007 (0.120) & 0.365 (0.219) & -0.142 (0.989) & 0.294 (0.206) & 0.161 (0.127) \\
 & $\beta_{21}$ & 0.50 & -0.002 (0.095) & -0.119 (0.093) & -0.051 (0.395) & -0.086 (0.082) & -0.103 (0.081) \\
\addlinespace[0.15em]
6 & $\beta_{12}$ & -0.40 & 0.007 (0.118) & 0.396 (0.198) & -0.166 (0.908) & 0.324 (0.192) & 0.191 (0.125) \\
 & $\beta_{21}$ & 0.50 & -0.002 (0.102) & -0.138 (0.092) & -0.001 (0.445) & -0.103 (0.084) & -0.125 (0.089) \\
\bottomrule
\end{tabular*}
\end{table}

\begin{table}[!htbp]
\centering
\caption{Empirical coverage probability (average length) of 95\% confidence intervals
at \(N=2000\).}
\label{tab:supp_primary_cp_length_n2000}
\footnotesize
\setlength{\tabcolsep}{2.4pt}
\renewcommand{\arraystretch}{1.08}
\begin{tabular*}{\textwidth}{@{\extracolsep{\fill}}clr rrrrr@{}}
\toprule
Setting & Target & True & H-JADE & IVW & MR-Egger & MR-RAPS & \shortstack{Weighted\\ median} \\
\midrule
1 & $\beta_{12}$ & 0.00 & 0.910 (0.264) & 0.000 (0.331) & 0.950 (0.858) & 0.000 (0.334) & 0.500 (0.118) \\
 & $\beta_{21}$ & 0.50 & 0.935 (0.179) & 0.920 (0.156) & 0.960 (0.393) & 0.955 (0.138) & 0.705 (0.098) \\
\addlinespace[0.15em]
2 & $\beta_{12}$ & -0.40 & 0.925 (0.166) & 0.000 (0.383) & 0.805 (1.093) & 0.015 (0.379) & 0.395 (0.122) \\
 & $\beta_{21}$ & 0.50 & 0.930 (0.119) & 0.345 (0.235) & 0.650 (0.621) & 0.800 (0.179) & 0.570 (0.101) \\
\addlinespace[0.15em]
3 & $\beta_{12}$ & -0.40 & 0.920 (0.570) & 0.000 (0.439) & 0.910 (1.198) & 0.000 (0.447) & 0.040 (0.169) \\
 & $\beta_{21}$ & 0.50 & 0.910 (0.415) & 0.240 (0.295) & 0.920 (0.746) & 0.440 (0.269) & 0.230 (0.130) \\
\addlinespace[0.15em]
4 & $\beta_{12}$ & -0.40 & 0.925 (0.251) & 0.000 (0.409) & 0.950 (1.171) & 0.010 (0.408) & 0.255 (0.139) \\
 & $\beta_{21}$ & 0.50 & 0.935 (0.174) & 0.265 (0.251) & 0.810 (0.658) & 0.650 (0.204) & 0.405 (0.111) \\
\addlinespace[0.15em]
5 & $\beta_{12}$ & -0.40 & 0.955 (0.545) & 0.455 (0.692) & 0.945 (3.469) & 0.630 (0.651) & 0.580 (0.365) \\
 & $\beta_{21}$ & 0.50 & 0.935 (0.376) & 0.715 (0.314) & 0.925 (1.444) & 0.805 (0.291) & 0.690 (0.282) \\
\addlinespace[0.15em]
6 & $\beta_{12}$ & -0.40 & 0.955 (0.537) & 0.350 (0.683) & 0.935 (3.369) & 0.495 (0.648) & 0.475 (0.360) \\
 & $\beta_{21}$ & 0.50 & 0.930 (0.403) & 0.630 (0.333) & 0.915 (1.450) & 0.745 (0.315) & 0.595 (0.282) \\
\bottomrule
\end{tabular*}
\end{table}

\subsection{Directional testing comparison}
\label{supp:focusing_comparison}

We applied the released \texttt{Focused-main.R} implementation of
\citet{LiYe2025} to the same simulated summary data. For each direction,
variants were screened at exposure-association \(p<0.05\); variants with
\(p\le10^{-5}\) were eligible as clumping index variants, and clumping required
pairwise \(r^2\le0.001\) within the simulated LD blocks. The reverse direction
used the same procedure after exchanging the two traits. For a post-clumping
candidate panel of size \(p_s\), the focused set satisfies
\[
\frac{|\widehat\gamma_{\mathrm{out}}|}
{\operatorname{se}(\widehat\gamma_{\mathrm{out}})}
\le \tau_f,
\qquad
\frac{|\widehat\gamma_{\mathrm{exp}}|}
{\operatorname{se}(\widehat\gamma_{\mathrm{exp}})}
\ge \tau_s,
\qquad
\tau_s=\Phi^{-1}(1-1/p_s).
\]
We used Fo-IVW with \(\tau_f=1.2\) and 1.5 and Fo-Median with
\(\tau_f=1.5\); Fo-Median used 300 bootstrap resamples. The released routine
was used without modifying its treatment of focused sets containing at most
two variants.

Table~\ref{tab:focusing_testing_results} reports
the rejection rates for all settings and both
causal directions, including the null cases
excluded from Figure~\ref{fig:focusing_power}.
Setting~1R is Setting~1 with the two directional effects interchanged,
\((\beta_{12},\beta_{21})=(0.5,0)\).

\begin{table}[!htbp]
\centering
\caption{Empirical rejection rates for the directional null hypotheses.}
\label{tab:focusing_testing_results}
\footnotesize
\setlength{\tabcolsep}{2.5pt}
\renewcommand{\arraystretch}{1.04}
\begin{tabular*}{\textwidth}{@{\extracolsep{\fill}}rclrrrrr@{}}
\toprule
\(N\) & Setting & Target & True & H-JADE & Fo-IVW(1.2) & Fo-IVW(1.5) & Fo-Median(1.5) \\
\midrule
2000 & 1 & $\beta_{12}$ ($2\to1$) & 0.00 & 0.090 & 0.030 & 0.025 & 0.025 \\
 &  & $\beta_{21}$ ($1\to2$) & 0.50 & 0.995 & 0.125 & 0.220 & 0.275 \\
\addlinespace[0.12em]
 & 2 & $\beta_{12}$ ($2\to1$) & -0.40 & 1.000 & 0.280 & 0.515 & 0.565 \\
 &  & $\beta_{21}$ ($1\to2$) & 0.50 & 1.000 & 0.190 & 0.395 & 0.475 \\
\addlinespace[0.12em]
 & 3 & $\beta_{12}$ ($2\to1$) & -0.40 & 0.795 & 0.120 & 0.095 & 0.165 \\
 &  & $\beta_{21}$ ($1\to2$) & 0.50 & 0.930 & 0.105 & 0.130 & 0.130 \\
\addlinespace[0.12em]
 & 4 & $\beta_{12}$ ($2\to1$) & -0.40 & 0.975 & 0.195 & 0.375 & 0.495 \\
 &  & $\beta_{21}$ ($1\to2$) & 0.50 & 0.985 & 0.180 & 0.275 & 0.410 \\
\addlinespace[0.12em]
 & 5 & $\beta_{12}$ ($2\to1$) & -0.40 & 0.840 & 0.380 & 0.315 & 0.295 \\
 &  & $\beta_{21}$ ($1\to2$) & 0.50 & 0.955 & 0.505 & 0.485 & 0.425 \\
\addlinespace[0.12em]
 & 6 & $\beta_{12}$ ($2\to1$) & -0.40 & 0.845 & 0.335 & 0.295 & 0.250 \\
 &  & $\beta_{21}$ ($1\to2$) & 0.50 & 0.945 & 0.365 & 0.350 & 0.335 \\
\addlinespace[0.12em]
 & 1R & $\beta_{12}$ ($2\to1$) & 0.50 & 0.995 & 0.115 & 0.180 & 0.285 \\
 &  & $\beta_{21}$ ($1\to2$) & 0.00 & 0.070 & 0.015 & 0.020 & 0.055 \\
\midrule
5000 & 1 & $\beta_{12}$ ($2\to1$) & 0.00 & 0.070 & 0.045 & 0.015 & 0.035 \\
 &  & $\beta_{21}$ ($1\to2$) & 0.50 & 1.000 & 0.235 & 0.500 & 0.620 \\
\addlinespace[0.12em]
 & 2 & $\beta_{12}$ ($2\to1$) & -0.40 & 1.000 & 0.625 & 0.880 & 0.920 \\
 &  & $\beta_{21}$ ($1\to2$) & 0.50 & 1.000 & 0.470 & 0.790 & 0.860 \\
\addlinespace[0.12em]
 & 3 & $\beta_{12}$ ($2\to1$) & -0.40 & 0.975 & 0.115 & 0.205 & 0.345 \\
 &  & $\beta_{21}$ ($1\to2$) & 0.50 & 0.995 & 0.115 & 0.205 & 0.270 \\
\addlinespace[0.12em]
 & 4 & $\beta_{12}$ ($2\to1$) & -0.40 & 1.000 & 0.390 & 0.685 & 0.795 \\
 &  & $\beta_{21}$ ($1\to2$) & 0.50 & 1.000 & 0.315 & 0.520 & 0.705 \\
\addlinespace[0.12em]
 & 5 & $\beta_{12}$ ($2\to1$) & -0.40 & 0.985 & 0.245 & 0.430 & 0.360 \\
 &  & $\beta_{21}$ ($1\to2$) & 0.50 & 0.990 & 0.280 & 0.545 & 0.550 \\
\addlinespace[0.12em]
 & 6 & $\beta_{12}$ ($2\to1$) & -0.40 & 0.990 & 0.175 & 0.290 & 0.280 \\
 &  & $\beta_{21}$ ($1\to2$) & 0.50 & 0.995 & 0.225 & 0.525 & 0.450 \\
\addlinespace[0.12em]
 & 1R & $\beta_{12}$ ($2\to1$) & 0.50 & 1.000 & 0.215 & 0.415 & 0.530 \\
 &  & $\beta_{21}$ ($1\to2$) & 0.00 & 0.055 & 0.020 & 0.010 & 0.050 \\
\bottomrule
\end{tabular*}
\end{table}

\subsection{Point-estimation comparison with LHC-MR}
\label{supp:lhcmr_comparison}

Each data set contained 5000 input variants. We used the
likelihood point-estimation stage of the publicly available
LHC-MR implementation, with \texttt{nStep = 2},
\texttt{SP\_single = 3}, \texttt{SP\_pair = 50}, and
\texttt{SNP\_filter = 10}. The resulting pair analysis
retained 500 variants. H-JADE was fitted both to these
same 500 variants and to the full 5000-variant panel.
LHC-MR block-jackknife inference was not run in this
comparison.

The independent-source scenario used independent trait-specific genetic sources.
The shared-heritable-component scenario additionally included the shared
source in the LHC-MR model. The generating causal effects were
\((\beta_{12},\beta_{21})=(-0.4,0.5)\). Table~\ref{tab:lhcmr_individual_estimates}
reports all five data sets in each scenario; ``Full'' and ``Matched'' denote
H-JADE fitted to the full and retained panels, respectively.

\begin{table}[!htbp]
\centering
\caption{Individual point estimates in the LHC-MR comparison.}
\label{tab:lhcmr_individual_estimates}
\footnotesize
\setlength{\tabcolsep}{3.3pt}
\begin{tabular}{lrrrrrrr}
\toprule
& & \multicolumn{3}{c}{\(\beta_{12}=-0.4\)} &
\multicolumn{3}{c}{\(\beta_{21}=0.5\)}\\
\cmidrule(lr){3-5}\cmidrule(lr){6-8}
Scenario & Data set & Full & Matched & LHC-MR &
Full & Matched & LHC-MR\\
\midrule
Independent sources & 1 & -0.381 & -0.259 & -0.262 & 0.501 & 0.405 & 0.514 \\
 & 2 & -0.388 & -0.765 & -0.661 & 0.487 & 0.606 & 0.603 \\
 & 3 & -0.346 & -0.268 & -0.361 & 0.477 & 0.474 & 0.545 \\
 & 4 & -0.434 & -0.394 & -0.383 & 0.544 & 0.536 & 0.479 \\
 & 5 & -0.458 & -0.309 & -0.336 & 0.539 & 0.390 & 0.507 \\
\addlinespace[0.2em]
Shared component & 1 & -0.298 & -0.195 & -0.229 & 0.510 & 0.408 & 0.514 \\
 & 2 & -0.319 & -0.640 & -0.511 & 0.507 & 0.643 & 0.540 \\
 & 3 & -0.269 & -0.280 & -0.219 & 0.493 & 0.547 & 0.483 \\
 & 4 & -0.372 & -0.317 & 0.321 & 0.569 & 0.539 & 0.365 \\
 & 5 & -0.357 & -0.206 & -0.355 & 0.538 & 0.420 & 0.669 \\
\bottomrule
\end{tabular}
\end{table}

The 50-start likelihood optimisation dominated the LHC-MR running time.
Median end-to-end point-estimation times were 7.94 and 7.84 hours in the
two considered scenarios. Median H-JADE point-fit times
were 0.30 and 0.45 seconds on the full panels and 0.01 and 0.03 seconds on the
matched panels. These timings exclude H-JADE moving-block-bootstrap inference
and LHC-MR block-jackknife inference.

\subsection{Sensitivity to local dependence}
\label{supp:local_dependence}

Within each LD block, the trait-\(k\) direct effects were generated as
\[
\alpha_{jk}
=
\sum_{h=0}^{5}c_h(\theta_k)\varepsilon_{j-h,k},
\qquad
c_h(\theta)
=
\frac{\theta^h}
{\{\sum_{r=0}^{5}\theta^{2r}\}^{1/2}},
\]
where \(\{\varepsilon_{jk}\}_{j=1}^N\) are independent innovations following the same
two-component Gaussian mixture as the primitive direct effects in primary
Setting~6. 
The process was restarted independently at each LD-block boundary, with
pre-block innovations generated to preserve the marginal variance near the
boundary. The two innovation sequences were independent. Within each Monte
Carlo replication, the same primitive innovations and GWAS noise were used
across the four dependence configurations. The case \((\theta_1,\theta_2)=(0,0)\) agrees with
primary Setting~6 replication by replication.

This construction departs from the covariance equalisation obtained under
independent primitive direct effects. Writing
\(\mathbf s_j=\ell_j^{-1/2}\sum_{r=1}^N R_{jr}\boldsymbol\alpha_{r\cdot}\),
the MA(5) dependence gives
\[
\operatorname{Cov}(\mathbf s_j)
=
\mathbf D_j
=
\operatorname{diag}\{\sigma_{\alpha,1}^2 f_{j1},
                     \sigma_{\alpha,2}^2 f_{j2}\},
\]
where
\[
f_{jk}
=
\frac{1}{\ell_j\sigma_{\alpha,k}^2}
\sum_{r=1}^N\sum_{q=1}^N
R_{jr}R_{jq}\operatorname{Cov}(\alpha_{rk},\alpha_{qk})
\]
measures the change in source variance induced by local dependence and LD; under
independent primitive effects, \(f_{jk}=1\). Thus \(\mathbf D_j\) is no longer
the common matrix \(\mathbf D_\alpha\) used in the primary model. Because the
two source processes remain independent, each \(\mathbf D_j\) is diagonal and
\(\overline{\mathbf D}=N^{-1}\sum_j\mathbf D_j\) remains diagonal. Hence the
average signal covariance has the form
\(\mathbf R_s=\mathbf A\overline{\mathbf D}\mathbf A^\top\),
so whitening based on \(\mathbf R_s\) still orthogonalises the common mixing
directions. What is lost is the rowwise standardisation:
\(\overline{\mathbf D}^{-1/2}\mathbf s_j\) has a variant-specific diagonal
covariance rather than \(\mathbf I_2\). The fourth-order correction derived
under common source covariance is therefore no longer exact, and the serial
dependence also changes the long-run variability of the sample moment estimators.

Table~\ref{tab:ma5_moment_diagnostics} summarises the induced departure at
\(N=5000\). The variance factor is the average of \(f_{jk}\) across variants.
The strongest first-source setting has lag-one correlation 0.767, while the
correlation between the two aggregated source processes remains close to zero.
As \(\theta_1\) increases from 0 to 0.8, the first-source variance factor rises
from 1.00 to 3.37 and its excess kurtosis falls from 17.31 to 5.64. These two
changes work in different directions: positive MA dependence combined with
positive LD increases the variance of the aggregated genetic signal, while the
moving average makes the source less strongly non-Gaussian. When
\((\theta_1,\theta_2)=(0.5,0.5)\), the two variance-factor profiles are
identical across variants; at \((0.8,0.5)\), their ratio has coefficient of
variation 0.147. The source-specific part of the rowwise scaling departure is
therefore smaller than the raw serial correlations alone suggest. The
median whitening condition number was 1.58 at \((0,0)\) and 2.37 at
\((0.8,0.5)\).

\begin{table}[!htbp]
\centering
\caption{Dependence and source-moment diagnostics for the MA(5) experiment at \(N=5000\).}
\label{tab:ma5_moment_diagnostics}
\footnotesize
\setlength{\tabcolsep}{2.7pt}
\begin{tabular}{crrrrrrr}
\toprule
& \multicolumn{2}{c}{Lag-1 corr.} &
\multicolumn{2}{c}{Variance factor} &
\multicolumn{2}{c}{Excess kurtosis} & Cross-source \\
\cmidrule(lr){2-3}\cmidrule(lr){4-5}\cmidrule(lr){6-7}
\((\theta_1,\theta_2)\) & Source 1 & Source 2 & Source 1 & Source 2 &
Source 1 & Source 2 & corr. \\
\midrule
$(0,0)$ & 0.000 & 0.000 & 1.000 & 1.000 & 17.314 & 14.862 & 0.0002 \\
$(0.2,0.5)$ & 0.200 & 0.500 & 1.346 & 2.171 & 12.759 & 6.950 & 0.0004 \\
$(0.5,0.5)$ & 0.500 & 0.500 & 2.171 & 2.171 & 7.876 & 6.950 & 0.0009 \\
$(0.8,0.5)$ & 0.767 & 0.500 & 3.369 & 2.171 & 5.638 & 6.950 & 0.0015 \\
\bottomrule
\end{tabular}
\end{table}

\section{Additional real-data information}
\label{supp:real_data_information}

\subsection{GWAS sources}
\label{supp:real_data_sources}

Table~\ref{tab:supp_realdata_gwas} lists the five GWAS datasets used
in the LDL-C--CHD, ApoB--CHD and TG--T2D applications. All three lipid
GWAS were conducted in UK Biobank; the CHD and strict T2D GWAS came from
FinnGen. Each trait pair therefore combines UK Biobank with FinnGen
summary statistics, limiting the scope for sample overlap between
the two GWAS in that pair. The original GWAS sample sizes are reported
below. 

\begin{table}[!htbp]
\centering
\caption{GWAS datasets used in the real-data analyses.}
\label{tab:supp_realdata_gwas}
\footnotesize
\setlength{\tabcolsep}{3.2pt}
\begin{tabular*}{\textwidth}{@{\extracolsep{\fill}}llllr}
\toprule
Trait & OpenGWAS ID & Primary data source & Ancestry & Sample size \\
\midrule
LDL-C & \texttt{ieu-b-110} & UK Biobank & European & 440,546 \\
ApoB & \texttt{ieu-b-108} & UK Biobank & European & 439,214 \\
TG & \texttt{ieu-b-111} & UK Biobank & European & 441,016 \\
CHD & \texttt{finn-b-I9\_CHD} & FinnGen & European & 218,792 \\
T2D & \texttt{finn-b-E4\_DM2\_STRICT} & FinnGen & European & 212,351 \\
\bottomrule
\end{tabular*}
\end{table}

\subsection{Estimation procedures and complete results}
\label{supp:real_data_full_results}

For H-JADE, alleles were harmonised between the two GWAS and variants with
incompatible alleles were removed.
The harmonised variants were then matched to a European-ancestry LD-score
reference \citep{buliksullivan2015atlas,bulik2015ldsc}. 

The GWAS-VCF associations and standard errors are reported per effect
allele. For SNP $j$, let $f_j$ denote its allele frequency in the
European genotype reference and define
$d_j=\{2f_j(1-f_j)\}^{1/2}$.
After matching the reference and harmonised allele pairs, the same
$d_j$ was applied to both traits:
\begin{equation}
\widehat x_{jk}
=\frac{d_j\widehat\Gamma_{jk}}{\sqrt{\ell_j}},
\qquad
\widetilde\sigma_{jk}
=\frac{d_j\widehat\sigma_{jk}}{\sqrt{\ell_j}},
\qquad k\in\{1,2\}.
\label{eq:supp_realdata_genotype_scaling}
\end{equation}
Here $\ell_j$ is the LD score, and
$\widehat\Gamma_{jk}$ and $\widehat\sigma_{jk}$ denote the per-allele
GWAS association estimate and its standard error. The resulting
noise-covariance diagonal entries are $\widetilde\sigma_{jk}^2$.
This conversion changes the genotype unit without altering the
phenotype-specific outcome scales. Cross-trait sampling-error
covariances were set to zero, reflecting the use of distinct GWAS
cohorts. H-JADE used the resulting variant-specific noise covariances
in its second- and fourth-order moment corrections.

Standard errors for H-JADE were estimated using 200
chromosome-stratified moving block bootstrap replications, with block
length $\operatorname{round}(3N^{1/3})$ for a final panel of $N$ SNPs.
Two-sided $p$-values and 95\% confidence intervals were calculated
from a normal approximation using the bootstrap standard deviation. The four direction-specific estimators (IVW, MR-Egger, MR-RAPS and the
weighted median) were fitted separately for each exposure--outcome
ordering. Exposure-associated variants with $p<10^{-5}$ were clumped
against the European reference ($r^2=0.001$, 10,000 kb), followed by
allele harmonisation. MR-RAPS used its default settings
(over-dispersion, Huber loss and no shrinkage).
Table~\ref{tab:supp_real_data_results} provides the estimates and
corresponding inference for all six directions.

\begin{landscape}
\begin{table}[p]
\centering
\caption{Complete estimation results in the three real-data applications.}
\label{tab:supp_real_data_results}
\footnotesize
\renewcommand{\arraystretch}{0.94}
\setlength{\tabcolsep}{2.3pt}
\begin{threeparttable}
\begin{tabular*}{\linewidth}{@{\extracolsep{\fill}}lllrrrrc}
\toprule
Pair & Direction & Method & Variants & Estimate & SE & $p$-value & 95\% CI \\
\midrule
LDL-C--CHD & LDL-C \(\rightarrow\) CHD & H-JADE & 689,477 & 0.458 & 0.087 & $1.61\times 10^{-7}$ & \([0.287,0.629]\) \\
 &  & IVW & 302 & 0.511 & 0.059 & $<10^{-16}$ & \([0.396,0.626]\) \\
 &  & MR-Egger & 302 & 0.757 & 0.080 & $<10^{-16}$ & \([0.600,0.914]\) \\
 &  & MR-RAPS & 302 & 0.574 & 0.0469 & $<10^{-16}$ & \([0.482,0.666]\) \\
 &  & Weighted median & 302 & 0.618 & 0.062 & $<10^{-16}$ & \([0.495,0.740]\) \\
\cmidrule(lr){2-8}
 & CHD \(\rightarrow\) LDL-C & H-JADE & 689,477 & -0.0257 & 0.0091 & 0.00481 & \([-0.0435,-0.0078]\) \\
 &  & IVW & 86 & 0.093 & 0.0431 & 0.0311 & \([0.0084,0.177]\) \\
 &  & MR-Egger & 86 & 0.147 & 0.081 & 0.0732 & \([-0.0118,0.306]\) \\
 &  & MR-RAPS & 86 & 0.0172 & 0.0152 & 0.256 & \([-0.0125,0.0470]\) \\
 &  & Weighted median & 86 & -0.0144 & 0.0053 & 0.00656 & \([-0.0248,-0.0040]\) \\
\midrule
ApoB--CHD & ApoB \(\rightarrow\) CHD & H-JADE & 689,374 & 0.338 & 0.078 & $1.52\times 10^{-5}$ & \([0.185,0.491]\) \\
 &  & IVW & 337 & 0.522 & 0.052 & $<10^{-16}$ & \([0.421,0.623]\) \\
 &  & MR-Egger & 337 & 0.634 & 0.070 & $<10^{-16}$ & \([0.496,0.771]\) \\
 &  & MR-RAPS & 337 & 0.532 & 0.0396 & $<10^{-16}$ & \([0.455,0.610]\) \\
 &  & Weighted median & 337 & 0.538 & 0.063 & $<10^{-16}$ & \([0.415,0.660]\) \\
\cmidrule(lr){2-8}
 & CHD \(\rightarrow\) ApoB & H-JADE & 689,374 & -0.0003 & 0.0107 & 0.978 & \([-0.0212,0.0206]\) \\
 &  & IVW & 86 & 0.121 & 0.055 & 0.0288 & \([0.0125,0.230]\) \\
 &  & MR-Egger & 86 & 0.178 & 0.104 & 0.0905 & \([-0.0258,0.383]\) \\
 &  & MR-RAPS & 86 & 0.0202 & 0.0171 & 0.238 & \([-0.0134,0.054]\) \\
 &  & Weighted median & 86 & -0.0126 & 0.0053 & 0.0173 & \([-0.0229,-0.0022]\) \\
\midrule
TG--T2D & TG \(\rightarrow\) T2D & H-JADE & 689,563 & 0.074 & 0.133 & 0.577 & \([-0.186,0.334]\) \\
 &  & IVW & 490 & 0.341 & 0.0462 & $1.54\times 10^{-13}$ & \([0.251,0.432]\) \\
 &  & MR-Egger & 490 & -0.077 & 0.062 & 0.218 & \([-0.199,0.0452]\) \\
 &  & MR-RAPS & 490 & 0.391 & 0.0400 & $<10^{-16}$ & \([0.312,0.469]\) \\
 &  & Weighted median & 490 & 0.182 & 0.052 & $4.83\times 10^{-4}$ & \([0.080,0.284]\) \\
\cmidrule(lr){2-8}
 & T2D \(\rightarrow\) TG & H-JADE & 689,563 & 0.088 & 0.0128 & $5.87\times 10^{-12}$ & \([0.063,0.113]\) \\
 &  & IVW & 177 & 0.0390 & 0.0128 & 0.00237 & \([0.0139,0.064]\) \\
 &  & MR-Egger & 177 & 0.0169 & 0.0276 & 0.540 & \([-0.0371,0.071]\) \\
 &  & MR-RAPS & 177 & 0.0428 & 0.0028 & $<10^{-16}$ & \([0.0372,0.0483]\) \\
 &  & Weighted median & 177 & 0.0282 & 0.0058 & $9.51\times 10^{-7}$ & \([0.0169,0.0395]\) \\
\bottomrule
\end{tabular*}
\begin{tablenotes}
\scriptsize
\item ``Variants'' denotes the common harmonised analysis panel for H-JADE
and the direction-specific set of selected, clumped and harmonised
instruments for the other methods. Reported $p$-values below $10^{-16}$
are displayed as $<10^{-16}$. 
\end{tablenotes}
\end{threeparttable}
\renewcommand{\arraystretch}{1}
\end{table}
\end{landscape}

\setlength{\textfloatsep}{10pt plus 2pt minus 2pt}
\setlength{\intextsep}{10pt plus 2pt minus 2pt}
\subsection{Focusing comparison}
\label{supp:liye_realdata_comparison}

We applied the Fo-IVW and Fo-Median tests using the implementation of
\citet{LiYe2025}. For each exposure--outcome ordering, variants with
exposure-association $p$-values below 0.05 were entered into PLINK
clumping (10,000 kb; $r^2=0.001$), with clump index variants required
to have $p\le 10^{-5}$. The candidate set comprised the index SNPs
with available, allele-harmonised outcome associations. Outcome data
were matched by SNP identifier, without proxies.

Let $n_{\mathrm{cand}}$ denote the number of candidates for a given
direction. Following the published implementation, the
strong-instrument threshold was
$\tau_s=\Phi^{-1}(1-n_{\mathrm{cand}}^{-1})$.
We used $\tau_f\in\{1.2,1.5\}$ for Fo-IVW and $\tau_f=1.5$ for
Fo-Median. For the latter, we set \texttt{weight.order = -1} and used
10,000 null bootstrap replicates. The focused sets were determined by the exposure- and
outcome-association thresholds described in
Supplementary Section~\ref{supp:focusing_comparison}.
Table~\ref{tab:supp_focus_primary} gives the corresponding set sizes
and test $p$-values.

\begin{table}[!htbp]
\centering
\caption{Direction-specific focusing tests in the real-data analyses.}
\label{tab:supp_focus_primary}
\footnotesize
\setlength{\tabcolsep}{2.7pt}
\begin{threeparttable}
\begin{tabular*}{\textwidth}{@{\extracolsep{\fill}}lrrrrrr}
\toprule
 & & \multicolumn{2}{c}{Focused SNPs} & \multicolumn{3}{c}{Test $p$-values} \\
\cmidrule(lr){3-4}\cmidrule(lr){5-7}
Direction & Candidates & $\tau_f=1.2$ & $\tau_f=1.5$ &
Fo-IVW (1.2) & Fo-IVW (1.5) & Fo-Median (1.5) \\
\midrule
LDL-C $\rightarrow$ CHD & 296 & 166 & 198 & 0.183 & 0.0352 & 0.249 \\
CHD $\rightarrow$ LDL-C & 81 & 38 & 46 & 0.642 & 0.101 & 0.262 \\
ApoB $\rightarrow$ CHD & 317 & 191 & 223 & 0.0166 & $7.98\times10^{-4}$ & 0.00100 \\
CHD $\rightarrow$ ApoB & 81 & 39 & 50 & 0.261 & 0.168 & 0.239 \\
TG $\rightarrow$ T2D & 475 & 244 & 289 & 0.0453 & $4.20\times10^{-5}$ & $<10^{-4}$ \\
T2D $\rightarrow$ TG & 167 & 76 & 87 & 0.121 & 0.0927 & 0.0447 \\
\bottomrule
\end{tabular*}
\begin{tablenotes}
\scriptsize
\item ``Candidates'' counts harmonised clump index SNPs before focusing;
``Focused SNPs'' counts those satisfying both selection thresholds.
Parentheses in method headings indicate $\tau_f$. For Fo-Median,
$<10^{-4}$ denotes zero exceedances in 10,000 bootstrap draws.
\end{tablenotes}
\end{threeparttable}
\end{table}

\end{document}